\pdfoutput=1 % Needed for arxive
\documentclass[11pt]{article}

\usepackage[utf8]{inputenc}
\usepackage[margin=1in]{geometry}
\usepackage{physics, amsmath, amssymb, amsthm}
\usepackage{bm}
\usepackage{hyperref}
\usepackage[noabbrev, nameinlink, capitalise]{cleveref}
\usepackage{graphicx}
\usepackage{authblk}  % Optional: Remove for the submission to avoid author info
\usepackage{stmaryrd}
\usepackage{comment}
\usepackage{tikz}
\usepackage{pgfplots}
\pgfplotsset{compat=1.9}
\usetikzlibrary{shapes,arrows,patterns,positioning,shapes.geometric,arrows.meta,decorations.markings}
\usepackage{definitions}

\newcommand{\jk}[1]{\textcolor{cyan}{JK:~#1}}
\newcommand{\ac}[1]{\textcolor{blue}{AC:~#1}}
\newcommand{\pg}[1]{\textcolor{darkred}{PG:~#1}}

\title{Quasi-optimal sampling from Gibbs states \\via non-commutative optimal transport metrics}

\date{\today} % Omitting the date
\author[1,2]{{\'A}ngela Capel\thanks{ac2722@cam.ac.uk}}
\author[1]{Paul Gondolf\thanks{paul.gondolf@uni-tuebingen.de}}
\author[3]{Jan Kochanowski\thanks{jan.kochanowski@inria.fr}}
\author[3]{Cambyse Rouz\'{e}\thanks{rouzecambyse@gmail.com \\  authors listed in alphabetical order}}

\affil[1]{Fachbereich Mathematik, Universit\"{a}t T\"{u}bingen, 72076 T\"{u}bingen, Germany}
\affil[2]{Dep. of App. Mathematics and Theo. Physics, University of Cambridge, United Kingdom}
\affil[3]{Inria, Télécom Paris - LTCI, Institut Polytechnique de Paris, 91120 Palaiseau, France}

\begin{document}
\maketitle

\begin{abstract}
    We study the problem of sampling from and preparing quantum Gibbs states of local commuting Hamiltonians on hypercubic lattices of arbitrary dimension. We prove that any such Gibbs state which satisfies a clustering condition that we coin decay of matrix-valued quantum conditional mutual information (MCMI) can be quasi-optimally prepared on a quantum computer. We do this by controlling the mixing time of the corresponding Davies evolution in a normalized quantum Wasserstein distance of order one. To the best of our knowledge, this is the first time that such a non-commutative transport metric has been used in the study of quantum dynamics, and the first time quasi-rapid mixing is implied by solely an explicit clustering condition. Our result is based on a weak approximate tensorization and a weak modified logarithmic Sobolev inequality for such systems, as well as a new general weak transport cost inequality. If we furthermore assume a constraint on the local gap of the thermalising dynamics, we obtain rapid mixing in trace distance for interactions beyond the range of two, thereby extending the state-of-the-art results that only cover the nearest neighbour case. We conclude by showing that systems that admit effective local Hamiltonians, like quantum CSS codes at high temperature, satisfy this MCMI decay and can thus be efficiently prepared and sampled from.
\end{abstract}
\thispagestyle{empty}
% Start a new page for the Table of Contents
\newpage
\tableofcontents
\thispagestyle{empty}
\newpage
\addtocounter{page}{-2}
\section{Introduction}\label{sec:introduction}

The problem of Gibbs state preparation is fundamental in statistical mechanics. Given a Hamiltonian $H$ on a quantum system and an inverse temperature $\beta< \infty$, the Gibbs state $\sigma^\beta = e^{-\beta H} / \Tr[e^{-\beta H}]$ describes the properties of the system at its thermal equilibrium \cite{Alhambra.2023}. Thus, Gibbs states and their fundamental properties are essential for the study of quantum systems, for example in the contexts of simulation of many-body systems \cite{Molnar.2015}, their use as topological quantum memories \cite{LandonCardinal.2013} or their thermalization processes \cite{Riera.2012,Mueller.2015}. 

Gibbs sampling has been for a long time a cornerstone in statistical mechanics. It is an extremely relevant problem for the fields of physics and computer science, both in the context of classical and quantum mechanics, in part due to its applications in multiple scenarios \cite{Apeldoorn.2020,Hinton.1983}. Classical Markov Chain Monte Carlo (MCMC) methods constitute one of the canonical tools for sampling from classical Gibbs states of spin systems \cite{Levin.2008}. These methods are shown to be efficient at high-enough temperatures \cite{Martinelli.1999} and are generally believed to be efficient in practice \cite{Brooks.2011}. However, there is no clear consensus on how to best extend such algorithms to quantum Gibbs sampling. Many relevant classical algorithms for Gibbs sampling have appeared in the past few years \cite{Chen.2024a, Anari.2021, Anari.2021a, Alaoui.2022}. Here we take a further step and aim at proving the efficiency of quantum extensions thereof.

A good Gibbs sampler is expected to prepare the Gibbs state in polynomial time. In the past few years, many quantum algorithms inspired by the classical Monte Carlo have been proposed in \cite{Temme.2011,Rall.2023,Wocjan.2023}, among others, but for long without any provable guarantee or only validated under strong theoretical assumptions \cite{Shtanko.2021,Chen.2021}. This has drastically changed with the very recent appearance of \cite{Chen.2023a,Chen.2023}, where a quantum algorithm to prepare quantum Gibbs states was proposed and subsequently shown to be efficient in \cite{Rouze.2024b}. This has inspired an important collection of works in the context of quantum Gibbs sampling, with algorithms not only based on dissipation \cite{Chen.2023,Gilyen.2024,Bardet.2024,Rall.2023,Ding.2024}, but also in some other techniques \cite{Gilyen.2019,Jiang.2024, Kashyap.2024}. A detailed discussion on some relevant literature for quantum Gibbs sampling is presented in \cref{sec:a-comparison-with-classical/quantum-Gibbs-sampling}.

In this paper, we investigate quantum Gibbs sampling with Davies generators \cite{Davies.1979,Davies.1976} associated with local commuting Hamiltonians. Local commuting Hamiltonians hereby constitute a class of systems that non-trivially extends beyond the classical regime \cite{Bostanci.2024, Hwang.2024, Aharonov.2018, Aharonov.2011} and include highly entangled systems like CSS codes (e.g. the Toric code) and quantum double models \cite{Kitaev.2003,Calderbank.1996,Steane.1996}. Davies Lindbladians constitute the canonical objects to model the Markovian dissipation of a quantum system weakly coupled with an infinite-dimensional thermal bath and can be regarded as the natural quantum analogue of Glauber dynamics  \cite{Martinelli.1999}. Hence they represent a natural tool for quantum Gibbs sampling. Many recent efforts to develop efficient Gibbs samplers have been done in the context of commuting Hamiltonians, such as for 1D translation-invariant systems at any positive temperature \cite{Bardet.2023,Bardet.2024, Kochanowski.2024}, and high-dimensional systems at high-enough temperature \cite{Capel.2020, Kochanowski.2024}. In high dimensions at low temperature, efficient Gibbs samplers are only known to exist for Kitaev's quantum double models in 2D \cite{Alicki.2009,Ding.2024a,Lucia.2023,Komar.2016}. 

The efficiency of Gibbs samplers is determined by the speed of convergence, or mixing, of the corresponding Lindbladian towards its thermal equilibrium, estimated through the notion of \textit{mixing time}. Generically the algorithmic sample complexity can be upper bounded by the system size times its mixing time  \cite{Li.2022,Rall.2023}. Both classically and quantumly, the most frequent way of estimating this mixing time is through the spectral gap of the Lindbladian \cite{Bardet.2017,Kastoryano.2016}. A positive spectral gap provides an upper bound on the mixing time that scales linearly with the system size, in a regime known in the literature as \textit{fast mixing}, or \textit{poly-time mixing} \cite{Hwang.2024}, but this is generally believed to be an overestimation. A more accurate estimate is provided by the existence of a strictly positive (or in the system size polylogarithmically-decaying) modified logarithmic Sobolev inequality \cite{Kastoryano.2013}, which yields a mixing time that scales polylogarithmically on the system size, in a regime known as \textit{rapid mixing}. In this regime the algorithmic sample complexity scales, hence, linearly in systems size up to a polylogarithmic correction, which is usually referred to as \emph{optimal sampling}. Whereas some quantum Gibbs samplers associated with non-commuting Hamiltonians have been recently proven to have fast mixing \cite{Rouze.2024b}, the only examples of rapid mixing can be found in the context of commuting Hamiltonians \cite{Capel.2020,Bardet.2023,Bardet.2024,Kochanowski.2024}, as far as we are aware. This is the reason for our focus on the latter in the current manuscript. 

\section{Main results}\label{sec:main-results}
\begin{figure}[h!]
    \centering
    \begin{tikzpicture}[font=\scriptsize, scale=.9]
        \def\toolcolor{SlateBlueDark}
        \def\assumptioncolor{GoldenrodDark}
        \def\resultcolor{TealDark}

        % Draw assumption boxes
        \boxElement{3.55}{6.05}{7.7}{2.45}{\toolcolor}{MCMI: \eqref{eq:definition-mcmi} }{ Matrix-valued \\ quantum Con-\\ ditional Mutual \\ Information \\ $H_\sigma(A:C|D)$
        };
        
        \boxElementAlt{6.53}{6.12}{4.65}{2.3}{\assumptioncolor}{Unif. MCMI-decay: \ref{defi:definition-mcmi}}{ \cref{thm:uniform-decay-of-the-mcmi}: Marginal commuting systems at high $T$. \\
        \cref{subsec:mcmi-decay-from-complete-analyticity-in-classical-systems}: Classically implied by \textit{complete}\\ \textit{analyticity} \cite{Dobrushin.1985}.
        };

        \boxElement{12}{6.1}{4}{2.3}{\assumptioncolor}{Poly local gap: \ref{subsec:weak-mlsi-for-the-global-davies-semigroup}}{$\lambda(\cL^D_A)=\Omega(|A|^{-\mu})$\\
        We expect it to hold at uniform high $T$ \cite{Kastoryano.2016}.
        };
        
        % DRAW TOOL BOXES
        \boxElement{10.5}{4.5}{3.2}{1.2}{\toolcolor}{Coarse-graining}{\cref{subsec:entropy-factorization-and-weak-at}, \\\cref{subsec:a-coarse-graining-of-the-hypercubic-lattice}
        };

        \boxElement{7}{4.5}{2.6}{1.2}{\toolcolor}{Weak AT}{\cref{thm:weakAT}, \\ \cref{thm:a-weak-approximate-tensorization-for-davies-channels}
        };

        \boxElement{2}{4.5}{4.2}{1.2}{\toolcolor}{Exp almost cMLSI}{\cref{lem:exponential-almostcmlsi},\\ \cref{lem:extended:relation-davies-at-finite-and-infinite-temperature}, \cref{lem:extended:mlsi-alike-inequality-for-infinite-temperature-local-davies}
        };
        
        \boxElement{4.5}{2.25}{3.5}{1.2}{\toolcolor}{Exp weak MLSI}{\cref{cor:quasipolynomial-weak-mlsi}, \\ \cref{thm:extended:weak-mlsi}
        };

        \boxElement{12}{2.25}{4}{1.2}{\toolcolor}{Poly weak MLSI}{\cref{cor:polylogarithmic-weak-mlsi}, \\ \cref{thm:extended:wmlsi-under-gap}
        };
        
        \boxElement{8.5}{2.25}{3}{1.2}{\toolcolor}{TC-inequality}{\cref{thm:weak-transport-cost-inequality}, \\ \cref{thm:extended:weak-transport-cost-inequality}
        };

        % DRAW RESULT BOXES
        \boxElement{6}{0}{5}{1.3}{\resultcolor}{Quasi-rapid $W_1$ mixing}{Main result: \cref{thm:wassersteinmixing}, see also \cref{thm:detailed:wassersteinmixing}
        };

        \boxElement{-0.2}{0}{5}{1.3}{\resultcolor}{Efficient Gibbs Sampler}{Main result: \cref{thm:main-sampling-result}, \\ see also \cref{thm:quasi-optimal-sampling-from-Gibbs-states-that-satisfy-MCMI-decay}. 
        };
        
        \boxElement{12}{0}{4}{1.3}{\resultcolor}{Rapid mixing}{ \cref{thm:rapidmixing}, see also \cref{thm:extended:rapid-mixing}};

        % LEGEND
        \boxElement{-1}{7.5}{2.8}{0.75}{\assumptioncolor}{Assumption}{Satisfied if...};
        \boxElement{-1}{6.5}{2.8}{0.75}{\toolcolor}{Tool}{};
        \boxElement{-1}{5.5}{2.8}{0.75}{\resultcolor}{Result}

        % ARROWS ZEROTH LEVEL
        \draw[\resultcolor, thick, ->] (6, 0.8) -- (4.8, 0.8);
        \draw[\resultcolor, thick] (12, 0.45) -- (11.5, 0.45);
        \draw[\resultcolor, thick, dotted] (11.5, 0.45) -- (11, 0.45);
        \draw[\resultcolor, thick, ->] (5.5, 0.45) -- (4.8, 0.45);
        \draw[\resultcolor, thick, dotted] (6, 0.45) -- (5, 0.45);

        % ARROWS FIRST LEVEL
        \draw[\toolcolor, thick, ->] (8.25, 1.95) -- (8.25, 1.5);
        \draw[\toolcolor, thick] (6.9, 2.25) -- (8.25, 1.95);
        \draw[\toolcolor, thick, plus arrow] (9.6, 2.25) --(8.25, 1.95);
        
        \draw[\toolcolor, thick, ->] (14, 2.25) -- (14, 1.5);

         % ARROWS SECOND LEVEL
        \draw[\toolcolor, thick, ->] (6.25, 4.15) -- (6.25, 3.7);
        \draw[\toolcolor, thick] (4.65, 4.5) -- (6.25, 4.15);
        \draw[\toolcolor, thick, plus arrow] (8.05, 4.5) -- (6.25, 4.15);

        \draw[\toolcolor, thick, ->] (8.45, 4.5) -- (9.6, 3.6);

        \draw[\toolcolor, thick, -] (9.4, 4.55) -- (14, 4.1);
        \draw[\toolcolor, thick, ->] (14, 4.1) -- (14, 3.6);
        
        % Horizontal arrow second level
        \draw[\toolcolor, thick, ->] (10.5, 5.25) -- (9.6, 5.25);

        % ARROWS THIRD LEVEL
        \draw[\assumptioncolor, thick, plus arrow] (10.1, 6.12) -- (10.1, 5.25);
        
        \draw[\assumptioncolor, thick, plus arrow] (14, 6.1) -- (14, 4.1);
    \end{tikzpicture}
    \caption{Logical structure of the results in this article. In orange, we represent the assumptions considered; in purple, the main technical lemmas, of independent interest; and in green, the main technical results regarding the mixing times of the evolution and their application in the context of Gibbs sampling. Note that the main result is implied by either the quasi-rapid $W_1$ mixing, or the rapid mixing.}
    \label{fig:paper-structure}
\end{figure}
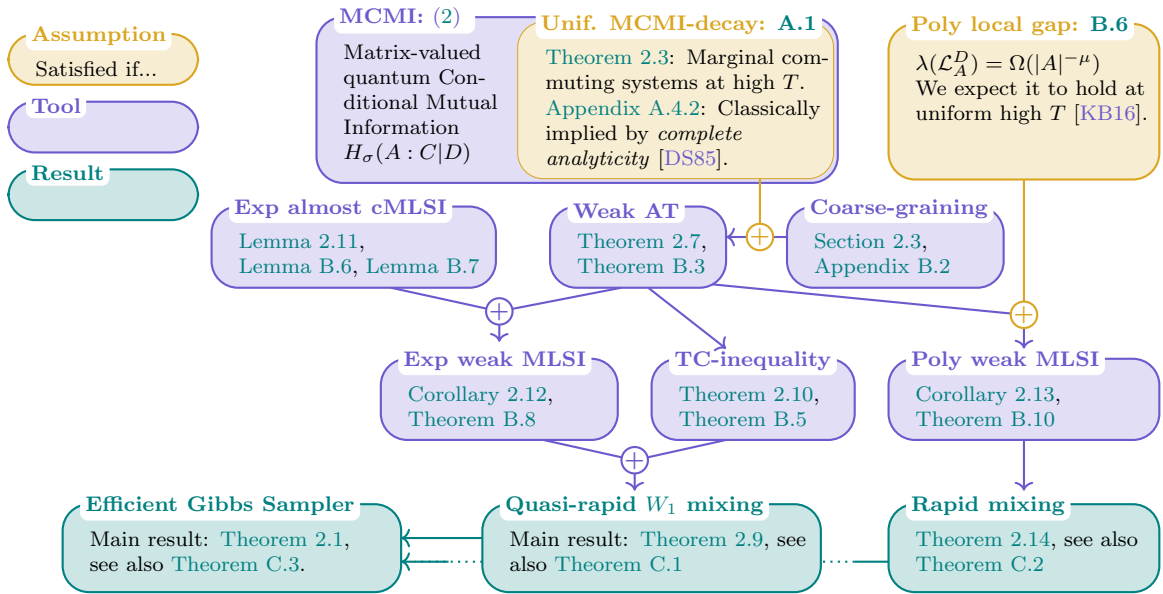

In this work, we demonstrate the quasi-rapid mixing in normalized Wasserstein distance of Davies generators associated with local commuting Hamiltonians, assuming that their Gibbs state satisfies a specific clustering condition. By additionally assuming a polynomially decaying gap of the local Davies generator, we strengthen our result to rapid mixing in trace distance.

To our knowledge, this is the first instance where quasi-rapid mixing is derived solely from a static and explicit notion of decay of correlations in the steady-state or Gibbs state. Furthermore, while mixing times based on normalized Wasserstein distances have been studied in the classical setting \cite{Alaoui.2022}, we believe this is the first time that mixing with respect to normalized quantum Wasserstein distances has been considered in the quantum setting, thereby further connecting quantum optimal transport theory and Gibbs sampling \cite{DePalma.2022,Rouze.2019, DePalma.2021, Depalma.2024, Beatty.2024}.

We introduce the \textit{matrix-valued quantum conditional mutual information (MCMI)} in \cref{def:mcmi-main}, whose decay is more general than that of both the covariance and the conditional mutual information—the former shown to be sufficient for rapid thermalization in one-dimensional lattices \cite{Kochanowski.2024}. Our first main result requires only the uniform decay of the MCMI across arbitrary 4-partitions of the underlying lattice:

\begin{thm}[Quasi-optimal preparation of MCMI-decaying Gibbs states, informal]\label{thm:main-sampling-result}
    Let $\sigma$ be a Gibbs state of a local commuting Hamiltonian that satisfies uniform exponential decay of its MCMI. Then there exists a quantum circuit with circuit complexity and runtime
    $$\cO\left(N,\text{quasi-log}(N), \textup{quasi-poly}\left(\frac{1}{\epsilon}\right)\right) = o(N^2)$$
    that prepares a state $\rho$ that is $\epsilon$-close in normalized $W_1$ distance to $\sigma$.
\end{thm}

To establish the quasi-rapid mixing time and achieve efficient sampling, we prove a suitable \emph{weak modified logarithmic Sobolev inequality} (weak MLSI) for the Davies evolution corresponding to a local commuting Hamiltonian (see \cref{cor:quasipolynomial-weak-mlsi}). This yields exponential decay of the relative entropy between any time-evolved initial state and the Gibbs state, with a prefactor depending polynomially on the system size. Following recent works \cite{Capel.2020, Bardet.2024, Kochanowski.2024}, we employ a ‘divide and conquer’ strategy, using uniform decay of the MCMI for the global Gibbs state and leveraging the locality of the Davies generators to estimate global mixing in terms of local mixings plus an additive error term.

More precisely we employ the uniform decay of the MCMI to prove a general entropy factorization (\cref{lem:weak-approximate-tensorization}), which we extend to an approximate tensorization on $\mathbb{Z}^D$ (\cref{thm:weakAT}). This result reduces the relative entropy of two states supported on the global lattice to a sum of conditional relative entropies in smaller subregions, with an additive error controlled by the decay of the MCMI. This is the crucial ingredient to prove the weak MLSI (\cref{cor:quasipolynomial-weak-mlsi}). Finally, to apply this weak MLSI to our Wasserstein distance-based mixing time, we derive a general \emph{weak transport cost} inequality (\cref{thm:weak-transport-cost-inequality}).

Notably, unlike previous results on MLSIs of Davies dynamics in many-body quantum spin systems \cite{Bardet.2023, Bardet.2024, Kochanowski.2024}, our main result (\cref{thm:wassersteinmixing}, see also \cref{thm:detailed:wassersteinmixing}) shows that quasi-rapid mixing is implied solely by the decay of the MCMI, without additional assumptions on quantities like the local gap. This is significant because it is the first time that a dynamical property as strong as quasi-rapid mixing can be derived for physically relevant quantum spin systems solely from a static condition on the system’s equilibrium (the Gibbs state), without further weaker dynamical assumptions.

Moreover, under further assumptions—specifically, an at most polynomially decaying gap of the local Davies generators, which we expect to hold at uniformly high-temperature %due to \cite{Kastoryano.2016}
—we also show rapid mixing of these dynamics with respect to the trace distance (see \cref{thm:rapidmixing}).

In the following sections, we provide a detailed overview of these results and the required lemmas, while omitting most technical details and proofs. A formal introduction to the necessary mathematical concepts, along with the full proofs, can be found in \cref{sec:theoretical-framework}, \cref{sec:proof-techniques-extended}, and \cref{sec:main-results-extended}. We conclude with \cref{sec:examples}, where we present examples of systems for which the required decay of correlations—specifically, the decay of the MCMI—holds, making our main results applicable.

For the readers' convenience, we collect all main results, important technical lemmas and assumptions in \cref{fig:paper-structure}.

\subsection{Notation and preliminaries}
This work considers quantum spin systems on finite hypercubic lattices $\Lambda\subset\Z^D$. The finite-dimensional Hilbert space of said systems is $\cH_\Lambda:=\bigotimes_{x\in\Lambda}\cH_x$, where each site contains a qudit $\cH_x\simeq\C^d$. The operator norm on the set of bounded operators $\mathcal{B}(\cH)$ is denoted as $\|\cdot\|_\infty$ and the trace norm as $\|\cdot\|_1$. The set of all quantum states, i.e. non-negative, trace-normalized operators on $\cH$, is denoted as $\cS(\cH)$. The partial trace is $\tr_{A}:\cB(\cH_\Lambda)\to \cB(\cH_{\bar{A}})$, where $\bar{A}:=\Lambda\setminus A$ is the complement region of $A\subset \Lambda$. A $(\kappa,r)-$local, commuting, $J-$bounded Hamiltonian $H_\Lambda\in\cB(\cH)$ is a self-adjoint operator $H_\Lambda=\sum_{A\subseteq\Lambda}h_A$, such that $\norm{h_A}_\infty\leq J$, $h_A=0$, whenever $|A|>\kappa,$ or $\diam(A)>r$, and for each $A,B\subset \Lambda$, $[h_A,h_B]=0$, where $[X,Y]:=XY - YX$ is the commutator of two operators $X, Y \in \cB(\cH)$. We define the boundary of a sublattice $R \subseteq \Lambda$, tied to the connectivity of the Hamiltonian as $\partial R := \{k \in \Lambda : \exists A \subseteq \Lambda, h_A \ne 0, A \cap R \ne \emptyset, k \in A\}$ and write $R\partial := R \cup \partial R$. For a sublattice $R\subseteq\Lambda$ we define the \emph{local Hamiltonian} as $H_R:=\sum_{A\subseteq R} h_A$ and set the \emph{growth constant} $g$ of the system $(\Lambda,H_\Lambda)$ as  $g:= \max_{k \in \Lambda} |\{A \subseteq \Lambda : h_A \ne 0, k \in A\}|$. Hence it satisfies $\norm{H_R}_\infty\leq gJ|R|$. The local, resp. global Gibbs state at inverse temperature $\beta>0$, $R\subset\Lambda$, resp. $R = \Lambda$ is given by $\sigma^R:=\frac{e^{-\beta H_R}}{\Tr[e^{-\beta H_R}]}$ and its reduced state is defined as $\sigma_R:=\tr_{\bar{R}}[\sigma^\Lambda]$. The relative entropy between a state $\rho$ and a full rank state $\sigma>0$ is defined as $D(\rho\|\sigma) := \tr[\rho(\log\rho-\log\sigma)]$. The generator of the semigroup dynamic we are going to investigate in this paper is called the Davies semigroup, which for a fixed local commuting Hamiltonian $H_\Lambda$ and inverse temperature $\beta$ is given by 
\begin{equation}\label{eq:definition-davies-generator}
    \cL_\Lambda^D(\rho) := \sum_{k \in \Lambda} \cL_k^D 
    \quad \text{with} \quad
    \cL_k^D(\rho) := \sum\limits_{\alpha, \omega} \chi^{\beta,\omega}_{\alpha,k}\,\Big(  S_{\alpha,k}^{\omega}\rho S_{\alpha,k}^{\omega,\dagger}-\frac{1}{2}\,\big\{ \rho, S_{\alpha,k}^{\omega,\dagger}S_{\alpha,k}^{\omega} \big\} \Big) \, . 
\end{equation}
The $S^\omega_{\alpha, k}$ depend on $H_{\Lambda}$ through $e^{itH_{\Lambda}} S_{\alpha, k} e^{-itH_\Lambda} = \sum_{\omega} e^{it\omega} S^{\omega}_{\alpha, k}$ for all $t \in \R$, where $\{S_{\alpha, k}\}_{\alpha}$ labels a set of self-adjoint operators supported on $k$ that form a Kraus-decomposition of the partial trace $\tr_k$. The prefactors $\chi^{\beta,\omega}_{\alpha, k}$ satisfy the KMS condition $\chi_{\alpha,k}^{\beta,-\omega}=e^{-\beta\omega}\,\chi_{\alpha,k}^{\beta,\omega}$ and we assume them to be uniformly bounded from above and below as $0 < \chi_{\min}^\beta\le \chi_{\alpha,k}^{\beta,\omega}\le \chi_{\max}^\beta$. We define the local Davies generators as $\cL_A^D := \sum_{k \in A} \cL_k^D$ and note that for all $A \subseteq \Lambda$ the generated dynamics converge to projecting channels, we denote by $E_A = \lim\limits_{t \to \infty} e^{t \cL_A}$.
 
\subsection{Uniform decay of the MCMI}\label{sec:MCMI}

In classical spin systems, there is a strong connection between the decay of correlations in the thermal state and the mixing time of the thermalizing dynamics \cite{Martinelli.1999}. Specifically, Glauber dynamics, which models the thermalization of a discrete spin system, rapidly approaches the equilibrium Gibbs measure if, and only if, correlations between spatially separated regions decay exponentially with the distance between them. A similar connection has been established in the quantum case, where notions of decay of correlations have been shown to be closely related to mixing \cite{Bluhm.2024, Capel.2020, Kochanowski.2024}. This work builds on these previous results by introducing the \emph{matrix-valued quantum conditional mutual information (MCMI)} as a correlation measure closely tied to the mixing properties of the so-called Davies dynamics. This quantity, previously considered in the study of quantum many-body systems \cite{Kuwahara.2020}, is closely related to quantum conditional mutual information and the mixing condition studied in \cite{Bardet.2024, Bluhm.2022,Bluhm.2024}, where the mixing property was identified as a condition to imply rapid mixing of Davies dynamics on 1D lattices.

\begin{defi}[Matrix valued quantum conditional mutual information (MCMI)]\label{def:mcmi-main} 
    Given a quantum state $\sigma\in \cS(\cH_{ABCD})$ on some tensor product Hilbert space $\cH_{ABCD}=\cH_A\otimes\cH_B\otimes\cH_C\otimes\cH_D$, then its MCMI is defined as
    \begin{align} \label{eq:definition-mcmi}
        \mathbf{H}_\sigma(A:C|D) := \log\sigma_{ACD}+\log\sigma_D-\log\sigma_{AD}-\log\sigma_{CD},
    \end{align} and we will frequently write $H_\sigma(A:C|D):=\norm{\mathbf{H}_\sigma(A:C|D)}_\infty$ for its operator norm.
\end{defi} \noindent 
It is a general notion of mixing that controls both:
\begin{enumerate}
    \item the conditional mutual information  $I_\sigma(A:C|D) := \tr[\sigma \mathbf{H}_\sigma(A:C|D)] \leq H_\sigma(A:C|D)$;
    \item the mutual information and thus also covariance $I_\sigma(A:C):=\tr[\sigma_{AC}\mathbf{H}_\sigma(A:C|\emptyset)]\leq H_\sigma(A:C|\emptyset)$
\end{enumerate}
For $\sigma$ a Gibbs state of a local Hamiltonian at inverse temperature $\beta$, the norm of the MCMI is believed to decay exponentially in $\dist(A,C)$ at high-enough temperature. This was, erroneously,  thought proven in \cite{Kuwahara.2020} for some time. Although we do not resolve this open question, we prove that the conjecture holds for \emph{marginal commuting} Hamiltonians: We say that $H_\Lambda = \sum_{A \subseteq \Lambda} h_A$ is \emph{marginal commuting} (see \cite[Definition 3.5]{Bluhm.2024}) if the algebra $\cA$ generated by $\{h_A\}_{A \subset \Lambda}$ is commuting and closed under arbitrary partial traces, i.e. for all $A \subseteq \Lambda$, $\Id_A \otimes \tr_A[\cA] \subseteq \cA$. This condition, at high-enough temperature, is a particular case of applicability of the following theorem:
\begin{thm}[Uniform decay of the MCMI]\label{thm:uniform-decay-of-the-mcmi}
    Let $\sigma$ be the Gibbs state to some $(\kappa,r)$-local and $J$-bounded Hamiltonian on a graph $\Lambda$ that has growth constant $g$, then there exist some constants $K,\xi>0$ such that for any partition $\Lambda=A\sqcup B\sqcup C\sqcup D$
    \begin{align} \label{eq:decay-mcmi}
        H_\sigma(A:C|D)\leq K|\Lambda|\exp(-\dist(A,C)/\xi),
    \end{align} if $H_\Lambda$ admits a strong local effective Hamiltonian in the sense of \cref{subsec:the-effective-hamiltonian-as-tool-to-proof-correlation-decay}, see also Definition 3.1 of \cite{Bluhm.2024}.
    In this case, we say that the state $\sigma$ satisfies \emph{uniform decay of the MCMI}.
\end{thm}
\begin{rmk}
In \cite{Bluhm.2024}, it was shown that all marginal commuting systems at suitably high temperature satisfy this condition and hence, by the theorem above, also uniform MCMI decay, see 
\cref{subsec:mcmi-decay-from-commutings-marginals-at-high-temperature} and \cref{thm:MCMI-ecay-for-marginal-commuting-systems-at-high-temperature}. These notably include all quantum CSS codes, and all classical Hamiltonians with a Gibbs state that satisfies complete analyticity \cite{Cesi.2001, Dobrushin.1985, Dobrushin.1987}, see \cref{subsec:mcmi-decay-from-complete-analyticity-in-classical-systems}.    
\end{rmk}

\subsection{Entropy factorization and a weak AT}\label{subsec:entropy-factorization-and-weak-at}
Statements on (strong) approximate tensorization (AT)  relate conditional relative entropies on lattices to ones on subsystems with multiplicative error terms that quantify the closeness of the fixed point to being a product state. Such statements are central in both the classical and quantum setting when proving MLSIs and efficient bounds on mixing times of certain dynamics. The issue, however, is that, unlike in the classical setting, so far no general iterable approximate tensorization statement is known in the quantum setting. The ones that are known are sufficient when considering the fixed point to be a tensor product \cite{Capel.2018a}, systems on 1-dimensional lattices \cite{Bardet.2023,Bardet.2024}, or nearest-neighbour interactions \cite{Capel.2020,Kochanowski.2024} but do not allow to go beyond these settings. 
In this work, we circumvent this problem by considering a \emph{weak approximate tensorization (weak AT)} which will suffice for the quasi-rapid mixing of the Davies dynamics in $W_1$-distance.

To state our weak AT we first want to introduce the concept of conditional relative entropy which will serve as a useful tractable proxy for the Davies conditional relative entropy we ultimately wish to study. 

\begin{defi}[Conditional relative entropy \cite{Capel.2018a}]
    Given a lattice $\Lambda$ the conditional relative entropy on $A\subset \Lambda$ between two states $\rho,\sigma\in\cS(\cH_\Lambda)$ is defined as
    \begin{align}
        D_A(\rho\|\sigma):=D(\rho\|\sigma)-D(\rho_{\Bar{A}}\|\sigma_{\Bar{A}}).
    \end{align}
\end{defi}
We are now able to show the following novel relation between the conditional relative entropies:
\begin{lem}[Weak entropy factorization]\label{lem:weak-approximate-tensorization}
   Let $\cH = \cH_A \otimes \cH_B \otimes \cH_C \otimes \cH_D$ and $\rho, \sigma \in \cS(\cH)$ with $\sigma > 0$, then 
   \begin{equation}\label{eq:weak-approximate-tensorization}
       D_{ABC}(\rho \Vert \sigma) \le D_{AB}(\rho \Vert \sigma) + D_{BC}(\rho \Vert \sigma) + \norm{\mathbf{H}_\sigma(A:C|D)}_\infty \, . 
   \end{equation}
\end{lem}
The proof only requires the data-processing-inequality and Hölder inequality for Schatten-$p$ norms and can be found in \cref{lem:extended:weak-entropy-factorization}. Unlike the previously known entropy factorizations \cite{Capel.2018a,Capel.2020}, our new results allow us to have a system $D$ that we carry as a conditioning through the estimate which is key in the iteration of this statement, yielding a weak AT statement for lattices of arbitrary dimension later on. Note that classically (i.e. when all involved states commute) one can derive a multiplicative correction, i.e. a strong \emph{entropy factorization} of the form
\begin{equation}
\label{equ:classicalentropyfactorization}
    (1 - \norm{\exp(\mathbf{H}_\sigma(A:C|D)) - \Id}_\infty) D_{ABC}(\rho \Vert \sigma) \le D_{AB}(\rho \Vert \sigma) + D_{BC}(\rho \Vert \sigma) \, . 
\end{equation}
Establishing a similar relation in the quantum setting would allow us to lift all weak inequalities to their strong counterparts making for example the requirement of uniform polynomial local gap as a prior for rapid mixing obsolete (c.f. \cref{thm:rapidmixing}). However, establishing such an inequality for quantum systems where $D \ne \emptyset$ remains still open. 

To extend the above entropy factorization to a (weak) approximate tensorization statement on a lattice $\Lambda_L:=\llbracket-L, L\rrbracket^D\subset\mathbb{Z}^D$, we require a suitable \emph{coarse-graining} of $\Lambda$ w.r.t. which we split the conditional relative entropy. For $D=1,2$ such were given in \cite{Brandao.2018}. Effectively a suitable coarse-graining is a family of subsets $\{C_{a,i}\}_{a,i}$ that covers $\Lambda$, are all suitably local, and have overlaps between neighbouring levels $a$. 
\begin{lem}
For any $D\in\mathbb{N}$, there exists a suitable coarse-graining. Explicitly, given $r\leq k,c,\ell\leq 2L+1$ such that $\ell>2D(k+c)$, there exists a coarse-graining $\{C_{a,i}\}_{a\in \llbracket D\rrbracket,i}$ of $\Lambda_L=\llbracket -L,L\rrbracket^D$ made up of $D+1$ different levels indexed by $a$, such that
\begin{enumerate}
    \item $\bigcup_{a,i}C_{a,i}=\Lambda$ and each site $x\in\Lambda$ is included in at most $D+1$ sets $C_{a,i}$,
    \item $\{C_{a,i}\}_{i}$ is a collection of mutually disjoint subsets $\forall a\in\llbracket D\rrbracket$,
    \item $|C_{a,i}|\leq \ell^D$,
    \item Each level $a$ has a suitable overlap $c$ with the level coming before and after.
\end{enumerate}
\end{lem}

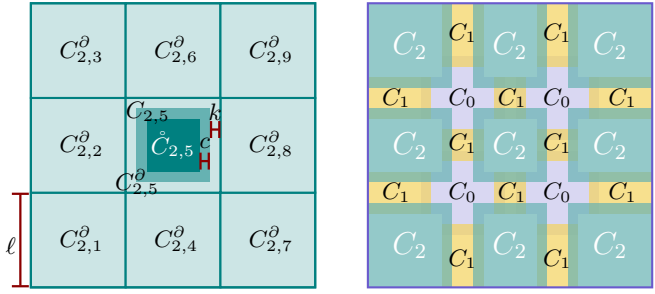
\begin{figure}[h]
    \centering
    \begin{minipage}{0.6\textwidth}
    \begin{tikzpicture}[scale=0.14, font=\footnotesize]
    \def\colorBoarder{TealDark}
    \def\colorCpartial{TealLight}
    \def\colorC{TealMid}
    \def\colorCcirc{TealDark}

    \begin{scope}
        
        % Loop to create outer squares
        \foreach \x/\y/\i in {1/1/1, 1/10/2, 1/19/3, 10/1/4, 10/19/6, 19/1/7, 19/10/8, 19/19/9} {
            \drawInitialSquare{\x}{\y}{\footnotesize$C_{2,\i}^\partial$}{}{\colorCpartial}{\colorBoarder};
        }
    
        % Draw the central square with additional inner structure
        \drawInitialSquare{10}{10}{5}{
            % Inner nested squares
            \fill[\colorC] (0.5,0.5) rectangle (7.5,7.5);
    
            \fill[\colorCcirc] (1.5,1.5) rectangle (6.5,6.5);
    
            % Labels for inner structures
            \node at (0.6,0.4) {$C_{2,5}^\partial$};
            \node at (1.6,7.1) {$C_{2,5}$};
            \node at (4,4) {\textcolor{white}{$\mathring{C}_{2,5}$}};
        }{\colorCpartial}{\colorBoarder};
        
        % Axis labels and arrows
        \node at (-1.4,4.5) {$\ell$};
        \draw [|-|, darkred, line width=0.3mm] (-0.5,0.5) -- (-0.5,9.5);
    
        % Additional distance markers
        \draw [|-|, darkred, line width=0.3mm] (17.5,15.5) -- (18.5,15.5);
        \node at (18,17.2) {$k$};
        \draw [|-|, darkred, line width=0.3mm] (16.5,12.5) -- (17.5,12.5);
        \node at (17,14.2) {$c$};
    
        \draw[\colorBoarder,thick] (0.5,0.5) rectangle (27.5,27.5);
    \end{scope}

    \def\colorBoarder{SlateBlueDark}
    \def\colorCzero{SlateBlueMid}
    \def\colorCone{GoldenrodMid}
    \def\colorCtwo{TealMid}

    \begin{scope}[xshift=32cm]
        % Draw C_0
        \fill[\colorCzero,opacity=0.4] (0.5,0.5) rectangle (27,27);

        % Loops to erase complement of C_0 without complement of C_1
        \foreach \x/\w/\ox/\ow in {0.5/5/0/-1, 13.5/1/1/-1, 22.5/5/1/0} {
                \drawConeColumns{\x}{0}{\w}{white}{1};
                \drawConeColumns{\x}{1}{\w}{white}{1};
        }

        % Draw C_1
        \foreach \x/\w/\ox/\ow in {0.5/6/0/-1, 12.5/3/1/-1, 21.5/6/1/0} {
            \drawConeColumns{\x}{0}{\w}{\colorCone}{0.7};
            \drawConeColumns{\x}{1}{\w}{\colorCone}{0.7};
        }

        % Loop to erase complement of C_1
        \foreach \x/\ox/\ow in {2.5/-2/0, 11.5/0/0, 20.5/0/2} {
            \foreach \y/\oy/\oh in {2.5/-2/0, 11.5/0/0, 20.5/0/2} {
                \drawCtwoRectangleColor{\x}{\y}{5}{5}{\ox}{\oy}{\ow}{\oh}{white};
            }
        }
        
        % Loop to create C_2
        \foreach \x/\ox/\ow in {1.5/-1/0, 10.5/0/0, 19.5/0/1} {
            \foreach \y/\oy/\oh in {1.5/-1/0, 10.5/0/0, 19.5/0/1} {
                \drawCtwoRectangleColorOp{\x}{\y}{7}{7}{\ox}{\oy}{\ow}{\oh}{\colorCtwo};
                \drawCtwoRectangleText{\x}{\y}{7}{7}{\ox}{\oy}{\ow}{\oh}{\textcolor{white}{$C_2$}};
            }
        }

        % Loop to create C_1 labels
        \foreach \x/\w/\ox/\ow in {0.5/7/0/-1, 11.5/5/1/-1, 20.5/7/1/0} {
            \node[anchor=center] at (9.5, {\x + \ox + (\w + \ow)/2 - 0.3}) {$C_1$};
            \node[anchor=center] at (18.5, {\x + \ox + (\w + \ow)/2 - 0.3}) {$C_1$};
            \node[anchor=center] at ({\x + \ox + (\w + \ow)/2 - 0.3}, 9.5) {$C_1$};
            \node[anchor=center] at ({\x + \ox + (\w + \ow)/2 - 0.3}, 18.5) {$C_1$};
        }
        
        \foreach \x/\ox/\ow in {0.5/-2/0, 14/0/0, 27.5/0/2} {
            \foreach \y/\oy/\oh  in {0.5/-2/0, 14/0/0, 27.5/0/2} {
                \begin{scope}
                    \clip (0.5,0.5) rectangle (27.5,27.5);
                \end{scope}
                
            }
        }

        % Draw C_0 labels
        \node at (9.5, 9.5) {$C_{0}$};
        \node at (18.5, 9.5) {$C_{0}$};
        \node at (18.5, 18.5) {$C_{0}$};
        \node at (9.5, 18.5) {$C_{0}$};

        \draw[\colorBoarder, thick] (0.5, 0.5) rectangle (27.5, 27.5);
    \end{scope}
    
    \end{tikzpicture}
    \end{minipage}
    \hspace{-0.6cm}
    \begin{minipage}{0.39\textwidth}
    \vspace{-0.26cm}
    \caption{An example of the coarse-graining of a $D=2$ dimensional lattice. On the left the largest regions in the top level $a=2$ of the hierarchy with side length $\ell$. On the right the different overlapping levels, labeled by $a=2,1,0$, in different colors. Each of the local cells $C_{a,i}$ has size $|C_{a,i}|\leq \ell^2$.}
    \label{fig:main:covering}
    \end{minipage}
\end{figure}

An explicit construction of the covering may be found in \cref{subsec:a-coarse-graining-of-the-hypercubic-lattice} with its key properties summarized in \cref{lem:extended:properties-of-lattice-coarse-graining} and a figure detailing the construction for the case $D = 2$ in \cref{fig:main:covering}.

With the entropy factorisation as well as the lattice coarse-graining in place, we are now able to prove one of the core results of this paper, namely the weak approximate tensorization of Gibbs states of local commuting Hamiltonians that exhibit uniform decay of the MCMI.

\begin{thm}[Weak approximate tensorization]\label{thm:weakAT}
    Given a suitable coarse-graining of the lattices $\Lambda$ with constants $(k,c,l)$ as described above, and corresponding Davies channels $\{E_{C_{a, i}} : a \in \llbracket0, D\rrbracket, i \in \cI_a\}$ corresponding to the Gibbs state $\sigma$ at temperature $\beta$ of a local commuting Hamiltonian, the following inequality holds for all $\rho\in\cS(\cH_{\Lambda})$:    \begin{align}\label{eq:approximate-tensorization}
        D(\rho\|\sigma) &\leq \underset{a=0}{\overset{D}{\sum}} \; \underset{i_a \in \mathcal{I}_a}{\sum} D (\rho \| E_{C_{a,i_a}} (\rho) ) +  \underset{a=1}{\overset{D}{\sum}} \, \zeta_a (\sigma), \quad \quad \zeta_a (\sigma) := \norm{\mathbf{H}_\sigma(X_a : Z_a | W_a)}_\infty \, ,
    \end{align}
    where $W_a \sqcup X_a \sqcup Y_a \sqcup Z_a =: \overline{C^a} \sqcup \mathring{C}_a \sqcup (C_a \backslash \mathring{C}_a) \sqcup (C^a \backslash C_a)$ with $d(X_a,Z_a)= c \geq r$, for $a \in \llbracket1, D\rrbracket$. The explicit definition of these sets can be found in \cref{subsec:a-coarse-graining-of-the-hypercubic-lattice}. If $\sigma$ further satisfies uniform decay of MCMI with constants $K$ and $\xi$, it holds for any $\rho\in\cS(\cH_{\Lambda})$ that
    \begin{align}\label{eq:approximate-tensorization-with-explicit-estimate}
        D(\rho\|\sigma) &\leq \underset{a=0}{\overset{D}{\sum}} \; \underset{i_a \in \mathcal{I}_a}{\sum} D (\rho \| E_{C_{a,i_a}} (\rho)) + D2^D K |\Lambda| e^{-c/\xi} \, .
    \end{align}
    We refer to \eqref{eq:approximate-tensorization} as a \emph{$(1,c_2)$-weak AT} w.r.t. the coarse-graining $\{C_{a,i}\}_{a,i}$ where
    $$c_2 = \sum_{a = 1}^D \,\zeta_a (\sigma)\leq K2^DK|\Lambda|e^{-\frac{c}{\xi}} \, .$$
\end{thm}
The proof works by induction over $a$ using the entropy factorization \cref{lem:weak-approximate-tensorization} and the coarse-graining introduced before, separating one level of the hierarchy in every step. It may be found in \cref{subsec:weak-mlsi-for-the-global-davies-semigroup}.

\subsection{Quasi-rapid mixing of Lipschitz observables}\label{sec:quasi-rapid-mixing-of-lipschitz-observables}
In this section, we prove the quasi-rapid mixing of the Davies dynamics w.r.t the quantum Wasserstein distance of order 1. We recall its definition. 
\begin{defi}[$W_1$-distance from \cite{DePalma.2021}]
    For two quantum states $\rho, \sigma \in \cS(\cH_\Lambda)$, the quantum Wasserstein distance of order 1 is given as the dual norm to a Lipschitz norm.
    \begin{equation}
       \norm{\rho-\sigma}_{W_1} := \max\limits_{H \in \cB_{\sa}(\cH_{\cI}), \norm{H}_L \le 1} \tr[H (\rho-\sigma)] \, ,
    \end{equation}
    where the Lipschitz-norm of $H \in \cB_{\sa}(\cH_\Lambda)$ is defined as
    \begin{equation}
        \norm{H}_L := 2\max_{k \in \Lambda}\min_{\tilde H \in \cB_{\operatorname{sa}}(\cH_{\Lambda\backslash\{k\}})} \norm{H - \Id_{k} \otimes \tilde H}_\infty \, . 
    \end{equation}
\end{defi}
One can show that for two states $\rho,\sigma\in\cS(\cH_\Lambda)$, it satisfies $\frac{1}{2}\|\rho-\sigma\|_1\leq \|\rho-\sigma\|_{W_1}\leq |\Lambda|\|\rho-\sigma\|_1$, and hence it is an extensive norm that indicates how well two states differ locally \cite{DePalma.2021}. 

Quite naturally we define the $\epsilon$-mixing time for a primitive QMS with fixed point $\sigma$ w.r.t this normalized quantum Wasserstein distance as
\begin{align}\label{eq:wasserstein-mixing-time}
    t^{W_1}_\textup{mix}(\epsilon) := \inf\Big\{t\ge 0 : \forall \rho\in \cS(\cH_{\Lambda}) , \, \left\|e^{t\cL_{\Lambda}}(\rho)-\sigma\right\|_{W_1}\le |\Lambda|\epsilon\Big\} \, .
\end{align}
Note that mixing times based on normalized Wasserstein distances have also been considered in recent classical Gibbs sampling, see e.g. \cite{Alaoui.2022} in which the authors prove rapid mixing of Glauber dynamics for certain spin-glass models at high temperature in normalized Wasserstein-$2$ distance.

The rest of this section is dedicated to proving the first main result of this work:
\begin{thm}[Quasi rapid Wasserstein mixing]\label{thm:wassersteinmixing}
    Let $\{\cL^D_{\Lambda_L}\}_L$ be a family of Davies Lindbladians corresponding to a family of $(\kappa,r)$-local, commuting, $J$-bounded Hamiltonians $\{H_{\Lambda_L}\}_L$, each of which has growth constant $g$. Consider $\epsilon>0$ and denote by $N=|\Lambda_L|$ the size of the lattice.
    Then, if the invariant states of $\cL^D_{\Lambda_L}$, i.e. the Gibbs states $\sigma^{\Lambda_L}$, all satisfy uniform exponential decay of its matrix-valued quantum conditional mutual information (MCMI) as defined in \eqref{eq:definition-mcmi}, the semigroups $\{e^{t\cL_{\Lambda_L}^D}\}_{t \ge 0}$ satisfy
    \begin{align}
        t^{W_1}_{\mix}(\epsilon)=\textup{quasi-poly}\left(\frac{1}{\epsilon^2}\poly\log\frac{N}{\epsilon^2}\right)=\textup{quasi-poly}(\epsilon^{-1})_{\epsilon\to 0}\textup{quasi-log}(N)_{N\to \infty}.
    \end{align}
    For a precise upper bound on $t^{W_1}_{\mix}(\epsilon)$, see \cref{thm:detailed:wassersteinmixing}. 
\end{thm}
In the following, we give an overview of the main strategy and tools required to prove this result. The proof may be found in  \cref{subsec:the-w1-mixing-from-mcmi-decay} and consists of the following two ingredients.
\begin{enumerate}
    \item Relating the $W_1$ distance to the relative entropy. We do this via a general \emph{weak transport cost} (weak TC) inequality implied by the weak AT \eqref{eq:approximate-tensorization}. This is an inequality relating the $W_1$ distance between any state and a fixed point of a QMS that satisfies a (weak) AT to their conditional relative entropy.
    \item Showing sufficient decay of the relative entropy: We prove a $(\alpha,\epsilon)$ weak MLSI, in which the cMLSI constant verifies $\alpha(\epsilon)^{-1}=\cO(\exp\poly\log\frac{N}{\epsilon})$. This requires the combination of the (weak) AT with a cMLSI alike estimate relating the Davies conditional relative entropy $D(\rho \Vert E_A(\rho))$ to the entropy production of $\cL^D_{A\partial}$ at an exponential cost in $|A\partial|$.
\end{enumerate}

Let us begin with the statement of the weak transport cost inequality for Gibbs states of local commuting Hamiltonians:

\begin{thm}[Weak transport cost inequality]\label{thm:weak-transport-cost-inequality} 
    Let $\sigma$ be the Gibbs state at inverse temperature $\beta$ of a local commuting Hamiltonian on a lattice $\Lambda$ that satisfies a $(1,c_2)$-weak AT w.r.t. some coarse-graining $\{A_i\}^{n_A}_{i=1}$ of $\Lambda$, like in \eqref{eq:approximate-tensorization}. 
    Then it satisfies a $(b_1,b_2)$\emph{-weak transport cost} inequality with $b_1=2\sqrt{2n_A}\max_i|A_i\partial|$ and $b_2=|\Lambda|\sqrt{2c_2}$, i.e.:
    \begin{equation}\label{eq:weak-transport-cost-inequality}
        \norm{\rho - \sigma}_{W^1} \le \max_{i} 2\sqrt{2} |A_i\partial| \sqrt{n_A D(\rho \Vert \sigma)} + |\Lambda| \sqrt{2 c_2}
    \end{equation}
\end{thm} \noindent
The proof of this result may be found in \cref{thm:extended:weak-transport-cost-inequality} of \cref{subsec:a-weak-transport-cost-inequality-for-davies-channels}. The second step estimates the Davies conditional relative entropy on some local region $A$ with the entropy production on a slightly larger region:
\begin{lem}\label{lem:exponential-almostcmlsi} 
    Let $A\subseteq\Lambda$ and let $E_A:=\lim_{t\to\infty}e^{t\cL^D_A}$ be the conditional expectation of a local (on $A$) Davies generator at inverse temperature $\beta$ corresponding to a local, commuting, $J-$bounded Hamiltonian that on the lattice $\Lambda$ has growth constant $g$, then  
    \begin{align}
        D(\rho\|E_A(\rho))\leq e^{2gJ(1+2\beta|A\partial\partial|)}(\chi^0_{\min})^{-1}\EP_{\cL^D_{A\partial}}(\rho),
    \end{align} where $\chi^0_{\min}>0$ is the constant from \eqref{eq:definition-davies-generator}.
\end{lem}
This is not quite a bound on the cMLSI constant of the QMS generated by $\cL_A$, because the entropy production on the RHS is the one of $\cL_{A\partial}$ and not $\cL_A$. However, this bound, together with the weak AT in \cref{thm:weakAT} will suffice to obtain a suitable global MLSI constant. 
The idea behind the proof of this lemma (c.f. \cref{subsec:a-mlsi-alike-inequality-for-local-davies-at-every-temperature}) is to relate both the Davies conditional relative entropy and the entropy production of the Davies generators at inverse temperature $\beta$ to the ones at $0$. Then at $0$ temperature, the Davies semigroup is closely related to the depolarizing one allowing us to connect the two only by slightly enlarging the domain of the generator in the entropy production (c.f. \cref{subsec:weak-mlsi-for-the-global-davies-semigroup}). A direct consequence of this lemma is the following weak MLSI.
\begin{cor}[Quasi-poly weak MLSI, \cref{thm:extended:weak-mlsi}] \label{cor:quasipolynomial-weak-mlsi}
    In the context of \cref{thm:wassersteinmixing} it holds that
    \begin{align*}
        D(e^{t\cL_\Lambda^D}(\rho)\|\sigma)\leq e^{-\alpha(\epsilon)t}D(\rho\|\sigma)+\epsilon \quad \text{where}\quad  \frac{1}{\alpha(\epsilon)}=\cO\left(\exp(\poly\log\frac{N}{\epsilon})\right).
    \end{align*}
\end{cor}
The proof combines the weak AT \eqref{eq:approximate-tensorization} with the weak transport cost inequality of \cref{thm:weak-transport-cost-inequality} and then uses a standard Grönwall argument to get to the result. For the detailed proof see \cref{thm:extended:weak-mlsi}.

\subsection{Rapid mixing under polynomial local gap}\label{sec:rapid-mixing-under-polynomial-gap}

Having stronger assumptions on the gap of the local Davies generators, i.e. faster local mixing unsurprisingly yields tighter mixing bounds.
The gap of the local Davies generator on $A\subset\Lambda$ is defined as the distance between the largest and second to largest eigenvalue of $\cL_A^D$. A formal definition is given in \eqref{eq:definition-gap}. In \cite{Gao.2022a, Rouze.2019} it is shown to lower bound the local cMLSI constant of $\cL_A$ as $\alpha_c(\cL_A)\geq \Omega\left(\frac{\lambda(\cL_A)}{|A|}\right)$. Assuming a polynomial local gap with degree $\mu\in\mathbb{N}_0$ amounts to 
\begin{align} 
    \lambda(\cL_A) \geq \Omega(|A|^{-\mu}) \quad\text{and hence}\quad \alpha_c(\cL_A) \geq \Omega(|A|^{-1-\mu}) \quad \forall A\subset \Lambda,
\end{align} 
which is equivalent to assuming a tightened version of \cref{lem:exponential-almostcmlsi} (or \cref{lem:extended:gap-to-mlsi}). Note that technically an exact gap is not necessary and a result where the region of the generator is enlarged would suffice (analogous to \cref{lem:exponential-almostcmlsi}), however, we will here and in the following always assume a gap for simplicity. With this assumption on the gap we can improve the weak cMLSI as follows:

\begin{cor}[Polylogarithmic weak MLSI]\label{cor:polylogarithmic-weak-mlsi} 
    In the notation of \cref{thm:wassersteinmixing} under the additional assumption of a local gap that is at most polynomially decaying with degree $\mu\in\mathbb{N}_0$ it holds that
    \begin{align}
         D(e^{t\cL^D_\Lambda}(\rho)\|\sigma)\leq e^{-\alpha(\epsilon)t}D(\rho\|\sigma)+\epsilon,
    \end{align} where $\frac{1}{\alpha(\epsilon)}=\cO\left(\left(\log\frac{N}{\epsilon}\right)^{D(1+\mu)}\right)$.
\end{cor}
This weak MLSI combined with either Pinsker's inequality or \cref{thm:weak-transport-cost-inequality} directly gives the following strengthened mixing time bounds.

\begin{thm}[Mixing under poly gap]\label{thm:rapidmixing}
    Let $\{\cL^D_{\Lambda_L}\}_L$ be a family of Davies Lindbladians corresponding to a family of $(\kappa,r)$-local, commuting, $J$-bounded Hamiltonians $\{H_{\Lambda_L}\}_L$ with uniform growth constant $g$. Assume that the local gap is at most polynomially decreasing with degree $\mu$. Consider $\epsilon>0$ and denote by $N=|\Lambda_L|$ the size of the lattice.
    Then, if the invariant state of $\cL^D_{\Lambda_L}$, i.e. the Gibbs state $\sigma^{\Lambda_L}$, satisfies uniform exponential decay of its matrix-valued quantum conditional mutual information (MCMI) as defined in \eqref{eq:definition-mcmi}, the semigroup $\{e^{t\cL_{\Lambda_L}^D}\}_{t \ge 0}$ satisfy
   \begin{align*}
           &t^{1}_{\text{mix}}(\epsilon)=\cO\Big(\Big(\log\frac{N}{\epsilon^2}\Big)^{1+D(1+\mu)}\Big)_{\underset{\epsilon\to0}{N\to \infty}},\quad
           &t^{W_1}_{\text{mix}}(\epsilon) = \cO\Big(\Big(\log\Big(\frac{1}{\epsilon^2}\log\frac{N}{\epsilon^2}\Big)\Big)^{1+D(1+\mu)}\Big)_{\underset{\epsilon\to0}{N\to \infty}}.
   \end{align*}
   For the non-asymptotic expressions for these mixing times see \cref{thm:extended:rapid-mixing}.
\end{thm}
Polynomial assumptions on the local gap, have been used before to show rapid thermalization of Davies dynamics, e.g. for commuting quantum systems in 1D \cite{Bardet.2024} under a condition equivalent to non-conditional MCMI-decay. In \cite{Kochanowski.2024} this was superseded, however, removing the required assumption on the local gap. 
\begin{rmk}\label{rmk:very-high-temperature-constant-almost-cmlsi}
    Note that instead of a uniform polynomial local gap, one can also require very high temperature, i.e. $\beta \sim \frac{1}{|\Lambda|}$ leading to a constant correction in \cref{lem:exponential-almostcmlsi}, giving a result analogous to the one above only relying on the uniform decay of the MCMI. 
\end{rmk}
To now get the main result \cref{thm:main-sampling-result} we can use any Lindblad simulation circuit to simulate $e^{t^{W_1}_{\text{mix}}(\varepsilon)\cL^D_\Lambda}$, which by the above, on any input state is close to $\sigma$ in normalized $W_1$ distance. Using the one from \cite[Theorem III.2]{Chen.2023} and our above mixing time bounds yields directly \cref{thm:main-sampling-result}, though any Lindblad simulation circuit, e.g. also \cite{Li.2022} may be used. We call this algorithm \emph{quasi-optimal} since it scales as $\cO(N)$ up to quasi-logarithmic corrections, following the convention to denote sampling algorithms which scale as $\cO(N)$ up to logarithmic corrections as `optimal'. For more details see \cref{thm:quasi-optimal-sampling-from-Gibbs-states-that-satisfy-MCMI-decay}. 

\section{A comparison with quantum Gibbs sampling}\label{sec:a-comparison-with-classical/quantum-Gibbs-sampling}

Here we further investigate the connections of the current submission with the vast recent literature in quantum Gibbs sampling. 

In the presence of a local commuting Hamiltonian, the natural candidate way to design quantum Gibbs samplers is employing a Davies generator, which is local in this case. In \cite{Kastoryano.2016}, the authors showed that there is an equivalence between the existence of a positive spectral gap for the Lindbladian (which yields fast mixing) and a certain form of strong clustering in the Gibbs state of the Hamiltonian. This generalizes to the quantum setting the works \cite{Stroock.1992,Stroock.1992a}. However, this is not enough to yield rapid mixing. \cite{Bardet.2023,Bardet.2024} showed the existence of a positive (logarithmically-decaying with the system size) MLSI for translation-invariant 1D commuting local Hamiltonians at any positive temperature, and thus rapid mixing. This result was improved to constant MLSI in \cite{Kochanowski.2024}, where rapid mixing was also shown for nearest neighbour Hamiltonians on hypercubic lattices at high-enough temperature, as well as for $b$-ary trees with small correlation length at high-enough temperature. This improves upon \cite{Capel.2020}, where another efficient Gibbs sampler was provided for the former setting in terms of the so-called Schmidt generators, based on \cite{Bravyi.2005}.

The very recent papers \cite{Jiang.2024,Gilyen.2024} present quantum generalizations of Glauber and Metropolis dynamics, inspired by the breakthrough \cite{Temme.2011}. Whereas \cite{Jiang.2024} uses quantum phase estimation (QPE) in its algorithm, \cite{Gilyen.2024} is built on the Lindbladian introduced in \cite{Chen.2023a,Chen.2023}, which can be regarded as a modification of Davies generator so that it is quasi-local for non-commuting Hamiltonians, in contrast to the usual Davies. A family of quantum Gibbs samplers satisfying the KMS detailed balance condition, which in particular includes the construction from \cite{Chen.2023}, was presented in \cite{Ding.2024}. Moreover, the concurrent \cite{Rouze.2024b,Bakshi.2024} show the existence of a positive spectral gap, and thus fast mixing, for the former Lindbladians at high-enough temperature, yielding efficiency for the Gibbs samplers. All these works appeared after long-term efforts to implement Davies generators for non-commuting Hamiltonians in the most accurate way possible, as an exact implementation is impossible. \cite{Wocjan.2023} dealt with this problem by assuming a rounding promise on the Hamiltonian, which was subsequently removed in \cite{Rall.2023} by using randomized rounding. 

Additionally to all these approaches based on Davies (or similar) generators, there are some others based on Grover approaches \cite{Poulin.2009,Chowdhury.2017}, quantum imaginary time evolution \cite{Motta.2019} or quantum Singular Value Transforms \cite{Gilyen.2019}. Some other approaches based on dissipation are \cite{Zhang.2023}, where the quantum Gibbs sampler is designed with simple local update rules, or \cite{Fang.2024}, which is constructed from an energy functional inspired by the hypocoercivity of (classical) kinetic theory. Other works related to preparation of quantum Gibbs states are \cite{Kashyap.2024,Harrow.2020,Fawzi.2024}, among many others. 

\section{Outlook}\label{sec:outlook}
The findings of this manuscript open up a series of natural questions in the contexts of entropy factorizations, mixing times, and decay of correlations on Gibbs states.  As mentioned after \cref{lem:weak-approximate-tensorization}, a strong entropy factorization is known to hold classically, see \eqref{equ:classicalentropyfactorization}. Such a result presents an upper bound for the conditional relative entropy in $ABC$ in terms of the sum of two conditional relative entropies in $AB$ and $BC$, respectively, and a multiplicative error term, without the need for an additional additive error term.  We strongly believe that something analogous should also hold in the quantum regime in quite some generality; however, a proof for this is still missing. Nevertheless, such a result is at least known to hold in the case that the second state is a tensor product \cite{Capel.2018a}. Such a strong entropy factorization would directly lead to `strong' versions of the MLSI and TC inequality, with which one could directly conclude rapid mixing from the decay of MCMI by allowing decomposition of the lattice into local regions of constant size, following the argument of e.g. \cite{Martinelli.1999, Capel.2020, Kochanowski.2024}.

Refocusing on the main approach developed in this paper, based on the existence of a weak AT, the ingredients required to conclude rapid mixing are decay of the MCMI and a suitable local gap. One would hope that a polynomially-bounded local gap, or more precisely an almost local gap (i.e. one can allow for constant size enlargements of the regions in the generator analogously to \cref{lem:exponential-almostcmlsi}) should hold for the local Davies generators, assuming that the global Gibbs state satisfies some clustering condition, such as decay of the MCMI, or that it stems from a local CSS code. Proving this specifically for 2D Toric codes at every positive temperature, which we expect to hold, would imply an exponential improvement in the mixing time recently proven in \cite{Ding.2024a}, inspired in the original \cite{Alicki.2009}, and subsequently extended to 2D quantum double models in \cite{Komar.2016} and \cite{Lucia.2023}. 

Finally, another relevant line of research is that of the MCMI decay for more general models, i.e. general (commuting) Hamiltonians, not only marginal commuting ones, which should be of independent interest, given the connections of the MCMI to the notions of conditional mutual information and mutual information, and their range of applicability. Furthermore, it would be interesting to see if the methods translate also to the non-commuting setting directly relating an information-theoretic measure - the MCMI - to the speed of the convergence of the samplers, e.g. the one in \cite{Chen.2023}.

\section{Acknowledgments}
J.K. and P.G. thank Samuel Scalet for helpful discussions regarding effective Hamiltonians, resulting in a much-simplified presentation of \cref{thm:MCMI-ecay-for-marginal-commuting-systems-at-high-temperature}. Both also would like to thank Sebastian Stengele for disclosing the marginal commuting property of CSS codes to them and for insights into this topic. A.C. and P.G. acknowledge the support of the Deutsche Forschungsgemeinschaft (DFG, German Research Foundation) - Project-ID 470903074 - TRR 352.  A.C. and P.G. also acknowledge funding by the Federal Ministry of Education and Research (BMBF) and the Baden-Württemberg Ministry of Science as part of the Excellence Strategy of the German Federal and State Governments. This project was funded within the QuantERA II Programme which has received funding from the EU’s H2020 research and innovation programme under the GA No 101017733.
C.R. acknowledge financial support from the ANR project QTraj (ANR-20-CE40-0024-01) of the French National Research Agency (ANR) as well as France 2030 under the French National Research Agency award number “ANR-22-PNCQ-0002”.

\bibliographystyle{alphaurl}
\bibliography{bibliography}

\newpage 

\appendix

\section{Theoretical Framework}\label{sec:theoretical-framework}

\subsection{Basic notation}\label{subsec:basic-notation}
We work with finite-dimensional Hilbert spaces, denoted by $\cH$ or $\cK$, and their associated algebras of bounded operators, $\cB(\cH)$ and $\cB(\cK)$, respectively. Throughout this discussion, we use several standard notations: the matrix trace is denoted by $\tr[\,\cdot\,]$, the Hilbert-Schmidt inner product by $\braket{\cdot, \cdot}$, and the Hilbert-Schmidt norm by $\norm{\,\cdot\,}_{2}$. More generally, we use $\norm{\,\cdot\,}_p$ for $p \in [1, \infty]$ to denote the Schatten-$p$-norms, with the operator norm given by $\norm{\,\cdot\,}_\infty$. Operators are represented by capital Latin or lowercase Greek letters, depending on the context. For any operator $X \in \cB(\cH)$, we write $X^\dagger$ for its adjoint and denote the subset of self-adjoint operators by $\cB_{\sa}(\cH)$. The Hilbert-Schmidt adjoint of a linear map $\Phi: \cB(\cH) \to \cB(\cH)$ is written as $\Phi^*$. We reserve $\Id_{\cH}$ for the identity operator on $\cH$ and $\id$ for the identity map on $\cB(\cH)$, with $d_{\cH} = \dim(\cH)$ denoting the dimension of $\cH$. A quantum state (density operator) $\rho$ is a positive semidefinite operator with unit trace, and we denote the set of states on $\cH$ by $\cS(\cH)$.

For any full-rank $X \in \cB_{\sa}(\cH)$, we define two fundamental linear maps: the first is $\Gamma_X: \cB(\cH) \to \cB(\cH)$, given by $\Gamma_X(Y) := X^{\frac{1}{2}} Y X^{\frac{1}{2}}$, and the second is the modular operator $\Delta_X: \cB(\cH) \to \cB(\cH)$, defined as $\Delta_X(Y) = X Y X^{-1}$. In bipartite systems $AB$ with $\cH_{AB} = \cH_A \otimes \cH_B$, we use $\rho_A$ to denote the reduced state $\tr_B[\rho_{AB}]$, where $\tr_B$ represents the partial trace over system $B$. By convention, we may write $\rho_A$ as an operator in $\cB(\cH_{AB})$, to be understood as $\rho_A \otimes \Id_B$. The normalized partial trace $\E_B: \cB(\cH_{AB}) \to \cB(\cH_{AB})$ is defined as $\E_B(X_{AB}) = \tr_B[X_{AB}] \otimes \Id_B/d_B$.

For a bipartite state $\rho \in \cS(\mathbb{C}^n \otimes \cH)$ and a quantum channel $\Psi: \cB(\cH) \to \cB(\cH)$, we write $\Psi(\rho) := (\id \otimes \Psi)(\rho)$, where a quantum channel is understood to be a linear completely positive trace preserving map $\Phi: \cB(\cH) \to \cB(\cK)$. The Umegaki relative entropy \cite{Umegaki.1962}, a fundamental measure in quantum information theory, is defined for states $\rho$ and $\sigma$ as:
\begin{equation*}
    D(\rho\Vert\sigma) :=
    \begin{cases}
        \tr[\rho\log\rho - \rho\log\sigma], & \text{if } \ker(\sigma) \subseteq \ker(\rho), \\
        +\infty, & \text{otherwise},
    \end{cases}
\end{equation*}
where $\ker(\cdot)$ denotes the kernel of an operator. By Pinsker’s inequality, the relative entropy relates to the trace distance as:
\begin{equation}\label{eq:pinsker-inequality}
    \norm{\rho - \sigma}_1 \le \sqrt{2 D(\rho \Vert \sigma)} \, .
\end{equation}

For composite (or many-body) systems $\cH_{\cI} = \bigotimes_{i \in \cI} \cH_i$, we introduce the quantum Wasserstein distance of order 1 \cite{DePalma.2021}. For two quantum states $\rho, \sigma \in \cS(\cH_{\cI})$, this distance is defined as the dual to a Lipschitz norm:
\begin{equation}
    \norm{\rho-\sigma}_{W_1} := \max\limits_{H \in \cB_{\sa}(\cH_{\cI}), \norm{H}_L \le 1} \braket{H, \rho - \sigma} \, ,
\end{equation}
where the Lipschitz norm is defined as
\begin{equation}
    \norm{H}_L := 2 \max_{i \in \cI}\min_{\tilde H \in \cB_{\sa}(\cH_{\cI\backslash\{i\}})} \norm{H - \Id_{i} \otimes \tilde H}_\infty \, ,
\end{equation}
for $H \in \cB_{\sa}(\cH_{\cI})$. The Wasserstein distance can also be regarded as a norm on the set of traceless self-adjoint operators and is an extensive quantity, scaling with the size of $\cI$. This is reflected in the following relation with the trace distance, shown in \cite[Proposition 5]{DePalma.2021}: For any traceless $X \in \cB_{\sa}(\cH_{\cI})$ with $\tr_{\cJ}[X] = 0$, where $\cJ \subseteq \cI$, we have
\begin{equation}\label{eq:relation-wasserstein-trace-distance}
    \norm{X}_{W^1} \le |\cJ| \norm{X}_1 , .
\end{equation}

Conditional expectations on von Neumann algebras play a central role in our analysis. For a von Neumann subalgebra $\cN \subseteq \cB(\cH)$, a conditional expectation onto $\cN$ is a completely positive unital map $E^*_{\cN}: \cB(\cH) \to \cN$ satisfying:
\begin{enumerate}
    \item $E^*_{\cN}(X) = X$ for all $X \in \cN$;
    \item $E^*_{\cN}(VXW) = VE^*_{\cN}(X)W$ for all $V, W \in \cN$ and $X \in \cB(\cH)$.
\end{enumerate}
For such conditional expectations, the relative entropy exhibits special properties known as the chain rule and exact entropy factorization: given conditional expectation $E^*_{\cN}: \cB(\cH) \to \cN$ and $E^*_{\cM}:\cB(\cH) \to \cB(\cH)$ with $[E^*_{\cM}, E^*_{\cN}] = 0$, the chain rule states that for any state $\sigma = E_{\cN}(\sigma) \in \cS(\cH)$, and any state $\rho \in \cS(\cH)$,
\begin{equation}\label{eq:chain-rule-relative-entropy}
    D(\rho\Vert\sigma)
    = D(\rho\Vert E_{\cN}(\rho)) + D(E_{\cN}(\rho)\Vert\sigma) \, .
\end{equation}
The exact entropy factorisation on the other hand states that 
\begin{equation}\label{eq:exaxt-factorization-relative-entropy}
    D(\rho\Vert E_{\cN}E_{\cM}(\rho)) \le D(\rho\Vert E_{\cM}(\rho)) + D(\rho \Vert E_{\cN}(\rho)) \, . 
\end{equation}
For a full-rank state $\sigma > 0$, we define a family of inner products, $s \in [0, 1]$,
\begin{equation*}
    \braket{X, Y}_{s, \sigma} := \braket{X, \sigma^s Y \sigma^{1 - s}} \, ,
\end{equation*}
turning $\cB(\cH)$ into a Hilbert space with norm
\begin{equation*}
    \norm{X}^2_{1/s, \sigma} := \braket{X, X}_{s, \sigma} \, .
\end{equation*}
In the cases $s = 1/2$ and $s = 1$ ($s = 0$), $\braket{X, Y}_{s, \sigma}$ are also referred to as the KMS and GNS inner products, respectively. Note that for the KMS inner product, we will simply write $\braket{\cdot, \cdot}_\sigma$. Given the Hilbert space structure, we can define adjoints of linear maps $\Phi:\cB(\cH) \to \cB(\cH)$, calling a map $\Phi$ KMS- or GNS-symmetric if it is self-adjoint with respect to the respective inner product. Note that GNS symmetry implies KMS symmetry, as it implies commutativity with the modular operator, $[\Phi, \Delta_\sigma] = 0$. These inner products play a crucial role in the analysis of quantum Markov semigroups (QMS). A QMS $(\cP_t)_{t \geq 0}$ is a semigroup of completely positive, trace-preserving maps on $\cB(\cH)$ in Schrödinger and of completely positive, unital maps on $\cB(\cH)$ in Heisenberg picture. Note that one is the dual of the other w.r.t. the Hilbert-Schmidt inner-product. The unique generator of this semigroup, called a Lindbladian, is denoted by $\cL: \cB(\cH) \to \cB(\cH)$ and connected to the semigroup through the linear differential equation
\begin{equation*}
    \frac{d}{dt} \rho = \cL(\rho)\quad\text{with}\quad \rho(0) = \rho_0 \in \cS(\cH) \, . 
\end{equation*}
This motivates the notation $\cP_t = e^{t \cL}$, which we will use henceforth. A QMS is said to be KMS- or GNS-symmetric with respect to a full-rank state $\sigma \in \cS(\cH)$ if the Hilbert-Schmidt dual of its generator $\cL^*:\cB(\cH) \to \cB(\cH)$ exhibits the respective symmetry. For a generator of a QMS $\cL^*$ which is further GNS-symmetric w.r.t $\sigma > 0$ the kernel $\ker(\cL^*)$ is a von Neuman subalgebra with conditional expectation $E^*$ given by $E^*[X]:=\lim_{t\to\infty}e^{t\cL^*}[X]$  for any $X \in \cB(\cH)$.

A central question, and the primary focus of our work, concerns the speed of this convergence. Specifically, we investigate the convergence rate of a particular family of Lindbladians on quantum spin systems, measured in both the Wasserstein and trace distance.

\subsection{Mixing, rapid mixing, and logarithmic Sobolev inequalities for open quantum systems}\label{subsec:mixing-rapid-mixing-and-logarithmic-sobolove-inequalities-for-open-quantum-systems}
The study of convergence speeds for quantum Markov semigroups under different distance measures forms the core of this research. Here, we use the term "measure" informally as a way to compare states, rather than in its strict mathematical sense. While mixing phenomena are interesting even in simple systems, their study primarily focuses on composite systems. Therefore, we will work in the setting $\cH = \bigotimes_{i \in \cI} \cH_i$.
We will restrict ourselves to quantum Markov semigroups that are GNS-symmetric with respect to a full-rank state $\sigma > 0$, which guarantees both the convergence $e^{t\cL} \to E$ as $t \to \infty$ and the von Neumann algebra structure of $\ker(\cL^*)$. The mixing time in trace distance for $\varepsilon > 0$ is defined as
\begin{equation}
    t_{\mix}^{1}(\varepsilon) := \inf\{ t \ge 0 : \forall \rho \in \cS(\cH), \norm{e^{t \cL}(\rho) - E(\rho)}_1 \le \varepsilon\} \, .
\end{equation}
Analogously, we define the mixing time in Wasserstein distance as
\begin{equation}
    t_{\mix}^{W_1}(\varepsilon) := \inf\{ t \ge 0 : \forall \rho \in \cS(\cH), \norm{e^{t \cL}(\rho) - E(\rho)}_{W^1} \le |\cI| \varepsilon\} \, .
\end{equation}
Our analysis aims to understand how these mixing times depend on both system size and $\varepsilon$. We classify systems as \emph{mixing}, sometimes in literature also referred to as `fast mixing', when mixing times scale linearly with system size, while polylogarithmic scaling characterizes \emph{rapidly mixing} or \emph{rapid mixing} systems and quasi-logarithmic scaling characterized \emph{quasi-rapid mixing}. Previous research has largely focused on trace distance, commonly analyzing the generator's gap (c.f. \cite{Kastoryano.2016, Chen.2023, Chen.2024, Bravyi.2023})—the distance between the largest and second-largest eigenvalue of the Lindbladian. For KMS-symmetric Lindbladians, this gap is given by
\begin{equation}\label{eq:definition-gap}
    \lambda(\cL^*) := \inf\limits_{X \in \cB(\cH)} \frac{\braket{X, -\cL^*(X)}_{\sigma}}{\norm{(\id - E^*)(X)}_{2, \sigma}} \, .
\end{equation}
By Grönwall's Lemma, we obtain $\norm{e^{t\cL^*}(X) - E^*(X)}_{2, \sigma} \le e^{-t\lambda(\cL^*)/2}\norm{X - E^*(X)}_{2, \sigma}$. Using Hölder's inequality on its variational form, this yields an estimate for the trace distance:
\begin{equation*}
    \norm{\rho - \sigma}_1 \le \norm{\sigma^{-1}}_\infty \sup\limits_{X \in \cB_{\sa}(\cH_{\cI}), \norm{X}_\infty \le 1} \norm{e^{t\cL^*}(X) - E^*(X)}_{2, \sigma} \le 2 \norm{\sigma^{-1}}_\infty e^{-t\lambda(\cL^*)/2} \, .
\end{equation*}
Since $\norm{\sigma^{-1}}_\infty$ typically scales exponentially with $|\cI|$, the mixing times must be at least linear in $|\cI|$ to compensate, assuming the gap is independent of $|\cI|$.
Alternative approaches have focused on analyzing decay in relative entropy (c.f. \cite{Carbone.2015,Kastoryano.2013, Bardet.2017, Bardet.2024}), attempting to establish (weak) modified logarithmic Sobolev inequalities (wMLSI). These inequalities relate the relative entropy to the entropy production of the semigroup's generator, defined as
\begin{equation}\label{eq:definition-entropy-production}
    \EP_{\cL}(\rho) := -\frac{d}{dt}\Big\vert_{t = 0} D(e^{t\cL}(\rho) \Vert \sigma) = -\tr[\cL(\rho)(\log(\rho) - \log(\sigma))] \, .
\end{equation}
The entropy production is always non-negative, following from monotonicity of the relative entropy under quantum channels. We say that $\cL$ satisfies a (weak) modified logarithmic Sobolev inequality if there exist constants $c_1 > 0$ and $c_2 \ge 0$ such that for all $\rho \in \cS(\cH)$
\begin{equation}\label{eq:weak-mlsi}\tag{wMLSI}
    D(\rho \Vert E(\rho)) \le c_1 \EP_{\cL}(\rho) + c_2
\end{equation}
Applying Grönwall's Lemma yields
\begin{equation*}
    D(e^{t\cL}(\rho) \Vert E(\rho)) \le e^{-t/c_1} D(\rho \Vert \sigma) + c_2 \, ,
\end{equation*}
and then by Pinsker's inequality \eqref{eq:pinsker-inequality}, we obtain an estimate on the trace distance:
\begin{equation*}
    \norm{\rho - \sigma}_1 \le \sqrt{2 e^{-t/c_1}D(\rho \Vert \sigma) + 2c_2} \,, .
\end{equation*}
Since $D(\rho \Vert \sigma) \le \log \norm{\sigma^{-1}}_\infty$, in the best case—assuming constant $c_1$ and $c_2 = 0$—the mixing time need only be logarithmic in system size, classifying the system as rapid mixing. The novelty of our proofs lies in accepting positive $c_2$ which, however, decays with system size.
As with many quantum-mechanical capacity quantities, there exists a notion of complete MLSI (cMLSI), defined as the supremum of MLSI of $\id_R \otimes \cL$ over arbitrary reference systems $R$, denoted by $\alpha_c(\cL)$. This notation yields the relation
\begin{equation*}
    \alpha_c(\cL_A \otimes \id_B + \id_A \otimes \cL_B) \ge \min\{\alpha_c(\cL_A), \alpha_c(\cL_B)\} \, ,
\end{equation*}
which was proven in \cite{Gao.2020} and is known for classical MLSI. Interestingly enough a relation between the gap of a GNS-symmetric QMS and the cMLSI was recently proven in \cite{Gao.2022a, Gao.2022, Rouze.2024a} and we want to quote this estimate here as we later will use it in our analysis:
\begin{equation}\label{eq:cmlsi-estimate-via-gap}
    \alpha_c(\cL) > \frac{\lambda(\cL)}{2\log(10C_{\cb}(E))} \, ,
\end{equation}
where $C(E) := \inf\{c \ge 0 : \rho \le c E(\rho),\, \forall \rho \in \cS(\cH)\}$, $C_{\cb}(E) := \sup_{R} C(\id_R \otimes E)$ is the (complete) Pimsner-Popa index, respectively.

\subsection{Local commuting Hamiltonian and Davies generators on lattices}\label{subsec:local-commuting-hamiltonians-and-davies-generators-on-lattices}
In this paper, we consider $D$-dimensional hypercubes $\Lambda := \Lambda_L := \llbracket-L, L\rrbracket^D$ of side length $2L + 1$ consisting of $N = |\Lambda|$ sites, where each site hosts a qudit system $\cH_i = \C^d$. The Hilbert space of the complete system is thus $\cH_\Lambda = \bigotimes_{k \in \Lambda} \cH_k$, and for any subsystem $A \subseteq \Lambda$, we have $\cH_A = \bigotimes_{k \in A} \cH_i$. In isolation, the system's dynamics are fully characterized by its Hamiltonian $H_\Lambda \in \cB_{\sa}(\cH)$. When in thermal equilibrium with a heat bath at fixed inverse temperature $\beta > 0$, the system eventually assumes its thermal state, also known as the Gibbs state:
\begin{equation*}
    \sigma := \sigma^\Lambda := \frac{e^{-\beta H_\Lambda}}{\tr[e^{-\beta H_\Lambda}]} \, .
\end{equation*}
Our analysis focuses on a special class of Hamiltonians that are $(\kappa, r)$-local and commuting, with bounded interaction strength and connectivity. Specifically, there exists a representation
\begin{equation*}\label{eq:definition-local-hamiltonian}
    H_\Lambda = \sum\limits_{A \subseteq \Lambda} h_A \otimes \Id_{\overline{A}}
\end{equation*}
where $\overline{A} := \Lambda \backslash A$ denotes the complement in $\Lambda$ and $h_A \in \cB_{\sa}(\cH_A)$, satisfying the following conditions:
\begin{enumerate}
    \item Commutativity: All terms $h_A \otimes \Id_{\overline{A}}$, $A \subseteq \Lambda$ pairwise commute;
    \item $(\kappa, r)$-locality: $h_A = 0$ if either $|A| > \kappa$ or $\diam(A) > r$;
    \item Bounded interaction strength: $\max_{A \subseteq \Lambda} \norm{h_A}_\infty =: J < \infty$;
    \item Bounded connectivity: $\max_{k \in \Lambda} |\{A \subseteq \Lambda : h_A \ne 0, k \in A\}| =: g < \infty$.
\end{enumerate}
Note that the \emph{growth constant} $g$ is a function of $\kappa, r$ and $D$, and that $\kappa,r$ are not independent. Despite that, we will treat them in this work as separate parameters. \\
When considering a family of Hamiltonians on increasing lattice sizes, such as those arising from an interaction on $\Z^D$, all the above constants are implicitly assumed to be independent of system size. In what follows, we will suppress the identity and simply write $h_A$, where the subindex indicates the support of the Hamiltonian term. For any set $R \subseteq \Lambda$, we define its closure dictated by the interaction structure of the Hamiltonian as
\begin{equation*}
    R\partial := \{k \in \Lambda : \exists A \subseteq \Lambda, h_A \ne 0, A \cap R \ne \emptyset, k \in A\}
\end{equation*}
and its boundary as $\partial R := R\partial \backslash R$. For our analysis, we introduce local Hamiltonians and local Gibbs states. The local Hamiltonian of $R \subseteq \Lambda$ is defined as
\begin{equation*}
    H_R := \sum_{A \subseteq R} h_A
\end{equation*}
and is strictly supported in $R$. Its corresponding local Gibbs state is $\sigma^R := \frac{e^{-\beta H_R}}{\tr[e^{-\beta H_R}]}$, which should not be confused with the marginal Gibbs state on $R$ given by $\sigma_R := \tr_{\overline{R}}[\sigma]$. Using the locality and connectivity constraints, one readily obtains
\begin{align}\label{eq:norm-estimates-local-hamiltonian}
    \norm{H_R}_\infty &\le g J |R|  \, ,
    % \label{eq:norm-estimates-local-gibbs-state}
    % \norm{(\sigma^R)^{-1}}_\infty &\leq d^{|R|}e^{2\beta gJ},
    % Not correct as it is stated atm: Counter example H_R = 0, gives \sigma^R = 1/d_{|\Lambda|}
\end{align}

which will prove useful in subsequent derivations.

\medskip

We previously mentioned that, when a system is brought into contact with a heat bath, it is assumed to thermalize and eventually conform to the Gibbs state statistics. Assuming a specific interaction form between the bath and the system, i.e., a combined Hamiltonian of the form
\begin{equation}
    H = H_\Lambda + H^{\HB} + \sum\limits_{k \in \Lambda, \alpha} S_{\alpha, k} \otimes B_{\alpha, k} \, ,
\end{equation}
where $H^{\HB}$ is the bath Hamiltonian and $\{S_{\alpha, k} \otimes B_{\alpha, k}\}_{k \in \Lambda, \alpha}$ are self-adjoint single-site couplings one can derive an effective system dynamic through a weak coupling limit \cite{Spohn.1978}. Note that we will not only assume that the $\{S_{\alpha, k}\}_\alpha$ are self-adjoint but for every $k$ form a Kraus decomposition of the partial trace on the site $k$, i.e. $\sum\limits_{\alpha} S_{\alpha, k} \, \cdot \, S_{\alpha, k} = \tr_k[\,\cdot\,]$. An example would be the Pauli operators on $\C^2$. This choice ensures that in the trivial case $H_\Lambda = 0$ one recovers the depolarising semigroup. The effective dynamics obtained by the weak-coupling limits are governed by the so-called Davies Lindbladien
\begin{equation}\label{eq:davies-lindbladian}
     {\cL}^{D, H}_{\Lambda}(\rho) := -i[H_\Lambda,\rho] + \cL_\Lambda^D(\rho) \:=  -i[H_\Lambda,\rho] +\sum_{k\in\Lambda}\,\cL^{D}_{k}(\rho) \, ,
\end{equation}
for some local generators
\begin{equation}\label{eq:local-davies-lindbladian}
    \cL^{{D}}_{k}(\rho)=\sum_{\omega,\alpha}\,\chi^{\beta,\omega}_{\alpha,k}\,\Big(  S_{\alpha,k}^{\omega}\rho S_{\alpha,k}^{\omega,\dagger}-\frac{1}{2}\,\big\{ \rho, S_{\alpha,k}^{\omega,\dagger}S_{\alpha,k}^{\omega} \big\} \Big) \, .
\end{equation}
The sum in \eqref{eq:local-davies-lindbladian} ranges over the index $\alpha$ of the local basis $\{S_{\alpha,k}\}$ as well as the Bohr frequencies $\omega$ of the Hamiltonian $H_{k\partial}$, i.e. all pairwise differences of its eigenvalues that arise as a result of the decomposition
\begin{equation}\label{eq:bohr-frequency-decomposition-of-jump-operators}
    e^{-itH_\Lambda}\,S_{\alpha,k}e^{it H_\Lambda} = e^{-itH_{k\partial}}\,S_{\alpha,k}e^{itH_{k\partial}} = \sum_\omega e^{it\omega}S_{\alpha,k}^{\omega} \, ,
\end{equation}
for arbitrary $t \in \R$. Note that the validity of the first equality relies on the commuting, local nature of the Hamiltonian, leading to the cancellation of all Hamiltonian terms that do not intersect $k$, thereby localizing the Bohr frequencies (i.e., their dependence only on  $H_{k\partial}$). For better readability, we will not make the dependence on $k$ explicit, however. Lastly the $\chi^{\beta, \omega}_{\alpha, k}$ in \eqref{eq:local-davies-lindbladian} are the Fourier coefficients of the two-point correlation functions of the environment, which we assume to be uniformly bounded above and below: $0 < \chi_{\min}^\beta\le \chi_{\alpha,k}^{\beta,\omega}\le \chi_{\max}^\beta$. By assumptions made in the weak coupling limit, those coefficients satisfy the KMS condition, i.e. $\chi_{\alpha,k}^{\beta,-\omega}=e^{-\beta\omega}\,\chi_{\alpha,k}^{\beta,\omega}$ making the $\cL^D_k$, GNS-symmetric w.r.t. to the Gibbs state of the same temperature. Similarly to the local Hamiltonians, we can define the Lindbladian on a set $R \subseteq \Lambda$ by $\cL^D_R$ by restricting the sum in \eqref{eq:davies-lindbladian} to $R$. Unlike the local Hamiltonians, however, the local Lindbladiens act non-trivially on $R\partial$ instead of $R$. One can define semigroups from these local Lindbladiens that, due to their GNS-symmetry also converge to projections respectively conditional expectation in the Heisenberg picture, we will denote by
\begin{equation*}\label{eq:local-davies-conditional-expectations}
    E_A^* := \lim\limits_{t \to \infty} e^{t\cL_A^{D, *}} \quad\text{or equivalently}\quad E_A := \lim\limits_{t \to \infty} e^{t\cL_A^{D}} \, .
\end{equation*}
Note that these maps satisfy the following properties \cite{Bardet.2021,Bardet.2024,Kastoryano.2016}: 
 \begin{enumerate}
    \item $E_{\emptyset}=\id$ and $E_\Lambda=\tr[\,\cdot\,]\,{\sigma}^\Lambda$;
    \item for each $A\subset A\partial\subseteq B\subseteq \Lambda$, $\sigma^{B}$ is a fixed point of $E_A$;
    \item for each $A\subseteq B\subseteq \Lambda$, $E_AE_{B}=E_{B}E_A=E_B$;
    \item for any two regions $A,B\subseteq\Lambda$ such that $A\partial\cap B\partial=\emptyset$, $E_{A\sqcup B}=E_AE_B=E_BE_A$,
\end{enumerate}
that will become useful later. Note that here and in the following we will use $\sqcup$ to denote a disjoint union of sets. By using the representation 
\begin{equation}\label{eq:local-davies-projection-as-limit-of-petz-recoveries}
    E_A = \lim\limits_{t \to \infty} \cR_{\sigma, A}^n \, ,
\end{equation}
in terms of the Petz recovery
\begin{equation*}
    \cR_{A, \sigma}(X) := \sigma^{1/2} (\sigma_{\overline{A}}^{-1/2} \tr_A[X] \sigma_{\overline{A}}^{-1/2}) \otimes \Id_A  \,\sigma^{1/2} \, ,
\end{equation*}
proven in \cite[Theorem 1]{Bardet.2021} and the locality of the Hamiltonian one readily verifies
\begin{equation}\label{eq:composition-relation-local-davies-local-depolarising-channel}
    E_A \E_A = E_A \quad\text{and}\quad \E_{A\partial} E_A = \E_{A\partial} \, . 
\end{equation}

Although $\cL^D_A$ and $E_A$ for $A \subseteq \Lambda$, as well as $\sigma^A$ and $\sigma_A$, depend on the temperature, we will only make this dependence explicit where necessary to avoid a cluttering of indices. 

\subsection{Decay of correlations in Gibbs states}\label{subsec:decay-of-correlations-in-gibbs-states}
The theory of thermalization in classical spin systems \cite{Martinelli.1999} suggests a strong connection between equilibrium and non-equilibrium properties of quantum spin systems. In classical systems, a well-established principle links the convergence rate of local stochastic dynamics to the decay of correlations in the Gibbs equilibrium measure. Specifically, the Glauber dynamics, which models the thermalization of a discrete spin system, rapidly approaches the equilibrium Gibbs measure if and only if correlations between spatially separated regions exhibit exponential decay with distance. Here, we consider a different notion of the decay of correlations and relate it to more standard classical notions.

\subsubsection{Correlation measures and the MCMI}\label{subsec:correlation-measures-and-the-mcmi}
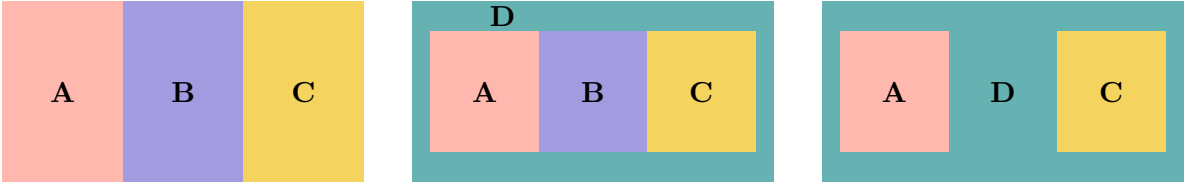
\begin{figure}[h!]
    \begin{center}
        \begin{tikzpicture}
            \def\colorD{TealMid}
            \def\colorA{CoralMid}
            \def\colorB{SlateBlueMid}
            \def\colorC{GoldenrodMid}
            \begin{scope}[scale=0.8]
                \fill[\colorA] (0, 0) rectangle (2, 3);
                \node at (1, 1.5) {\textbf{A}}; % Center of A
                \fill[\colorB] (2, 0) rectangle (4, 3);
                \node at (3, 1.5) {\textbf{B}}; % Center of B
                \fill[\colorC] (4, 0) rectangle (6, 3);
                \node at (5, 1.5) {\textbf{C}}; % Center of C
            \end{scope}

            \begin{scope}[xshift=0.33\linewidth, scale=0.8]
                \fill[\colorD] (0, 0) rectangle (6, 3);
                \node at (1.5, 2.75) {\textbf{D}}; % D
                \fill[\colorA] (0.3, 0.5) rectangle (2.1, 2.5);
                \node at (1.2, 1.5) {\textbf{A}}; % Center of A
                \fill[\colorB] (2.1, 0.5) rectangle (3.9, 2.5);
                \node at (3, 1.5) {\textbf{B}}; % Center of B
                \fill[\colorC] (3.9, 0.5) rectangle (5.7, 2.5);
                \node at (4.8, 1.5) {\textbf{C}}; % Center of C
            \end{scope}

            \begin{scope}[xshift=0.66\linewidth, scale=0.8]
                \fill[\colorD] (0, 0) rectangle (6, 3);
                \node at (3, 1.5) {\textbf{D}}; % Center of the full rectangle
                \fill[\colorA] (0.3, 0.5) rectangle (2.1, 2.5);
                \node at (1.2, 1.5) {\textbf{A}}; % Center of A
                \fill[\colorC] (3.9, 0.5) rectangle (5.7, 2.5);
                \node at (4.8, 1.5) {\textbf{C}}; % Center of C
            \end{scope}
        \end{tikzpicture}
        \caption{A lattice $\Lambda$ is partitioned into four distinct regions, such that $A$ and $C$ are separated by either $B$ or $D$ or both. $B$ is a system that is averaged over (through a partial trace) while $D$ we condition on.}\label{fig:ABCD}
    \end{center}
\end{figure}

The notion we want to introduce is that of the matrix-valued quantum conditional mutual information or MCMI for short. It has previously appeared as a tool to prove the decay of correlations in Gibbs states. More precisely the decay of the (conditional) mutual information \cite{Kuwahara.2020}. This result, unfortunately, contains a mistake, meaning we cannot recycle the decay of MCMI from there and have to rely on the toolbox of effective Hamiltonians, introduced later in \cref{subsec:the-effective-hamiltonian-as-tool-to-proof-correlation-decay}.
\begin{defi}[Uniform exponential decay of matrix-valued quantum conditional mutual information]\label{defi:definition-mcmi}
    Given any partition $\Lambda=A\sqcup B\sqcup C\sqcup D$ of the lattice, we define the \textit{matrix-valued quantum conditional mutual information} (MCMI) of a state $\sigma$ as
    \begin{align}\label{eq:extended:definition-mcmi}
        \mathbf{H}_\sigma(A:C|D) := \log\sigma_{ACD} + \log\sigma_{D} - \log\sigma_{AD} - \log\sigma_{CD} \, .
    \end{align}
    Then the Gibbs state $\sigma=\sigma^\Lambda$ is said to satisfy \textit{uniform exponential decay of its matrix-valued quantum conditional mutual information} with constants $K,\xi$ if there exist constants $K,\xi>0$ independent of the regions $A, B, C, D$ and such that
    \begin{align}\label{eq:extended:decay-mcmi}
        H_\sigma(A:C|D) := \norm{\mathbf{H}_\sigma(A:C|D)}_\infty 
        \leq K\abs{\Lambda}e^{-\frac{\dist(A,C)}{\xi}} \, .   
    \end{align}
\end{defi}
Note that under slight abuse of notation, we will be referring to both $\textbf{H}$ and $H$ as the MCMI. To illustrate this decay and its geometric aspects, let us examine three two-dimensional situations in \cref{fig:ABCD} which, despite their simplicity, capture all use cases later. The leftmost and the rightmost can be regarded as the extremal cases while the one in the middle in some sense interpolates the two. What all of them have in common and what is the crucial property is the separation of the regions $A$ and $C$. This can be achieved either by averaging as in the case of $B$ (operationally a partial trace) or through conditioning as in the case of $D$.

The system size-dependent prefactor we impose in our examples can be replaced by one that only depends on the minimum sizes of $A$ and $C$ (see \cref{sec:examples}). For our purposes, the linear dependence is, however, sufficient. As the work in \cite{Kuwahara.2020} suggests the MCMI is closely tied to other correlation measures. More precisely: Given MCMI decay, one in particular has the decay of the conditional mutual information
\begin{align*}
    I_\sigma(A:C|D) &:= -S(\sigma_{ACD}) - S(\sigma_{D}) + S(\sigma_{CD}) + S(\sigma_{AD})\\
    &= \tr[\sigma_{ACD} \,\textbf{H}_\sigma(A:C|D)],
\end{align*}
the mutual information 
\begin{align*}
    I_{\sigma}(A:C) &:= -S(\sigma_{AC}) + S(\sigma_A) + S(\sigma_C)\\
    &= \tr[\sigma_{AC} \,\textbf{H}_\sigma(A:C|\emptyset)],
\end{align*}
and in consequence also the decay of the covariance (see \cite{Bluhm.2022}). Note that $S(\sigma) := -\tr[\sigma \log \sigma]$ here denotes the von Neuman entropy. Besides its connection to quantum information theoretic quantities, the MCMI also is connected to a notion of decay of correlation from classical statistical physics, namely complete analyticity. This comes as no big surprise since complete analyticity in classical systems implies a logarithmic Sobolev inequality and thereby also rapid mixing \cite[Theorem 4.1]{Cesi.2001}.

\subsubsection{MCMI decay from complete analyticity in classical systems}\label{subsec:mcmi-decay-from-complete-analyticity-in-classical-systems}
Let us more closely investigate the connection to the \emph{complete analyticity} also called the Dobrushin-Shlosmann condition of classical spin systems \cite{Dobrushin.1985, Dobrushin.1987, Cesi.2001}. We will show that indeed complete analyticity implies uniform decay of the MCMI. Although we believe that also the reverse implication holds, we are still missing the proof as of the day of writing.\\
Note that in this section we will follow the notation of \cite{Martinelli.1999} mostly (re)defining the necessary objects ad-hoc. We will further suppress all temperature dependence to improve readability. Let us consider the single spin system $S = \{-1, 1\}$ with counting measure leading to a full state space $\Omega = \{\omega:\Z^D \to S\}$. We will write $\omega_V = \omega|_{V}$ for the restriction of $\omega \in \Omega$ to $V$. The interaction which completely characterises the model is given as $J:\{V \Subset \Z^D\} \to \R$ and we write $J_V$ instead of $J(V)$. Here and in the following $V \Subset \Z^D$ refers to finite subsets. The corresponding Hamiltonian on $V \Subset \Z^D$ is given by 
\begin{equation*}
    H_V(\omega) = -\sum\limits_{A:\; A \cap V \neq \emptyset} J_A \prod\limits_{x \in A} \omega(x) \, ,
\end{equation*}
and the Gibbs measure for the boundary condition $\tau \in \Omega$ as 
\begin{equation*}
    \mu_V^\tau(\omega) = 
    \begin{cases}
        (Z_V^\tau)^{-1} \exp[- H_V(\omega)] & \text{if } \omega_{\overline{V}} = \tau_{\overline{V}} \, ,\\
        0 & \text{else.}
    \end{cases}
\end{equation*}
Here and in the rest of this section a line over a set denotes the complement w.r.t. $\Z^D$, e.g. $\overline{V} = \Z^D \backslash V$ in the above definition. Interestingly for $V \subset \Lambda \Subset \Z^D$, we find that $\mu_V^\tau$ alternatively can be written as
\begin{equation}\label{eq:rewriting-gibbs}
    \mu_V^\tau(\omega) = 
    \begin{cases}
        (\tilde{Z}_V^\tau)^{-1} \exp[- H_\Lambda(\omega)] & \text{if } \omega_{\overline{V}} = \tau_{\overline{V}}\, ,\\
        0 & \text{else.}
    \end{cases}
\end{equation}
i.e. the Hamiltonian on $V$ can be replaced by the Hamiltonian on $\Lambda$ through a suitable change of normalisation $Z_V^\tau \to \tilde Z_V^\tau$. This is due to the conditioning with $\tau$ which as a consequence ensures the existence of a $c > 0$ only dependent on $\tau_{\overline{V}}$, s.t. for all $\sigma_V$, $H_V(\sigma_V \tau_{\overline{V}}) + c = H_\Lambda(\sigma_V\tau_{\overline{V}})$. The marginal of a Gibbs distribution on $\Delta \subset V$ is defined as $\mu_{V, \Delta}^\tau:\Omega \to \R$ with
\begin{equation}\label{eq:classical-marginal}
    \mu_{V, \Delta}^\tau(\omega) = \sum\limits_{\nu: \nu|_{\Delta} = \omega|_{\Delta}} \mu_{V}^\tau(\nu) \, .
\end{equation}
By the definition of \cite{Cesi.2001}, a system is \emph{completely analytic} if there exists $K > 0$, $\xi > 0$ such that for all $V \Subset \Z^D$, $x \in \partial V := \{x \in \overline{V} \;:\; \dist(x, V) \le r\}$, $\Delta \subset V$ and $\tau, \tau' \in \Omega$ with $\tau(y) = \tau'(y)$ for all $y \neq x$
\begin{equation}
    \norm{\frac{\mu_{V, \Delta}^\tau}{\mu^{\tau'}_{V, \Delta}} - 1}_\infty \le K e^{-\frac{\dist(x, \Delta)}{\xi}} \, . 
\end{equation}
Now let us try to make the connection to ~\cref{defi:definition-mcmi}. Let us first fix the equivalent of $\sigma^\Lambda$ of $\Lambda \Subset \Z^D$. Since quantumly we are considering only Hamiltonians on finite lattice $\Lambda$ let us fix a boundary condition $\nu \in \Omega$ and set classical analogue of our Gibbs measure $\sigma_\Lambda$ on $\Lambda$ to be $\mu^\nu = \mu_\Lambda^\nu:\Omega \to \R$. Marginals of this distribution are defined through \eqref{eq:classical-marginal} and for $\Sigma \subset \Lambda$ denoted by $\mu^\nu_\Sigma$. We further adapt the shorthand notation $\mu^\nu_{\Sigma|\Lambda} := \frac{\mu^{\nu}_{\Sigma\Lambda}}{\mu_{\Lambda}^\nu}$. Incorporating the implicit boundary condition explicitly, ~\cref{defi:definition-mcmi} states: There exist $\xi > 0$ and $K > 0$ such that for all partitions $\Lambda = A\sqcup B\sqcup C\sqcup D$
\begin{equation}
    \norm{\log\frac{\mu^\nu_{A|CD}}{\mu^\nu_{A|D}}}_\infty \le K \abs{\Lambda} e^{-\frac{\dist(A, C)}{\xi}} \, . 
\end{equation}
First note that it suffices to investigate the case when $C$ only contains a single site, as we can write
\begin{equation}\label{eq:classical-mcmi}
    \norm{\log\frac{\mu^\nu_{A|CD}}{\mu^\nu_{A|D}}}_\infty \le \sum\limits_{i = 1}^{|C|} \norm{\log\frac{\mu^\nu_{A|\{x_i\}D_i}}{\mu^\nu_{A|D_i}}}_\infty 
\end{equation}
where $D_1 = D$, $D_i = D_{i - 1} \sqcup C_{i - 1}$ and $C_i = \{x_i\}$ with $\{x_i\}_{i = 1}^{|C|}$ an enumeration of $C$. In the following, we will drop $\nu$ for better readability. By the argument above we are now in the setting of $|C| = 1$. Let us assume $\tau \in \Omega$ to be the optimiser such that $\norm{\log\frac{\mu_{A|CD}}{\mu_{A|D}}}_\infty = \log \frac{\mu_{A|D}(\tau)}{\mu_{A|CD}(\tau)}$. In the case that the optimum is achieved for the reversed fraction, the convexity of $x \mapsto \frac{1}{x}$ gives an inequality instead of equality in the second line of \eqref{eq:estimate} while also the roles of $\tau$ and $\tau'$  change, however, leaving the rest of the argument invariant. Let us denote by $\tau'$ the element of $\Omega$ which agrees with $\tau$ on $\overline{C}$ but has a flipped spin at site $C$. By $\log(x + 1) \le x$, the fact that $\mu_{C|D}(\tau) = 1 - \mu_{C|D}(\tau')$ with both $\mu_{C|D}(\tau)$ and $\mu_{C|D}(\tau')$ in $[0, 1]$, we get
\begin{equation}\label{eq:estimate}
    \begin{aligned}
        \log \frac{\mu_{A|D}(\tau)}{\mu_{A|CD}(\tau)} 
        &\le \frac{\mu_{A|D}(\tau)}{\mu_{A|CD}(\tau)} - 1 = \frac{\mu_{A|CD}(\tau)\mu_{C|D}(\tau) + \mu_{A|CD}(\tau')\mu_{C|D}(\tau')}{\mu_{A|CD}(\tau)} - 1\\
        &= \mu_{C|D}(\tau') \left(\frac{\mu_{A|CD}(\tau')}{\mu_{A|CD}(\tau)} - 1\right) \le \left|\frac{\mu_{A|CD}(\tau')}{\mu_{A|CD}(\tau)} - 1\right| = \left|\frac{\mu^{\tau'}_{AB, A}(\tau)}{\mu^\tau_{AB, A}(\tau)} - 1\right|\\
        &\le \norm{\frac{\mu^{\tau'}_{AB, A}}{\mu^\tau_{AB, A}} - 1}_\infty \, .
    \end{aligned}
\end{equation}
In the last equality, we used the fact that 
\begin{equation}\label{eq:identification}
    \mu_{AB, A}^\tau(\omega) = \mu_{A|CD}(\omega_A\tau_{\overline{A}}) \, ,
\end{equation}
which one readily obtains from \eqref{eq:rewriting-gibbs}. As the system satisfies complete analyticity, this last norm is uniformly bounded by $K e^{-\frac{\dist(A, C)}{\xi}}$ which for the general setting of \eqref{eq:classical-mcmi} allows us to conclude
\begin{equation*}
    \norm{\log\frac{\mu_{A|CD}}{\mu_{A|D}}}_\infty \le K |C| e^{-\frac{\dist(A, C)}{\xi}} \, . 
\end{equation*}

\subsubsection{The effective Hamiltonians as tool to proof correlation decay}\label{subsec:the-effective-hamiltonian-as-tool-to-proof-correlation-decay}
Let us conclude the section about the theoretical framework by introducing a tool which has proved useful when showing the decay of some correlation measures for explicit models. This tool is called effective Hamiltonians and has appeared in \cite{Kuwahara.2020} and was further developed in \cite{Bluhm.2024}. We will restrict our discussion to the case where  $\Lambda \subseteq \mathbb{Z}^D$ is a finite hypercube and address the Hamiltonian directly, whereas the results in \cite{Bluhm.2024} apply to interactions on general graphs. Next, we introduce an essential norm originally developed for interactions; here, we define it specifically for Hamiltonians with a fixed local representation (cf. \eqref{eq:definition-local-hamiltonian}): For $H_\Lambda$ with representation as in \eqref{eq:definition-local-hamiltonian} and $\mu > 0$ the interaction norm is defined as
\begin{equation}\label{eq:definition-interaction-norm}
    \inorm{H_\Lambda}_\mu := \sup_{x \in \Lambda} \sum\limits_{X \subseteq \Lambda : x \in X} \norm{h_X}_\infty e^{\mu \diam(X)} \, . 
\end{equation}
Slightly adapting the definition from \cite[Definition 3.1]{Bluhm.2024} we say that $H_\Lambda$ admits a \emph{strong local effective Hamiltonian at inverse temperature $\beta > 0$} if for every $A \subseteq \Lambda$ there exists a Hamiltonian with a local decomposition $\tilde H^A = \sum_{X \subseteq \Lambda} \tilde h^A_X$ such that 
\begin{enumerate}
    \item $\tilde h^A_X$ is supported in $X \cap A$ for every $X \subseteq \Lambda$;
    \item For $A' \subseteq \Lambda$, $\tilde h_X^A = \tilde h_X^{A'}$ for all $X \subseteq \Lambda$ for which $X \cap A' = X \cap A$;
    \item One has $\log d_A^{-1} \Id_A \otimes \tr_A[e^{-\beta H_\Lambda}] = \sum\limits_{X \subseteq \Lambda} \tilde h^A_X = \tilde H^A$.
\end{enumerate}
The existence of an effective Hamiltonian alone does not suffice to ensure the decay of the MCMI, as the above definition lacks information on the system’s locality. In general, this decay will require an additional locality assumption (see \cref{subsec:mcmi-decay-from-strong-effective-hamiltonian}). However, when the system Hamiltonian is also marginal commuting, \cite[Theorem 4]{Bluhm.2024} directly guarantees the existence of an effective Hamiltonian with explicit decay properties. This result allows us to establish uniform decay of the MCMI at high temperatures in \cref{subsec:mcmi-decay-from-commutings-marginals-at-high-temperature}. To clarify, we define what it means for $H_\Lambda$ to have commuting marginals: We say that $H_{\Lambda} = \sum_{A \subseteq \Lambda} h_A$ is \emph{marginal commuting} if there exists a commuting algebra $\cA$ that contains as a sub algebra the one generated by all the local terms $h_A$ and remains invariant under the application of $\E_A$ for any subset $A\subseteq \Lambda$, i.e., $\E_A[\cA] \subset \cA $. This definition is exactly \cite[Definition 3.5]{Bluhm.2024} and is named here to highlight that, under this assumption, arbitrary marginals of the Hamiltonian state commute. 

\section{Proof techniques - extended}\label{sec:proof-techniques-extended}

\subsection{A general weak entropy factorization}\label{subsec:a-general-weak-entropy-factorization}
This section is dedicated to the proof and a short discussion of a weak entropy factorization. More precisely a splitting of the conditional relative entropy defined for $\cH = \cH_A \otimes \cH_{\overline{A}}$ with $A \subseteq \Lambda $ as 
\begin{equation}\label{eq:definition-conditional-relative-entropy}
    D_A(\rho \Vert \sigma) := D(\rho \Vert \sigma) - D(\rho_{\overline{A}} \Vert \sigma_{\overline{A}})
\end{equation}
for $\rho, \sigma \in \cS(\cH_\Lambda)$. We can understand this as the relative entropy in the $A$ subsystem conditioned on the rest of the lattice. It made its first appearance in \cite{Capel.2018} where it also was connected to the Davies channel on $A$, more precisely the following important inequality was shown:
\begin{equation}\label{eq:davies-conditional-expectation-and-conditional-relative-entropy}
    D_A(\rho \Vert \sigma) \le D(\rho \Vert E_A(\rho)) \, . 
\end{equation}

The following lemma details the weak entropy factorization which is one of the key ingredients in our main results. 
\begin{lem}[Weak entropy factorization]\label{lem:extended:weak-entropy-factorization}
   Let $\cH = \cH_A \otimes \cH_B \otimes \cH_C \otimes \cH_D$ and $\rho, \sigma \in \cS(\cH)$ with $\sigma > 0$, then 
   \begin{equation}\label{eq:extended:weak-entropy-factorization}
       D_{ABC}(\rho \Vert \sigma) \le D_{AB}(\rho \Vert \sigma) + D_{BC}(\rho \Vert \sigma) + \norm{\mathbf{H}_\sigma(A:C|D)}_\infty \, . 
   \end{equation}
\end{lem}
\begin{proof}
    We have that 
    \begin{align*}
        D_{ABC}(\rho \Vert \sigma) &- D_{AB}(\rho\Vert \sigma) - D_{BC}(\rho \Vert \sigma) \\
        &= -D(\rho_{D}\Vert \sigma_D) - D(\rho_{ABCD} \Vert \sigma_{ABCD}) + D(\rho_{CD} \Vert \sigma_{CD}) + D(\rho_{AD} \Vert \sigma_{AD}) \\
        &\le -D(\rho_{D}\Vert \sigma_D) - D(\rho_{ACD} \Vert \sigma_{ACD}) + D(\rho_{CD} \Vert \sigma_{CD}) + D(\rho_{AD} \Vert \sigma_{AD}) \smalltag{DPI}\\
        &\le \tr[\rho_{ACD}(\log \sigma_{ACD} + \log \sigma_{D} - \log \sigma_{AD} - \log \sigma_{CD})] \smalltag{SSA}\\
        & \le  \norm{\mathbf{H}_\sigma(A:C|D)}_\infty \, . \smalltag{Hölder}
    \end{align*}
\end{proof}
The correction's additive nature is why we only manage to obtain the weak approximate tensorization for the Davies (\cref{subsec:a-weak-approximate-tensorization-for-davies-channels}) and as a consequence a weak modified logarithmic Sobolev inequality and a weak transport cost inequality (\cref{subsec:a-weak-transport-cost-inequality-for-davies-channels}). Lifting the above inequality to one which has a multiplicative correction would strengthen all results in this paper to their strong counterparts and further allow one to eliminate the requirement for a polynomial gap in \cref{subsec:rapid-mixing-from-mcmi-decay-and-polynomial-local-gap} and in the strengthening of the $W^1$ mixing in \cref{subsec:rapid-mixing-from-mcmi-decay-and-polynomial-local-gap}. At the date of writing, we only managed to prove the inequality in the fully classical case where we obtain
\begin{equation*}
        (1 - \norm{\exp(\mathbf{H}_\sigma(A:C|D)) - \Id}_\infty) D_{ABC}(\rho \Vert \sigma) \le D_{AB}(\rho \Vert \sigma) + D_{BC}(\rho \Vert \sigma) \, . 
\end{equation*}
Note that in the special case where $D$ is empty, i.e. $\dim \cH_ D = 1$, the authors in \cite{Capel.2018a} show a multiplicative correction. Their proof yields a slightly altered factor, classically equivalent to the above bound.

\subsection{A coarse-graining of the hypercubic lattice}\label{subsec:a-coarse-graining-of-the-hypercubic-lattice}
In this section, we present the construction of the coarse-graining w.r.t. which we apply the weak entropy factorization (c.f. \cref{subsec:a-general-weak-entropy-factorization}) to prove the weak approximate tensorization later (c.f. \cref{subsec:a-weak-approximate-tensorization-for-davies-channels}). This coarse-graining will be at the heart of its proof and our divide-and-conquer scheme involving both exact and weak approximate tensorizations at every step.

We do this by recalling the construction of a coarse-graining of the 2-dimensional hypercube due to \cite{Brandao.2018} and extending it to any dimension $D\in\mathbb{N}$.  We denote by $\Lambda_L := \llbracket-L,L\rrbracket^D \subset \Z^D$ the $D$-dimensional hypercube of side length $2L + 1$. Our decomposition relies on the fixation of three parameters:
\begin{enumerate}
    \item $c \in \N$ with $c \ge r$ the \emph{overlap length}, ensuring a sufficiently fast decay when using a weak approximate tensorization to consecutively cut out cells reducing the extensive dimension in every step. 
    \item $k \in \N$ with $k \ge r$ the \emph{buffer length}, separating cells of the same dimensionality such that the corresponding conditional expectations commute and hence allow us to use tensorization of the relative entropy.
    \item $\ell \in \N$ with $L \ge \ell$ and $\ell$ odd the `extensive' side length of the cells we decompose into. As we will consecutively reduce the `extensive' dimensions until we reach zero subtracting $\ell$ in every step by $2(k + c)$, we further require $\ell > 2D(k + c)$.
\end{enumerate}

\begin{rmk} \label{rmk:rectifyingdivisibility}
    In the first step of the decomposition, we aim to divide $\Lambda_L$ into hypercubes of side length $\ell$. This might not be possible due to the choice of $\ell$ we made prior as divisibility of $2L + 1$ by $\ell$ may not be guaranteed. We can, however, always embed $\Lambda_L$ into some $\Lambda_{L + r}$ with $r \in \llbracket 1, l - 1\rrbracket$ such that $2(L + r) + 1$ is now divisible by $\ell$, when $\ell$ is odd. Since we are interested in how certain quantities scale in $\ell$ we may always choose $\ell$ odd, however, we will overlook this subtlety. 
    
    The original Hamiltonian is likewise embedded by padding it with zeros, leading to a Gibbs state of the form $\sigma^{\Lambda_L} \otimes \pi_{\text{rest}} \in \cS(\cH_{\Lambda_{L + r}})$ where $\pi_{\text{rest}}$ is the maximally mixed state on $\Lambda_{L + r}\backslash\Lambda_{L}$. The $(\kappa, r)$-locality and interaction strength of both Hamiltonians, original and padded agree, due to the specific embedding. Furthermore, the condition of the global and local gaps and the decay of MCMI hold for the embedded if and only if they hold for the original Hamiltonian. This means we employ our proof for the embedded version and immediately get a result for the original one. 

    Consider the example of the MLSI, where we denote by $\cL$ the Davies Lindbladian in $\Lambda_L$ and by $\cL_{\text{rest}}$ that of $\Lambda_{L+r} \setminus \Lambda_L$. The separation between them is due to the embedding which in addition gives $[\cL, \cL_{\text{rest}}] = 0$. Assuming the embedded system has MLSI, we have $D(e^{t(\mathcal{L} + \mathcal{L}_{\text{rest}})}(\rho') \Vert \sigma \otimes \pi_{\text{rest}}) \le e^{-\frac{t}{c_1}} D(\rho' \Vert \sigma \otimes \pi_{\text{rest}})$. By setting $\rho' = \rho \otimes \pi_{\text{rest}}$ and utilizing commutativity of the Lindbladians, we obtain $D(e^{t\mathcal{L}}(\rho) \otimes \pi_{\text{rest}} \Vert \sigma \otimes \pi_{\text{rest}}) \le e^{-\frac{t}{c_1}} D(\rho \otimes \pi_{\text{rest}} \Vert \sigma \otimes \pi_{\text{rest}})$. Through the tensorization of relative entropy, we can conclude that $c_1$ also serves as an MLSI constant for the original system.
    This approach extends to other contexts (e.g. the decay of Wasserstein distance). We've established that all conditions imposed on the original system remain unchanged for the embedding, meaning the only change in the obtained bounds is in the system size. Notably, $|\Lambda_{L + r}| \le 2^D |\Lambda_L|$ since $\ell \le L$, meaning the correction only depends on the lattice dimension and does not affect the scaling with system size.
    Throughout the paper, $\Lambda_L$ is treated as general. However, in the proofs, we always assume without loss of generality that $\ell$ divides $(2L + 1)$. When providing explicit constants for the bounds, we include the correction factor $2^D$.
\end{rmk}
 
In the process of constructing the coarse-graining, we will define various cells of dimension $a$, where $a$ ranges from 0 to $D-1$. These cells should be understood as embedded within the corresponding $D$-dimensional hypercube for which ${e_1, \ldots, e_D}$ denote the canonical orthonormal basis. We begin by partitioning the $D$-dimensional hypercube $C^D := \Lambda_L$ into smaller $D$-hypercubes, each with side length $\ell$. We denote these partition elements as $\{C_{D,i}^\partial\}_{i \in \cI_{D}}$ with $\cI_D = \llbracket1,  ((2L+1)/\ell)^D\rrbracket$.
For each cell $C_{D,i}^\partial$, we define two subsets:
\begin{enumerate}
    \item The interior of $C_{D,i}^\partial$, excluding a boundary layer of buffer length $k$:
    \begin{equation*}
        C_{D,i} := \{v \in C_{D,i}^\partial \;:\; \operatorname{dist}(v, \overline{C_{D,i}^\partial}) > k\} \, ;
    \end{equation*}
    \item A further restricted interior, excluding a boundary layer of width $k+c$, i.e excluding buffer and overlap length:
    \begin{equation*}
        \mathring{C}_{D,i} := \{v \in C_{D,i}^\partial \;:\; \operatorname{dist}(v, \overline{C_{D,i}^\partial}) > k+c\} \, . 
    \end{equation*}
\end{enumerate}

\begin{figure}[h]
    \centering
    \def\colorBoarder{TealDark}
    \def\colorCpartial{TealLight}
    \def\colorC{TealMid}
    \def\colorCcirc{TealDark}
    
    \begin{tikzpicture}[scale=0.2]
        \begin{scope}
            % Loop to create outer squares
            \foreach \x/\y/\i in {1/1/1, 1/10/2, 1/19/3, 10/1/4, 10/19/6, 19/1/7, 19/10/8, 19/19/9} {
                \drawInitialSquare{\x}{\y}{$C_{2,\i}^\partial$}{}{\colorCpartial}{\colorBoarder};
            }
        
            % Draw the central square with additional inner structure
            \drawInitialSquare{10}{10}{5}{
                % Inner nested squares
                \fill[\colorC] (0.5,0.5) rectangle (7.5,7.5);
        
                \fill[\colorCcirc] (1.5,1.5) rectangle (6.5,6.5);
        
                % Labels for inner structures
                \node at (0.6,0.4) {\large $C_{2,5}^\partial$};
                \node at (1.6,7.1) {\large $C_{2,5}$};
                \node at (4,4) {\large \textcolor{white}{$\mathring{C}_{2,5}$}};
            }{\colorCpartial}{\colorBoarder};
            % Axis labels and arrows
            \node at (-1.3,4.5) {\Large $\ell$};
            \draw [|-|, darkred, thick, line width=0.6mm] (-0.5,0.5) -- (-0.5,9.5);
        
            % Additional distance markers
            \draw [|-|, darkred, thick, line width=0.4mm] (17.5,14.5) -- (18.5,14.5);
            \node at (18,15.5) {$k$};
            \draw [|-|, darkred, thick, line width=0.4mm] (16.5,13.5) -- (17.5,13.5);
            \node at (17,14.5) {$c$};
        
            % Main label at the top
            \node at (13.5,29.5) {\Large $\Lambda_L \subset \mathbb{Z}^2$};
            \draw[\colorBoarder,thick] (0.5,0.5) rectangle (27.5,27.5);
        \end{scope}
        
        %%%%%%%%%%%%%%%%%%%%%%%%%%%%%%%%%%%%%%%%%%%%%%%%%%%%%%%%%
        % second picture
        %%%%%%%%%%%%%%%%%%%%%%%%%%%%%%%%%%%%%%%%%%%%%%%%%%%%%%%%%
        \begin{scope}[xshift=32cm]
            \fill[\colorCpartial] (0.5, 0.5) rectangle (27.5, 27.5);
            \foreach \x/\ox/\ow in {1.5/-1/0, 10.5/0/0, 19.5/0/1} {
                \foreach \y/\oy/\oh in {1.5/-1/0, 10.5/0/0, 19.5/0/1} {
                    \drawCtwoRectangleColor{\x}{\y}{7}{7}{\ox}{\oy}{\ow}{\oh}{\colorC};
                    \drawCtwoRectangleColor{\x+1}{\y+1}{5}{5}{2*\ox}{2*\oy}{2*\ow}{2*\oh}{\colorCcirc};
                    \drawCtwoRectangleText{\x+1}{\y+1}{5}{5}{2*\ox}{2*\oy}{2*\ow}{2*\oh}{\textcolor{white}{$\mathring{C}_2$}};
                }
            }
            \foreach \x in {8, 14, 20} {
                \foreach \y in {8, 14, 20} {
                    % Convert floating-point to dimension and compare
                    \pgfmathsetmacro{\xcomp}{\x}
                    \pgfmathsetmacro{\ycomp}{\y}
                    \ifdim\xcomp pt=14pt
                        \ifdim\ycomp pt=14pt
                            % Both \x and \y are 14.5
                        \else
                            \node[anchor=center] at (\x, \y) {\footnotesize $C_2$};
                        \fi
                    \else
                        \node[anchor=center] at (\x, \y) {\footnotesize $C_2$};
                    \fi
                }
            }
            \node at (0,0) {\Large $\Lambda_L$};
            \node[rotate=270] at (29.6,13.5) {\Large $2L + 1$};
            \draw [|-|,darkred,thick,line width=0.6mm](28.5,0.5) -- (28.5,27.5);
            \draw[\colorBoarder, thick] (0.5, 0.5) rectangle (27.5, 27.5);
        \end{scope}
    \end{tikzpicture}
    \caption{Splitting of $\Lambda_L = C^D$ for $D=2$ as described in the text. On the left side, $\Lambda_L$  has been split into $9$ $2$-cells. For the cell in the middle $C_{2,5}$ and $\mathring{C}_{2,5}$ are explicitly represented. On the right side, $C_2$ is the union of the midblue and darkblue regions, and $\mathring{C}_2$ is the union of the dark blue regions.}
    \label{fig:covering}
\end{figure}
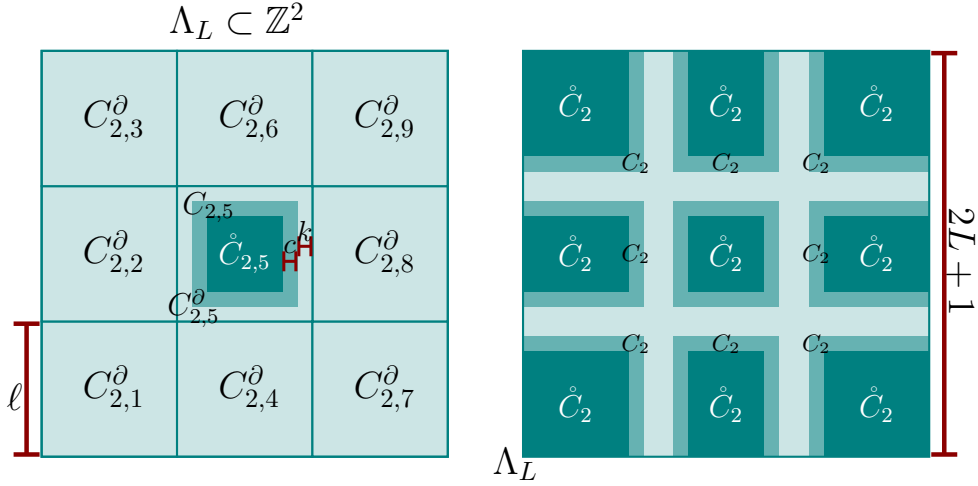

\noindent In the above $R \subseteq \Lambda_L$ is given by $\overline{R} = \Lambda_L \backslash R$. We define the aggregate sets:
\begin{equation*}
    C_D := \bigsqcup_{i \in \cI_D} C_{D,i} \qquad \mathring{C}_D := \bigsqcup_{i \in \cI_D} \mathring{C}_{D,i} , 
\end{equation*}
where $\bigsqcup$ denotes the disjoint union. Finally, we set $C^{D-1} := \Lambda_L \setminus \mathring{C}_{D}$. For a visual representation of this construction in two dimensions ($D=2$), refer to Figure~\ref{fig:covering}.

We continue the deconstruction by defining cells of lower extensive dimensions. For each dimension $a \in \llbracket1, D - 1\rrbracket$, we iteratively construct $a$-cells from $C^a$ as follows:
\begin{enumerate}
    \item First, we identify the set of vertices that ``join" each pair of neighbouring $\mathring{C}_{a + 1,i}$:
    \begin{equation*}
        C^\partial_a := \left\{v \in C^a \;:\; \exists j \in \mathbb{Z}, b \in \llbracket 1, D\rrbracket: v + je_{b} \in \overline{C^a}\right\} \, . 
    \end{equation*}
    \item By construction, $C^\partial_a$ is a disjoint union of fattened $a$-cells, which we denote as $C_{a,i}^\partial$ with index set $\cI_a$.
    \item We again define the two subsets:
    \begin{equation*}
        C_{a,i} := \{v \in C_{a,i}^\partial \;:\; \operatorname{dist}(v, \overline{C_{a,i}^\partial} \cap C^a) > k\} \, , \qquad \mathring{C}_{a,i} := \{v \in C_{a,i}^\partial \;:\; \operatorname{dist}(v, \overline{C_{a,i}^\partial} \cap C^a) > k+c\}
    \end{equation*}
    separated from the skeleton that remains after removal of $\bigsqcup_{b = a + 1}^D \mathring{C}_b$ by buffer and overlap plus buffer length respectively. We set the aggregated sets to be 
    \begin{equation*}
        C_a := \bigsqcup_{i \in \cI_a} C_{a,i} \,, \qquad \mathring{C}_a := \bigsqcup_{i \in \mathcal{I}_a} \mathring{C}_{a,i} \, .
    \end{equation*}
    \item Finally, we define the set for the next lower dimension: $C^{a-1} := C^a \setminus \mathring{C}_a$. 
\end{enumerate}

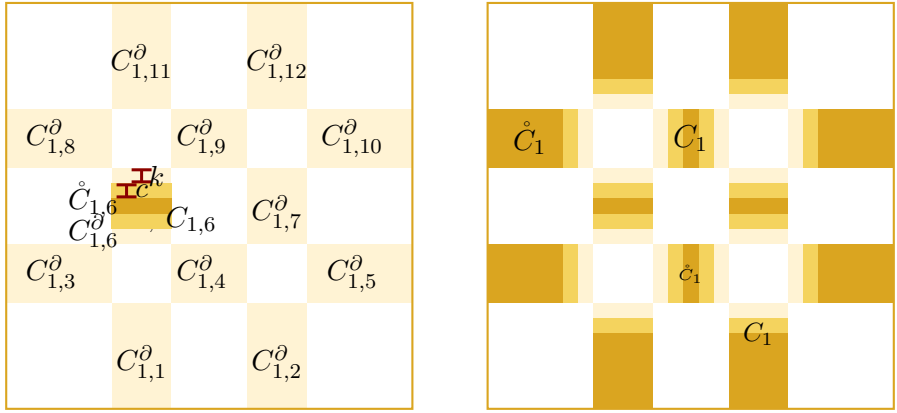
\begin{figure}[h]
    \centering 
    \begin{tikzpicture}[scale=0.2]
        \def\colorBoarder{GoldenrodDark}
        \def\colorCpartial{GoldenrodLight}
        \def\colorC{GoldenrodMid}
        \def\colorCcirc{GoldenrodDark}
        
        \begin{scope}

            % Loop to create the columns
            \foreach \x/\w in {0.5/7, 11.5/5, 20.5/7} {
                \drawConeColumns{\x}{1}{\w}{\colorCpartial}{1}
                \drawConeColumns{\x}{0}{\w}{\colorCpartial}{1}
            }
            
            % Loop to erase cutout
            \foreach \x/\ox/\ow in {2.5/-2/0, 11.5/0/0, 20.5/0/2} {
                \foreach \y/\oy/\oh in {2.5/-2/0, 11.5/0/0, 20.5/0/2} {
                    \drawCtwoRectangleColor{\x}{\y}{5}{5}{\ox}{\oy}{\ow}{\oh}{white};
                }
            }

            % Loop to create the labels
            \foreach \x/\ox/\ow [count=\i from 1] in {2/-1/0, 11/0/0, 20/0/1} {
                \foreach \y/\oy/\oh [count=\j from 1] in {2/-1/0, 11/0/0, 20/0/1} {
                    \pgfmathsetmacro{\idx}{int((\j - 1) * 5 + mod(\i - 1, 2) + 1)}
                    \pgfmathsetmacro{\idy}{int((\j - 1) * 5 + 2 + mod(\i - 1, 3) + 1)}
                    \begin{scope}
                        \clip (0.5,0.5) rectangle (27.5,27.5);
                        \node at (\x + 7.5 + 2*\ow, \y + 2.5 + \oy + \oh) {$C^\partial_{1, \idx}$};
                        \node at (\x + 2.5 + \ox + \ow, \y + 7.5 + 2 * \oh) {$C^\partial_{1, \idy}$};
                    \end{scope}
                    
                }
            }
            \begin{scope}[xshift=8cm,yshift=12cm]
                \fill[\colorC] (-0.5,0.5) rectangle (3.5,3.5);
                \fill[\colorCcirc] (-0.5,1.5) rectangle (3.5,2.5);
                
                \node at (-1.7,0.2) {$C_{1,6}^\partial$};
                \node at (4.8,1) {$C_{1,6}$};
                \node at (-1.7,2.4) {$\mathring{C}_{1,6}$};
                
                \draw [|-|,darkred,thick,line width=0.4mm](0.5,2.5) -- (0.5,3.5) ;
                \node at (1.5,3) {$c$};
                \draw [|-|,darkred,thick,line width=0.4mm](1.5,3.5) -- (1.5,4.5) ;
                \node at (2.5,4) {$k$};
            \end{scope}
            
            \draw[\colorBoarder,thick] (0.5,0.5) rectangle (27.5,27.5);
        \end{scope}
        
        %%%%%%%%%%%%%%%%%%%%%%%%%%%%%%%%%%%%%%%%%%%%%%%%%%%%%%%%%
        % second picture
        %%%%%%%%%%%%%%%%%%%%%%%%%%%%%%%%%%%%%%%%%%%%%%%%%%%%%%%%%
        
        \begin{scope}[xshift=32cm]

            % Loop to draw the outer parts
            \foreach \x/\w/\ox/\ow in {0.5/7/0/-1, 11.5/5/1/-1, 20.5/7/1/0} {
                \drawConeColumns{\x}{0}{\w}{\colorCpartial}{1};
                \drawConeColumns{\x}{1}{\w}{\colorCpartial}{1};
            }
            % Loop to draw the middle parts
            \foreach \x/\w/\ox/\ow in {0.5/6/0/-1, 12.5/3/1/-1, 21.5/6/1/0} {
                \drawConeColumns{\x}{0}{\w}{\colorC}{1};
                \drawConeColumns{\x}{1}{\w}{\colorC}{1};
            }
            %Loop to draw the inner parts
            \foreach \x/\w/\ox/\ow in {0.5/5/0/-1, 13.5/1/1/-1, 22.5/5/1/0} {
                \drawConeColumns{\x}{0}{\w}{\colorCcirc}{1};
                \drawConeColumns{\x}{1}{\w}{\colorCcirc}{1};
            }
            
            % Loop to erase cutout
            \foreach \x/\ox/\ow in {2.5/-2/0, 11.5/0/0, 20.5/0/2} {
                \foreach \y/\oy/\oh in {2.5/-2/0, 11.5/0/0, 20.5/0/2} {
                    \drawCtwoRectangleColor{\x}{\y}{5}{5}{\ox}{\oy}{\ow}{\oh}{white};
                }
            }
            
            \node at (3.3,18.5) {\textcolor{black}{$\mathring{C}_1$}};
            \node at (14,9.5) {\tiny \textcolor{black}{$\mathring{C}_1$}};
            \node at (14,18.5) {\large\textcolor{black}{${C}_1$}};
            \node at (18.5,5.5) {\small \textcolor{black}{${C}_1$}};
            \draw[\colorBoarder,thick] (0.5,0.5) rectangle (27.5,27.5);

        \end{scope}
    \end{tikzpicture}
    \caption{The figure shows the construction of $C_1^\partial$, contained in $C^1 = \Lambda_L\backslash \mathring{C}_1 $. On the left side, $C^1$ is the whole region, and $C_1^\partial$ has been split into several fattened $1$-cells respectively; we represent explicitly, $C_{1,6}$, $\mathring{C}_{1,6}$. On the right side, $\mathring{C}_1$ is the union of the dark green regions, $C_1  \backslash \mathring{C}_1$ is that of the medium green ones, and $C_1^\partial  \backslash C_1 $ is the union of the lighter green, dashed regions without the lighgreen squares in the centre, however.}
    \label{fig:covering2}
\end{figure}

This process is iterated for decreasing values of $a$, creating a hierarchical structure of cells of different extensive dimensions.
For a visual representation of this construction for $a=1$ in two dimensions ($D=2$), refer to Figure~\ref{fig:covering2}.

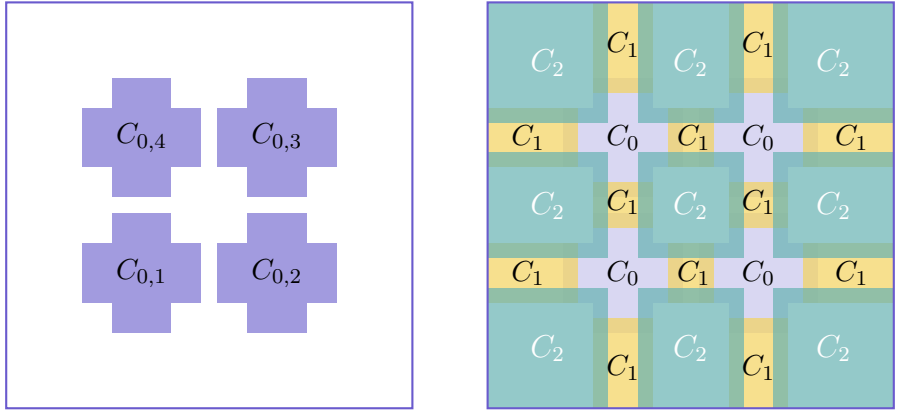
\begin{figure}[h!]
    \centering
    \def\colorBoarder{SlateBlueDark}
    \def\colorCzero{SlateBlueMid}
    \def\colorCone{GoldenrodMid}
    \def\colorCtwo{TealMid}
    
    \begin{tikzpicture}[scale=0.2]
        \fill[\colorCzero] (0.5,0.5) rectangle (27.5,27.5);
        
        % Loops to erase cutout
        \foreach \x/\w/\ox/\ow in {0.5/5/0/-1, 13.5/1/1/-1, 22.5/5/1/0} {
                \drawConeColumns{\x}{0}{\w}{white}{1};
                \drawConeColumns{\x}{1}{\w}{white}{1};
        }
        \foreach \x/\ox/\ow in {2.5/-2/0, 11.5/0/0, 20.5/0/2} {
            \foreach \y/\oy/\oh in {2.5/-2/0, 11.5/0/0, 20.5/0/2} {
                \drawCtwoRectangleColor{\x}{\y}{5}{5}{\ox}{\oy}{\ow}{\oh}{white};
            }
        }
        \node at (9.5, 9.5) {$C_{0,1}$};
        \node at (18.5, 9.5) {$C_{0,2}$};
        \node at (18.5, 18.5) {$C_{0,3}$};
        \node at (9.5, 18.5) {$C_{0,4}$};
        
        \draw[\colorBoarder,thick] (0.5,0.5) rectangle (27.5,27.5);
        %%%%%%%%%%%%%%%%%%%%%%%%%%%%%%%%%%%%%%%%%%%%%%%%%%%%%%%%%
        % second picture
        %%%%%%%%%%%%%%%%%%%%%%%%%%%%%%%%%%%%%%%%%%%%%%%%%%%%%%%%%

        \begin{scope}[xshift=32cm]
            % Draw C_0
            \fill[\colorCzero,opacity=0.4] (0.5,0.5) rectangle (27,27);

            % Loops to erase complement of C_0 without complement of C_1
            \foreach \x/\w/\ox/\ow in {0.5/5/0/-1, 13.5/1/1/-1, 22.5/5/1/0} {
                    \drawConeColumns{\x}{0}{\w}{white}{1};
                    \drawConeColumns{\x}{1}{\w}{white}{1};
            }

            % Draw C_1
            \foreach \x/\w/\ox/\ow in {0.5/6/0/-1, 12.5/3/1/-1, 21.5/6/1/0} {
                \drawConeColumns{\x}{0}{\w}{\colorCone}{0.7};
                \drawConeColumns{\x}{1}{\w}{\colorCone}{0.7};
            }

            % Loop to erase complement of C_1
            \foreach \x/\ox/\ow in {2.5/-2/0, 11.5/0/0, 20.5/0/2} {
                \foreach \y/\oy/\oh in {2.5/-2/0, 11.5/0/0, 20.5/0/2} {
                    \drawCtwoRectangleColor{\x}{\y}{5}{5}{\ox}{\oy}{\ow}{\oh}{white};
                }
            }
            
            % Loop to create C_2
            \foreach \x/\ox/\ow in {1.5/-1/0, 10.5/0/0, 19.5/0/1} {
                \foreach \y/\oy/\oh in {1.5/-1/0, 10.5/0/0, 19.5/0/1} {
                    \drawCtwoRectangleColorOp{\x}{\y}{7}{7}{\ox}{\oy}{\ow}{\oh}{\colorCtwo};
                    \drawCtwoRectangleText{\x}{\y}{7}{7}{\ox}{\oy}{\ow}{\oh}{\textcolor{white}{$C_2$}};
                }
            }

            % Loop to create C_1 labels
            \foreach \x/\w/\ox/\ow in {0.5/7/0/-1, 11.5/5/1/-1, 20.5/7/1/0} {
                \node[anchor=center] at (9.5, {\x + \ox + (\w + \ow)/2 - 0.3}) {$C_1$};
                \node[anchor=center] at (18.5, {\x + \ox + (\w + \ow)/2 - 0.3}) {$C_1$};
                \node[anchor=center] at ({\x + \ox + (\w + \ow)/2 - 0.3}, 9.5) {$C_1$};
                \node[anchor=center] at ({\x + \ox + (\w + \ow)/2 - 0.3}, 18.5) {$C_1$};
            }
            
            \foreach \x/\ox/\ow in {0.5/-2/0, 14/0/0, 27.5/0/2} {
                \foreach \y/\oy/\oh  in {0.5/-2/0, 14/0/0, 27.5/0/2} {
                    \begin{scope}
                        \clip (0.5,0.5) rectangle (27.5,27.5);
                    \end{scope}
                    
                }
            }

            % Draw C_0 labels
            \node at (9.5, 9.5) {$C_{0}$};
            \node at (18.5, 9.5) {$C_{0}$};
            \node at (18.5, 18.5) {$C_{0}$};
            \node at (9.5, 18.5) {$C_{0}$};

            \draw[\colorBoarder, thick] (0.5, 0.5) rectangle (27.5, 27.5);
        \end{scope}
        
    \end{tikzpicture}
    \caption{On the left side we show $C_0$, which is defined as $C^1 \backslash \mathring{C}_1$. On the right side, we show the coarse-graining in terms of the combined $C_0$, $C_1$ and $C_2$. We omit the corresponding $C_x^\partial$ and $\mathring{C}_x$, for $x=0,1,2$, for simplicity.}
    \label{fig:covering3}
\end{figure}

Lastly for $a=0$ we set $C^0 := C^1 \backslash \mathring{C}_1$ and  $\mathring{C}_{0,i} := C_{0,i} := C_{0,i}^\partial$ so that $C^0=\bigsqcup_{i\in\mathcal{I}_0} C_{0,i}$. This is shown on the left side of Figure~\ref{fig:covering3}. Let us summarise the properties of the above construction in the following lemma. 

\begin{lem}\label{lem:extended:properties-of-lattice-coarse-graining}
    The decomposition of $\Lambda_L$ described in the previous paragraphs satisfies the following properties:
    \begin{enumerate}
        \item For each $a\in\llbracket 0, D \rrbracket$, $\mathring{C}_a:=\bigsqcup_{i\in\mathcal{I}_a}\mathring{C}_{a,i}, {C}_a:=\bigsqcup_{i\in\mathcal{I}_a}{C}_{a,i}$, and ${C}^\partial_a:=\bigsqcup_{i\in\mathcal{I}_a}{C}^\partial_{a,i}$ are unions of disjoint sets, with size bounded as $|\mathring{C}_{a,i}|\leq|C_{a,i}|\leq |C^\partial_{a,i}|\leq \ell^D$.\footnote{This is a non-tight bound and it  holds that $|C_{a,i}|\leq [2(D-a)(k+c)]^{D-a}[\ell-2(D-a)(k+c)]^{a}\leq \ell^D$.} 
        \item $C^0=C_0=\mathring{C}_0 = \bigsqcup_{i\in\mathcal{I}_0} C^\partial_{0,i}$ is a disjoint union of ``fat'' $0$-cells, in the shape of $D$-dimensional ``crosses'', with each $C^\partial_{0,i}$ included in a hypercube of sidelength $2D(k+c)$ and distance $$\textup{dist}(C^\partial_{0,i},C^\partial_{0,j})\geq \ell-2D(k+c)>0$$ from each other. 
        \item The hierarchy $\{C_a\}_{a=0}^D$ induces a suitably overlapping coarse-graining of $\Lambda_L$, i.e. $\bigcup_{a=0}^{D}C_a=C^D=\Lambda_L$, and each site $x\in\Lambda_L$ is included in at most $D+1$ sets $\{C_{a,i}\}_{a,i}$.
        \item $c\leq\textup{dist}(C^a\backslash C_a,\mathring{C}_a)= \textup{dist}(C^a\backslash C_a,C^a\backslash C^{a-1})$ $\forall a \in \llbracket1, D\rrbracket$,
        \item $2k\leq \dist(C_{a,i},C_{a,j}) \quad \forall i,j\in \cI_a, \forall a\in\llbracket 0, D \rrbracket.$ 
    \end{enumerate}
\end{lem}
For a visualization, find an example of the case $D=2$ on the right side of Figure~\ref{fig:covering3}. 
\begin{proof}
    \begin{enumerate}
        \item We proceed by induction on $a$. For $a=D$, each of the sets $C^\partial_{D,i}$ are hypercubes with sidelength $\ell$, hence $C_{D,i},\mathring{C}_{D,i}$ are hypercubes of side length at most $\ell-k$ and $\ell-(k+c)$, respectively. Disjointness is clear.
        For $a<D$, it holds that 
        \begin{align*}
            C^\partial_{a} 
            &:= \Big\{v\in C^{a} \;:\; \exists j\in\mathbb{Z}, b\in \llbracket1, D\rrbracket:\, v + j e_{b}\in \overline{C^a}\Big\}\\
            &= \bigcup_{b=1}^D\left(C^a\cap\left\{\mathring{C}_{a+1}+\mathbb{Z}e_{b}\right\}\right) \\
            &= \bigcup_{a'=1}^D\left(C^a\cap\left\{\mathring{C}_{a+1}+\mathbb{Z}_{2(D - a)(k + c)}e_{b}\right\}\right). 
        \end{align*}

        More precisely: For each $\mathring{C}_{a + 1, i}$ and each canonical direction $b$, we construct a set $C_{a, j(i, b)}^\partial$ by translating $\mathring{C}_{a + 1, i}$ by $2(D - a)(k + c)$ along $e_b$. We then define $C_{a, j(i, b)}^\partial$ as this translated set minus its intersection with $\mathring{C}_{a + 1, i}$.
        We eliminate duplicates from the collection $\{C_{a, j(i, b)}^\partial\}_{(i, b) \in \cI_{a + 1} \times \llbracket 1, D\rrbracket}$ to obtain the index set $\cI_a$, which we use to obtain $C_a^\partial$. The sets $C_{a, i}^\partial$ are mutually disjoint by construction. Each $C_{a, i}^\partial$ has $D - a$ sides of length $2(D-a)(k + c)$, while the remaining sides retain the length $\ell - 2(D - a)(k + c)$ inherited from the sets composing $\mathring{C}_{a + 1}$. The disjointness of $C_{a, i}$ and $\mathring{C}_{a, i}$ as well as the bounds on their sizes immediately follow.
            
        \item The disjointedness statement in 2. follows equally from the proof in 1. The statement $C^0=\mathring{C}_0$ follows by their definitions.
        
        \item  The proof proceeds by finite induction with at most $D$ steps, starting from $a = D$. We consider the following cases: Base case: If $x \in C_D$, we are done. Inductive step: If $x \notin C_D$, then $x \in C^{D-1} \supseteq \Lambda_L \setminus C_D$. For each subsequent step $a$, we have three possibilities: 
        \begin{enumerate}
            \item If $a = 0$, then $x \notin \bigcup_{b=1}^D C_b$. Consequently, $x \in C^0 = C_0^\partial = C_0 = \mathring{C}_0$, and we are done.
            \item  If $x \in C_a$, we are done.
            \item  If neither (a) nor (b) holds, then $x \notin \bigcup_{b=a}^D C_b$. Hence, $x \in   \Lambda_L \setminus \bigcup_{b=a}^D C_b \subset C^{a-1}$, and the induction continues to the next step.
        \end{enumerate}
        The induction terminates after at most $D$ steps because there are only $D+1$ sets $C_a$. And since each is a disjoint union of $C_{a,i}$ every $x \in \Lambda_L$ is contained in at most $D+1$ of the $C_{a,i}$. 
        
        \item Recall that $k,c\geq r$, the range of the interactions. The claim follows via 
        \begin{align*}
            \dist(C^a\backslash C_a,\mathring{C}_a)
            &= \dist (C^a\backslash C_{a,i},\mathring{C}_{a,i}) = \dist(\overline{C_{a,i}}\cap C^a,\mathring{C}_{a,i}\cap C^a) \\ 
            &= \dist(\{x\in C^a : \textup{dist}(x,\overline{C^\partial_{a,i}})\leq k\},\{x\in C^a : \dist(x,\overline{C^\partial_{a,i}})>k+c\}) = c
        \end{align*} and it is easy to check that $C^a\backslash C^{a-1}=C^a\backslash(C^a\backslash \mathring{C}_a)=\mathring{C}_a$ for $a \in \llbracket1, D\rrbracket$. 
        \item This follows directly from the disjointness of $C_{a,i}$ and $C_{a,j}$, see $1.$ and their definition. 
    \end{enumerate}
\end{proof}

\subsection{A weak approximate tensorization for Davies channels}\label{subsec:a-weak-approximate-tensorization-for-davies-channels}
By combining the general weak entropy factorization from \cref{subsec:a-general-weak-entropy-factorization} with the coarse-graining in \cref{subsec:a-coarse-graining-of-the-hypercubic-lattice} this section contains a weak approximate tensorization (wAT) for the Davies channels with corrections that are the MCMIs of the splitting. Under the assumption that the Gibbs state exhibits exponential decay of its matrix-valued quantum conditional mutual information this correction term decays exponentially with the overlap length $l$ (see \cref{lem:extended:properties-of-lattice-coarse-graining}).

\begin{thm}[Weak approximate tensorization]\label{thm:a-weak-approximate-tensorization-for-davies-channels}
    Given the decomposition of the lattices $\Lambda$ with constants $(k\geq r,c,l)$ described in the previous section (c.f. \cref{subsec:a-coarse-graining-of-the-hypercubic-lattice}) and then in the context of \cref{subsec:local-commuting-hamiltonians-and-davies-generators-on-lattices}, the family $\{E_{C_{a, i}} \,:\, a \in \llbracket0, D\rrbracket, i \in \cI_a\}$ of Davies conditional expectations with respect to $\sigma$ satisfies the following inequality: for all $\rho \in \cS(\cH_\Lambda)$,
    \begin{align}\label{eq:extended:weak-approximate-tensorization}
        D(\rho\|\sigma) &\leq \underset{a=0}{\overset{D}{\sum}} \; \underset{i_a \in \mathcal{I}_a}{\sum} D (\rho \| E_{C_{a,i_a}} (\rho) ) +  \underset{a=1}{\overset{D}{\sum}} \, \zeta_a (\sigma) \, ,
    \end{align}
    with 
    \begin{equation}
        \zeta_a (\sigma) := \norm{\mathbf{H}_\sigma(X_a : Z_a | W_a)}_\infty \, , 
    \end{equation}
    and $W_a \sqcup X_a \sqcup Y_a \sqcup Z_a =: \overline{C^a} \sqcup \mathring{C}_a \sqcup (C_a \backslash \mathring{C}_a) \sqcup (C^a \backslash C_a)$ with $d(X_a,Z_a)= c \geq r$, for $a \in \llbracket1, D\rrbracket$. 
    Assuming further that the  Gibbs state satisfies exponential decay of MCMI with constants $K$ and $\xi$, then we can estimate \eqref{eq:extended:weak-approximate-tensorization} with 
    \begin{align}\label{eq:extended:weak-approximate-tensorization-explicit-estimates}
        D(\rho\|\sigma) &\leq \underset{a=0}{\overset{D}{\sum}} \; \underset{i_a \in \mathcal{I}_a}{\sum} D (\rho \| E_{C_{a,i_a}} (\rho)) + D K |\Lambda_L| e^{-c/\xi} \, .
    \end{align}
\end{thm}

\begin{proof}
    We proceed by induction on $a$, starting at $a = D$ and decreasing to zero. Using the notation and result from \cref{lem:extended:weak-entropy-factorization}, and conditioning on $W_D = \emptyset = \overline{C^D} = \overline{\Lambda}$, we obtain:
    \begin{align*}
        D(\rho \Vert \sigma) = D_{C^D}(\rho \Vert \sigma) &= D_{X_DY_DZ_D}(\rho \Vert \sigma) \le D_{X_DY_D}(\rho \Vert \sigma) + D_{Y_D Z_D}(\rho \Vert \sigma) + \norm{\mathbf{H}_\sigma(X_D : Z_D | \emptyset)}_\infty \\
        &= D_{C_D}(\rho \Vert \sigma) + D_{C^{D - 1}}(\rho \Vert \sigma) + \zeta_D(\sigma)\\
        &\overset{(1)}{\le} D(\rho \Vert E_{C_D}(\sigma)) + D_{C^{D - 1}}(\rho \Vert \sigma) + \zeta_{D}(\sigma)\\
        &\overset{(2)}{\le} \sum\limits_{i \in \cI_D} D(\rho \Vert E_{C_{D, i}}(\sigma)) + D_{C^{D - 1}}(\rho \Vert \sigma) + \zeta_D(\sigma).
    \end{align*}
    In step (1), we apply \eqref{eq:davies-conditional-expectation-and-conditional-relative-entropy}. Then, we use the fact that $C_D = \bigsqcup_{i \in \cI_D} C_{D, i}$ with $\dist(C_{i, D}, C_{j, D}) > k \ge r$, as per \cref{lem:extended:properties-of-lattice-coarse-graining}. This allows us to apply 3. of the properties of the Davies channels from \cref{subsec:local-commuting-hamiltonians-and-davies-generators-on-lattices}, i.e split $E_D = \prod_{i \in \cI_D} E_{C_{D, i}}$ into commuting constituents. Consequently, in step (2), we split the relative entropy according to \eqref{eq:exaxt-factorization-relative-entropy}. The induction proceeds with $D_{C^{D - 1}}(\rho \Vert \sigma)$. \par 
    The induction step follows a similar line of reasoning: 
    For $a \ne 0$, we again employ the weak entropy factorisation from \cref{lem:extended:weak-entropy-factorization}:
    \begin{align*}
        D_{C^a}(\rho \Vert \sigma) 
        &= D_{X_aY_aZ_a}(\rho \Vert \sigma) \le D_{X_a Y_a}(\rho \Vert \sigma) + D_{Y_aZ_a}(\rho \Vert \sigma) + \norm{H_\sigma(X_a:Z_a|W_a)}_\infty \\
        &= D_{C_a}(\rho \Vert \sigma) + D_{C^{a - 1}}(\rho \Vert \sigma) + \norm{\mathbf{H}_\sigma(X_a:Z_a|W_a)}_\infty\\
        &\le \sum\limits_{i \in \cI_a} D(\rho \Vert E_{C_{a, i}}(\rho)) + D_{C^{a - 1}}(\rho \Vert \sigma) + \zeta_{a}(\sigma) \, .
    \end{align*}
    For $a = 0$, we only need to estimate $D_{C^0}(\rho \Vert \sigma) = D_{C_0}(\rho \Vert \sigma) \le \sum_{i \in \cI_0} D(\rho \Vert E_{C_{0, i}}(\rho))$. This inequality holds because we first can employ \eqref{eq:davies-conditional-expectation-and-conditional-relative-entropy} and then use that $C_0 = \bigsqcup_{i \in \cI_0} C_{0, i}$, where the $C_{0, i}$ have mutual distances greater than $r$. Thus, 3. for the Davies channels from \cref{subsec:local-commuting-hamiltonians-and-davies-generators-on-lattices} applies, allowing us to decompose $E_{C_0}$ into a mutually commuting composition of the $E_{C_{0, i}}$ and then apply \eqref{eq:exaxt-factorization-relative-entropy}.
    The final inequality holds because again $C_a = \bigsqcup_{i \in \cI_a} C_{a, i}$ with $\dist(C_{a, i}, C_{a, j}) > r$ for $i \ne j$ by \cref{lem:extended:properties-of-lattice-coarse-graining}, allowing us to apply property 3. for Davies channels from \cref{subsec:local-commuting-hamiltonians-and-davies-generators-on-lattices} and, subsequently, \eqref{eq:exaxt-factorization-relative-entropy}. We thereby complete the induction. In Figure~\ref{fig:decomposition-lattice-wat} the decomposition of the lattice for the weak approximate tensorization is demonstrated for the case $D = 2$ and $a = D$ and $a = D - 1$ respectively.\par
    In the case of exponential decay of matrix-valued quantum conditional mutual information, we have:
    \begin{equation*}
        \zeta_a(\sigma) \le K |\Lambda| e^{-c/\xi} \, ,
    \end{equation*}
    independent of $a$, which immediately gives \eqref{eq:extended:weak-approximate-tensorization-explicit-estimates}. 
    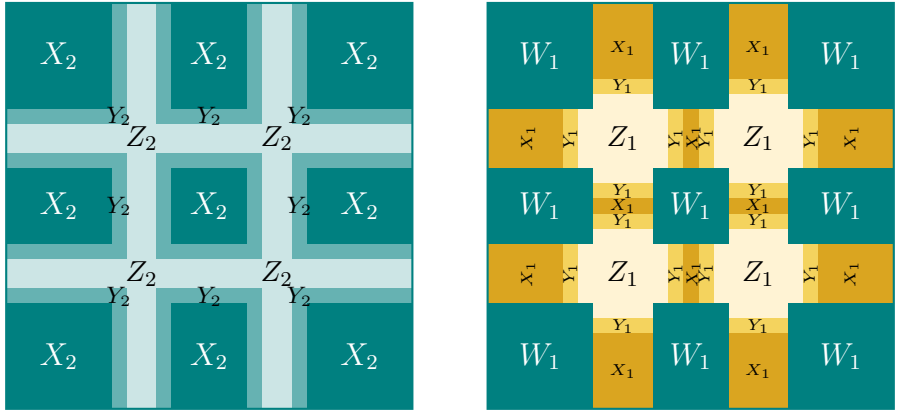
\begin{figure}
        \centering
        \def\colorBoarder{TealDark}
        \def\colorX{TealDark}
        \def\colorY{TealMid}
        \def\colorZ{TealLight}
        \begin{tikzpicture}[scale=0.2]
            \begin{scope}
                \fill[\colorZ] (0.5,0.5) rectangle (27.5,27.5);
                
                \foreach \x/\ox/\ow in {1.5/-1/0, 10.5/0/0, 19.5/0/1} {
                    \foreach \y/\oy/\oh in {1.5/-1/0, 10.5/0/0, 19.5/0/1} {
                        \drawCtwoRectangleColor{\x}{\y}{7}{7}{\ox}{\oy}{\ow}{\oh}{\colorY};
                        \drawCtwoRectangleColor{\x+1}{\y+1}{5}{5}{2*\ox}{2*\oy}{2*\ow}{2*\oh}{\colorX};
                        \drawCtwoRectangleText{\x+1}{\y+1}{5}{5}{2*\ox}{2*\oy}{2*\ow}{2*\oh}{\textcolor{white}{$X_2$}};
                    }
                }
                
                \foreach \x in {8, 14, 20} {
                    \foreach \y in {8, 14, 20} {
                        % Convert floating-point to dimension and compare
                        \pgfmathsetmacro{\xcomp}{\x}
                        \pgfmathsetmacro{\ycomp}{\y}
                        \ifdim\xcomp pt=14pt
                            \ifdim\ycomp pt=14pt
                                % Both \x and \y are 14.5
                            \else
                                \node[anchor=center] at (\x, \y) {\footnotesize $Y_2$};
                            \fi
                        \else
                            \node[anchor=center] at (\x, \y) {\footnotesize $Y_2$};
                        \fi
                    }
                }
                
                \foreach \x in {9.5, 18.5}{
                    \foreach \y in {9.5, 18.5} {
                        \node at (\x, \y) {$Z_2$};
                    }
                }
                \draw[\colorBoarder, thick] (0.5, 0.5) rectangle (27.5, 27.5);
            \end{scope}

            %%%%%%%%%%%%%%%%%%%%%%%%%%%%%%%%%%%%%%%%%%%%%%%%%%%%%%%%%
            % second picture
            %%%%%%%%%%%%%%%%%%%%%%%%%%%%%%%%%%%%%%%%%%%%%%%%%%%%%%%%%
            \def\colorW{TealDark}
            \def\colorX{GoldenrodDark}
            \def\colorY{GoldenrodMid}
            \def\colorZ{GoldenrodLight}
            \begin{scope}[xshift=32cm]
                \fill [\colorZ] (0.5,0.5) rectangle (27.5,27.5);
                
                % Loop to draw Y
                \foreach \x/\w/\ox/\ow in {0.5/6/0/-1, 12.5/3/1/-1, 21.5/6/1/0} {
                    \drawConeColumns{\x}{0}{\w}{\colorY}{1};
                    \drawConeColumns{\x}{1}{\w}{\colorY}{1};
                }
                
                %Loop to draw the X
                \foreach \x/\w/\ox/\ow in {0.5/5/0/-1, 13.5/1/1/-1, 22.5/5/1/0} {
                    \drawConeColumns{\x}{0}{\w}{\colorX}{1};
                    \drawConeColumns{\x}{1}{\w}{\colorX}{1};
                }
                
                \foreach \x/\ox/\ow in {1.5/-1/0, 10.5/0/0, 19.5/0/1} {
                    \foreach \y/\oy/\oh in {1.5/-1/0, 10.5/0/0, 19.5/0/1} {
                        \drawCtwoRectangleColor{\x+1}{\y+1}{5}{5}{2*\ox}{2*\oy}{2*\ow}{2*\oh}{\colorW};
                        \drawCtwoRectangleText{\x+1}{\y+1}{5}{5}{2*\ox}{2*\oy}{2*\ow}{2*\oh}{\textcolor{white}{$W_{1}$}};
                    }
                }

                % Loop to create Y-labels
                \foreach \x in {9.5, 18.5}{
                    \foreach \y in {9.5, 18.5} {
                        \node[anchor=center] at (\x, \y) {$Z_1$};
                        \node[rotate=90, anchor=center] at (\x - 3.5, \y) {\tiny $Y_{1}$};
                        \node[rotate=90, anchor=center] at (\x + 3.5, \y) {\tiny $Y_{1}$};
                        \node[anchor=center] at (\x, \y - 3.5) {\tiny $Y_{1}$};
                        \node[anchor=center] at (\x, \y + 3.5) {\tiny $Y_{1}$};
                    }
                }
                % Loop to create X labels
                \foreach \x/\w/\ox/\ow in {0.5/7/0/-1, 11.5/5/1/-1, 20.5/7/1/0} {
                    \node[anchor=center] at (9.5, {\x + \ox + (\w + \ow)/2 - 0.5}) {\tiny $X_{1}$};
                    \node[anchor=center] at (18.5, {\x + \ox + (\w + \ow)/2 - 0.5}) {\tiny $X_{1}$};
                    \node[anchor=center, rotate=90] at ({\x + \ox + (\w + \ow)/2 - 0.3}, 9.5) {\tiny $X_{1}$};
                    \node[anchor=center, rotate=90] at ({\x + \ox + (\w + \ow)/2 - 0.3}, 18.5) {\tiny $X_{1}$};
                }
                \draw[\colorBoarder,thick] (0.5,0.5) rectangle (27.5,27.5);
            \end{scope}
        \end{tikzpicture}
        \caption{The decomposition of the lattice used in the weak approximate tensorization for the case $D = 2$ in the steps $a = 2$ on the left and $a = 1$ on the right.}
        \label{fig:decomposition-lattice-wat}
    \end{figure}
\end{proof}

\subsection{A weak transport cost inequality for Davies channels}\label{subsec:a-weak-transport-cost-inequality-for-davies-channels}
Tailored to our weak approximate tensorization from the previous section \cref{subsec:a-weak-approximate-tensorization-for-davies-channels} but valid also for arbitrary coarse-grainings of the lattice with Davies channels, we derive in this section a weak transport cost (wTC) inequality which forms a crucial ingredient in the proof of the Wasserstein mixing result. Let us begin with an auxiliary lemma.

\begin{lem}\label{lem:extended:path-length-estimate}
    Let $\cL$ a generator of a GNS-symmetric quantum Markov semigroup with invariant state $\sigma > 0$ that satisfies \eqref{eq:weak-mlsi} with constants $c_1$ and $c_2$ and let $\rho \in \cS(\cH)$. Assume that $D(\rho \Vert \sigma) \ge c_2$. Then for $t \in (0, t_{c_2}(\rho))$ where $t_{c_2}(\rho) := \sup\{t \ge 0 : D(e^{t\cL}(\rho) \Vert \sigma) \ge c_2\}$
    \begin{equation}
        \int_{0}^t \sqrt{\EP_{\cL}(\rho_s)} ds \le 2 \sqrt{c_1 D(\rho \Vert \sigma)}  \quad\text{with}\quad \rho_s := e^{s\cL}(\rho) \, . 
    \end{equation}
\end{lem}
\begin{proof}
    Let us define $F:(0, t_{c_2(\rho)}) \to \R$ as
    \begin{equation}
        F(t) := \int_0^t \sqrt{\EP_{\cL}(\rho_s)} ds + 2\sqrt{c_1(D(\rho_t \Vert \sigma) - c_2)} \, . 
    \end{equation}
    This function is non-increasing in $t$, as 
    \begin{equation}
        \frac{d}{dt} F(t) = \sqrt{\EP_{\cL}(\rho_t)} - \frac{\sqrt{c_1} \EP_{\cL}(\rho_t)}{\sqrt{D(\rho_t \Vert \sigma) - c_2}} \le 0
    \end{equation}
    due to \ref{eq:weak-mlsi}. The claim then immediately follows.
\end{proof}

With this lemma in place, we can now move our attention to the proof of the wTC from the wAT. 

\begin{thm}[Weak transport cost inequality]\label{thm:extended:weak-transport-cost-inequality}
    In the context of \cref{subsec:local-commuting-hamiltonians-and-davies-generators-on-lattices}, assume that there exists $A = \{A_i\}_{i = 1}^{n_A}$ with $A_i \subseteq \Lambda$ an overlapping coarse-graining of $\Lambda$, i.e. $\bigcup_{i = 1}^{n_A} A_i = \Lambda$ with corresponding Davies projectors $E_{A_i}$ such that one has
    \begin{equation}
        D(\rho \Vert \sigma) \le \sum\limits_{i = 1}^{n_A} D(\rho\Vert E_{A_i}(\rho)) + c_2
    \end{equation}
    for all $\rho \in \cS(\cH_\Lambda)$, then the following holds
    \begin{equation} \label{eq:extended:weak-transport-cost-inequality}
        \norm{\rho - \sigma}_{W^1} \le \max_{i} 2\sqrt{2} |A_i\partial| \sqrt{n_A D(\rho \Vert \sigma)} + |\Lambda| \sqrt{2 c_2}
    \end{equation}
    for all $\rho \in \cS(\cH_\Lambda)$.
\end{thm}
\begin{proof}
    In the following let us denote $\cL^H_{A_i} := E_{A_i} - \id$. By non-negativity of the relative entropy, it holds that
    \begin{equation}\label{eq:mlsi-local-heat-bath}
        \begin{aligned}
            D(\rho \Vert E_{A_i}(\rho)) &\le D(\rho \Vert E_{A_i}(\rho)) + D(E_{A_i}(\rho) \Vert \rho) = \tr[(\id - E_{A_i})(\rho)(\log(\rho) - \log E_{A_i}(\rho))] \\
            &= \tr[(\id - E_{A_i})(\rho)(\log(\rho) - \log \sigma)] \\
            &=\EP_{\cL_{A_i}^H}(\rho) \, ,
        \end{aligned}
    \end{equation}
    where in the last equality we used that $\log E_{A_i}(\rho) - \log \sigma$ is a fixpoint of $E_A^*$ which immediately gives $\tr[(\id - E_{A_i})(\rho)(\log E_{A_i}(\rho) - \log\sigma)] = 0$. Using \eqref{eq:mlsi-local-heat-bath}, the assumption can be rewritten as 
    \begin{equation}\label{eq:wmlsi-global-heat-bath}
        D(\rho\Vert \sigma) \le \EP_{\cL^H_{\Lambda}}(\rho) + c_2
    \end{equation}
    where we defined the primitive GNS-symmetric generator $\cL^H_\Lambda := \sum_{i = 1}^{n_A} \cL^H_{A_i}$. As $E_{A_i}$ only acts non-trivially on $A_i \partial$, we find that $\tr_{A_i\partial} \circ \cL_{A_i}^H = 0$ which is a direct consequence of \eqref{eq:composition-relation-local-davies-local-depolarising-channel}. From this fact one immediately obtains
    \begin{equation}\label{eq:wasserstein-norm-local-heat-bath-generator}
        \norm{\cL_{A_i}^H(\rho)}_{W^1} \le |A_i\partial| \sqrt{2\EP_{\cL_{A_i}^H}(\rho)} \, ,
    \end{equation}
    as by \eqref{eq:relation-wasserstein-trace-distance} one has
    \begin{equation*}
        \norm{\cL_{A_i}^H(\rho)}_{W^1} \le |A_i \partial| \norm{\cL_{A_i}^H(\rho)}_1 \, . 
    \end{equation*}
    Now using Pinskers' inequality we get $\norm{\cL_{A_i}^H(\rho)}_1 \le \sqrt{2 D(\rho \Vert E_{A_i}(\rho))}$ which combined with \eqref{eq:mlsi-local-heat-bath} gives \eqref{eq:wasserstein-norm-local-heat-bath-generator}. We can now shift our attention to the main result. First assume that $D(\rho \Vert \sigma) \le c_2$, then the inequality holds trivially by $\norm{\cdot}_{W^1} \le |\Lambda| \norm{\,\cdot\,}_1$ \eqref{eq:wasserstein-norm-local-heat-bath-generator} and Pinsker's inequality \eqref{eq:pinsker-inequality}. If this is not the case we set $t_{c_2} := \sup\{ t \ge 0 : D(e^{t \cL^H_{\Lambda}}(\rho) \Vert \sigma) \ge c_2\}$ and get
    \begin{equation*}
        \norm{\rho - \sigma}_{W^1} \le \norm{\rho - \rho_{t_{c_2}}}_{W^1} + \norm{\rho_{t_{c_2}} - \sigma}_{W^1} \le \norm{\rho - \rho_{t_{c_2}}}_{W^1} + |\Lambda| \sqrt{2c_2}
    \end{equation*}
    where we defined $\rho_t := e^{t\cL^H_{\Lambda}}(\rho)$ for $t \ge 0$. The second inequality follows again by the above reasoning. For the first term, we write
    \begin{equation*}
        \norm{\rho - \rho_{t_{c_2}}}_{W^1} \le \int_0^{t_{c_2}}\norm{\cL^H_{\Lambda}(\rho_s)}_{W^1} ds \le \int_0^{t_{c_2}}\sum_{i = 1}^{n_A} \norm{\cL^H_{A_i}(\rho_s)}_{W^1} ds \, . 
    \end{equation*}
    By \eqref{eq:wasserstein-norm-local-heat-bath-generator} we get
    \begin{equation*}
        \int_0^{t_{c_2}}\sum_{i = 1}^{n_A} \norm{\cL^H_{A_i}(\rho_s)}_{W^1} ds \le \max\limits_{i} \sqrt{2} |A_i\partial| \int_{0}^{t_{c_2}} \sum_{i = 1}^{n_A} \sqrt{\EP_{\cL^H_{A_i}}(\rho_s)} ds
    \end{equation*}
    and lastly through concavity of $x \mapsto \sqrt{x}$, 
    \begin{equation*}
        \norm{\rho - \rho_{t_{c_2}}}_{W^1} \le \max\limits_{i} \sqrt{2} |A_i\partial| \sqrt{n_A} \int_{0}^{t_{c_2}}\sqrt{\EP_{\cL^H_\Lambda}(\rho_s)}  ds \, . 
    \end{equation*}
    Using \eqref{eq:wmlsi-global-heat-bath} the claim follows by \cref{lem:extended:path-length-estimate}.
\end{proof}

\subsection{A MLSI alike inequality for local Davies semigroups at every temperature}\label{subsec:a-mlsi-alike-inequality-for-local-davies-at-every-temperature}
This section is dedicated to collecting some results about Davies conditional expectations and entropy productions that will all be used in an inequality that almost resembles an MLSI for the local Davies Lindbladiens. More precisely the following inequality 
\begin{equation}\label{eq:mlsi-alike-inequality-for-local-davies}
    D(\rho \Vert E_A(\rho)) \le \cO(e^{\cO(|A\partial|)}) \EP_{\cL_{A\partial}}(\rho) \, , 
\end{equation}
which will be shown and used in the next section, i.e. \cref{subsec:weak-mlsi-for-the-global-davies-semigroup}. As our proof will rely on going through the infinite temperature Davies (i.e. $\lim\limits_{\beta \to 0} \cL_A^\beta = \cL_A^0$) we will keep the temperature explicit in the following. The strategy will be to relate the LHS and the entropy production on the RHS of \eqref{eq:mlsi-alike-inequality-for-local-davies} to their infinite temperature counterparts through \cref{lem:extended:relation-davies-at-finite-and-infinite-temperature} and then to proof \eqref{eq:mlsi-alike-inequality-for-local-davies} only for the infinite temperature Davies, where the semigroup is almost the depolarizing one.

\begin{lem}\label{lem:extended:relation-davies-at-finite-and-infinite-temperature}
    In the context of \cref{subsec:local-commuting-hamiltonians-and-davies-generators-on-lattices}, let $A \subseteq \Lambda$ and define $E_A^\beta = \lim\limits_{t \to \infty} e^{t\cL_A^{D, \beta}}$ and $E_A^0 = \lim\limits_{t \to \infty} e^{t\cL_A^{D, 0}}$. Then the following inequalities hold:
    \begin{equation}
        D(\rho \Vert E_A^\beta(\rho)) \le e^{2 gJ\beta|A\partial|} D(\rho \Vert E_A^0(\rho)) \, . 
    \end{equation}
    and 
    \begin{equation}
        \EP_{\cL^{D, 0}_A}(\rho) \le e^{2gJ(\beta|A\partial| + 1)} \EP_{\cL^{D, \beta}_A}(\rho)
    \end{equation}
\end{lem}
\begin{proof}
    Let $\sigma = E_A^\beta(d_\Lambda^{-1}\Id_\Lambda)$ and $\sigma'= E_A^0(d_{\Lambda}^{-1}\Id_\Lambda) = d_{\Lambda}^{-1}\Id_\Lambda $. Then $\cL_A^{D, \beta}$ is GNS-symmetric with respect to $\sigma$, and $\cL_A^{D, 0}$ is GNS-symmetric with respect to $\sigma'$. Applying \cite[Proposition 4.2]{Junge.2022}, we obtain: 
    \begin{align*}
        D(\rho \Vert E^\beta_A(\rho)) \le \Vert E_A^\beta(\Id_\Lambda)\Vert_\infty D(\rho \Vert E^0_A(\rho)) \, . 
    \end{align*}
    Observe that $E_A^\beta(e^{-\beta H_{A\partial}}) = e^{-\beta H_{A\partial}}$, hence $\Vert E_A^\beta(\Id_\Lambda)\Vert_{\infty} \le \norm{e^{\beta H_{A\partial}}}_\infty \norm{e^{-\beta H_{A\partial}}}_\infty \le e^{2 \beta g J |A\partial|}$. For the entropy productions, \cite[Proposition 4.3]{Junge.2022} yields:
    \begin{equation}
        \EP_{\cL^{D, 0}_A}(\rho) \le \max_\omega e^{|\omega|/2} \Vert (E_A^\beta(\Id_\Lambda))^{-1}\Vert_\infty  \EP_{\cL^{D, \beta}_A}(\rho)  
    \end{equation}
    The Bohr frequencies are localized and can be bounded above by $|\omega| \le \max_{k \in A}2\norm{H_{k\partial}}_\infty \le 2 gJ$ as per \eqref{eq:bohr-frequency-decomposition-of-jump-operators}. Furthermore, $\Vert (E_A^\beta(\Id_\Lambda))^{-1}\Vert_\infty \le \norm{e^{\beta H_{A\partial}}}_\infty \norm{e^{-\beta H_{A\partial}}}_\infty \le e^{2 \beta g J |A\partial|}$, completing the proof.
\end{proof}

To conclude let us present \eqref{eq:mlsi-alike-inequality-for-local-davies} for the infinite temperature Davies.

\begin{lem}\label{lem:extended:mlsi-alike-inequality-for-infinite-temperature-local-davies}
    In the context of \cref{subsec:local-commuting-hamiltonians-and-davies-generators-on-lattices}, let $A \subseteq \Lambda$ and $E_A^0 = \lim\limits_{t \to \infty} e^{t\cL_A^{D, 0}}$. Then:
    \begin{equation}
        \chi_{\min}^0 D(\rho\Vert E^0_A(\rho)) \le \EP_{\cL^{D,0}_{A\partial}}(\rho) \, . 
    \end{equation}
\end{lem}
\begin{proof}
    By the GNS-symmetry of $\cL_{A\partial}^0$, we can employ \cite[Theorem 5.10]{Carlen.2017} to express $\EP_{\cL_{A\partial}^{D,0}}(\rho)$ as a sum of non-negative inner products with positive coefficients, including $\chi_{\omega, k}^0$. Using the bound $\chi_{\omega, k}^0 \ge \chi_{\min}^0$, we obtain:
    \begin{equation*}
        \chi_{\min}^0 \EP_{\tilde{\cL}_{A\partial}^{D, 0}}  \le \EP_{\cL_{A\partial}^{D, 0}} \, ,
    \end{equation*}
    where 
    \begin{align*}
        \tilde{\cL}_{A\partial}^{D, 0} &= \sum\limits_{k \in A\partial}\sum\limits_{\omega, \alpha} \Big( S_{\alpha,k}^{\omega}\rho S_{\alpha,k}^{\omega,\dagger}-\frac{1}{2}\,\big\{ \rho, S_{\alpha,k}^{\omega,\dagger}S_{\alpha,k}^{\omega} \big\} \Big)\\
        &= \sum\limits_{t \ge 0} \sum\limits_{k \in A\partial} (\sum\limits_{\alpha}\Delta_{e^{itH}}\left( S_{\alpha, k} \Delta_{e^{-itH}}(\rho) S_{\alpha, k} \right) - \rho) \, . 
    \end{align*}
    
    The last simplification follows from \cite[Theorem 31]{Kastoryano.2016}. Define $\rho_t := \Delta_{e^{-itH}}(\rho)$. Using the gauge freedom of entropy production with respect to the choice of fixed point, and noting that $\Id_\Lambda/d_{\Lambda}$ is a fixed point, we have:
    \begin{align*}
        \EP_{\tilde{\cL}_{A\partial}^{D, 0}}(\rho) &= -\tr[\tilde{\cL}_{A\partial}^{D, 0}(\rho) (\log(\rho) - \log \Id/d)] = \tr[\tilde{\cL}_{A\partial}^{D, 0}(\rho) \log(\rho)] \\
        &= \sum\limits_{t \ge 0} \sum\limits_{k\in A\partial,}\tr[(\rho_t - \sum\limits_{\alpha}S_{\alpha, k} \rho_t S_{\alpha, k}) \log(\rho_t)] = \sum\limits_{t \ge 0} \EP_{\cL^{\text{depol}}_{A\partial}}(\rho_t) \ge \min_t \EP_{\cL^{\text{depol}}_{A\partial}}(\rho_t) \, . 
    \end{align*}
    Where $\cL_{A\partial}^{\text{depol}}(\,\cdot\,) = \sum_{k \in A\partial} (\sum_{\alpha}S_{\alpha, k}(\,\cdot\,)S_{\alpha, k} -\id) = \sum_{k \in A\partial} (\tr_k[\,\cdot\,] - \id)$. Since the depolarizing channel has a cMLSI of 1 \cite{Capel.2018a}, and using the chain rule of relative entropy \eqref{eq:chain-rule-relative-entropy}, we have:
    \begin{align*}
        \EP_{\cL^{\text{depol}}_{A\partial}}(\rho_t) &\ge D(\rho_t \Vert \E_{A\partial}(\rho_t)) = D(\rho_t \Vert d_\Lambda^{-1}\Id_\Lambda) - D(\E_{A\partial}(\rho_t) \Vert d_\Lambda^{-1}\Id_\Lambda)\\
        &= D(\rho_t \Vert  d_\Lambda^{-1}\Id_\Lambda) - D(\E_{A\partial}(E_{A}^0(\rho_t)) \Vert \E_{A\partial}( d_\Lambda^{-1}\Id_\Lambda))\\
        &\overset{\text{DPI}}{\ge} D(\rho_t \Vert  d_\Lambda^{-1}\Id_\Lambda) - D(E_{A}^0(\rho_t) \Vert  d_\Lambda^{-1}\Id_\Lambda)\\
        &= D(\rho_t \Vert E^0_A(\rho_t))
    \end{align*}

    We used that $\E_{A\partial} \circ E_A = \E_{A\partial}$ (which follows from $\E_{A\partial}\circ \cA_{A, \sigma} = \E_{A\partial}$ and \eqref{eq:local-davies-projection-as-limit-of-petz-recoveries}), the data-processing inequality (DPI), and the chain rule. Finally, observe that $[E_{A}^0, \Delta_{e^{-itH}}] = 0$ by the GNS-symmetry of $\cL_{A}^\beta$ which is preserved in the limit, therefore:
    \begin{equation*}
        D(\rho_t \Vert E^0_A(\rho_t)) = D(\Delta_{e^{-itH}}(\rho) \Vert \Delta_{e^{-itH}}(E_{A}^0(\rho))) = D(\rho \Vert E_{A}^0(\rho)) \, . 
    \end{equation*}
\end{proof}

\subsection{Weak MLSI for the global Davies semigroup}\label{subsec:weak-mlsi-for-the-global-davies-semigroup}
In this section, we are going to show two weak MLSIs for the global Davies semigroups relying in one case solely on the MCMI decay (c.f. \cref{thm:extended:weak-mlsi}), while in the other case (c.f. \cref{thm:extended:wmlsi-under-gap}), we replace the result from \cref{subsec:a-mlsi-alike-inequality-for-local-davies-at-every-temperature} with an assumption on the local gap of the generator to improve the scaling with system size. Hence the second result relies on the existence of MCMI and a uniformly polynomial local gap. More precisely we require the following \emph{uniform local polynomial gap}: There exists $\mu \in \N_0$ and $C > 0$ such that for all $A \subseteq \Lambda$
\begin{equation}
    \lambda(\cL_A) \ge C|A|^{-\mu} = \Omega(|A|^{-\mu}) \, . 
\end{equation}
The first result will be then used in \cref{sec:main-results-extended} to derive the quasi-rapid Wasserstein mixing under every temperature while the second one improves the scaling quasi-rapid Wasserstein mixing and further gives rapid mixing in trace distance under the assumption of a polynomial local gap. Let us begin with the first result only requiring MCMI decay and no additional assumptions on the local or global gap.

\begin{thm}[weak MLSI at every $\beta$]\label{thm:extended:weak-mlsi} 
    In the setting of \cref{subsec:local-commuting-hamiltonians-and-davies-generators-on-lattices} assume that the Gibbs state at inverse temperature $\beta$ satisfies MCMI decay with constants $K,\xi$, then the semigroup $\{e^{t\cL_\Lambda^D}\}_{t\geq 0}$ at that temperature satisfies the following weak MLSI for $\epsilon\geq KD2^{D}(2L+1)^D\exp(\frac{2r}{\xi}-\frac{L}{D\xi})= \cO(L^De^{-\cO(L)})$.
    \begin{align}
        D(\rho_t\|\sigma)\leq e^{-\alpha(\epsilon)t}D(\rho\|\sigma)+\epsilon
    \end{align} 
    with $\frac{1}{\alpha(\epsilon)}=(D+1)(\chi^0_{\min})^{-1}e^{2gJ}e^{4\beta gJ(2D(2r+\xi\log\frac{KD2^DN}{\epsilon})+1)^D}= \cO(\exp\{\cO((\log\frac{N}{\epsilon})^D)\})$, where $\rho_t = e^{t\cL_\Lambda^D}(\rho)$ for arbitrary $\rho \in \cS(\cH_\Lambda)$.
\end{thm}
Such a lower bound on the admissible $\epsilon$ is generic and to be expected. Likewise we can also interpret this as giving a minimal lattice size $N$ for which this MLSI eventually holds for a given fixed $\epsilon$.
\begin{proof}
    Using \cref{thm:a-weak-approximate-tensorization-for-davies-channels} w.r.t. a coarse-graining with parameters $(k = 2r-\frac{r}{D}, c, l)$ and both \cref{lem:extended:relation-davies-at-finite-and-infinite-temperature} and \cref{lem:extended:mlsi-alike-inequality-for-infinite-temperature-local-davies} yields
    \begin{align*}
        D(\rho\|\sigma) &\leq \sum_{a,i}D(\rho\|E^D_{C_{a,i}}(\rho)) + D2^DKNe^{-\frac{c}{\xi}}  \\ &\leq \sum_{a,i}e^{2\beta gJ|C_{a,i}\partial|}e^{2\beta gJ|C_{a,i}\partial\partial|}e^{2gJ}(\chi^0_{\min})^{-1}\EP_{\cL^D_{C_{a,i}\partial}}(\rho)+ D2^DKNe^{-\frac{c}{\xi}} \\ &\leq (D+1)(\chi^0_{\min})^{-1}e^{4\beta gJ(\ell+2r)^D+2gJ}\EP_{\cL^D_\Lambda}(\rho)+ D2^DKNe^{-\frac{c}{\xi}},
    \end{align*} where we used that $|C_{a,i}\partial|\leq |C_{a,i}\partial\partial|\leq (\ell+2r)^D$ by construction of this coarse-graining (c.f. \cref{lem:extended:properties-of-lattice-coarse-graining}).
    Picking $c=\max\{\xi\log\frac{KD2^DN}{\epsilon},r\}=\cO(\log\frac{N}{\epsilon})$ and $\ell=2D(k+c)+1=2D(2r+c)+1-2r =\cO(\log\frac{N}{\epsilon})$ gives a valid coarse-graining iff $c,\ell=2D(k+c)+1\leq 2L+1$, since $r\leq k\leq 2r\leq 2L+1$. This is equivalent to $c\leq \frac{L}{D}-k$, which is implies by $c\leq \frac{L}{D}-2r$ and thus $\epsilon\geq KD2^DN\exp(\frac{2r}{\xi}-\frac{L}{D\xi})\geq KD2^D(2L+1)^D\exp(\frac{2r}{\xi}-\frac{L}{D\xi})$. We get
    \begin{align}
        D(\rho\|\sigma) &\leq \frac{(D+1)e^{2gJ}}{\chi^0_{\min}}e^{4\beta gJ(2D(2r+\xi\log\frac{KD2^DN}{\epsilon})+1)^D}\EP_{\cL_\Lambda}(\rho)+\epsilon \\ &= \cO\left(\exp(\cO\left(\left(\log\frac{N}{\epsilon}\right)^D\right))\right)\EP_{\cL_\Lambda}(\rho)+\epsilon.
    \end{align} Applying Grönwalls lemma now yields the desired statement.
\end{proof}

For the second result, let us first convert the uniform polynomial local gap into a uniform polynomial cMLSI of the local generators $\cL_A$: 
\begin{lem}\label{lem:extended:gap-to-mlsi}
    In the context of \cref{subsec:local-commuting-hamiltonians-and-davies-generators-on-lattices} a uniform polynomial lower bound on the local gap with constants $C$ and $\mu$ implies a uniform lower bound on the cMLSI constant of the form
    \begin{equation}
        \alpha_c(\cL_A) \ge \frac{C}{2\log 10 + 2(2\beta g J + 3\log d) |A\partial|} |A|^{-\mu} \ge \Omega(|A|^{-\mu-1}) \, .
    \end{equation}
\end{lem}
\begin{proof}
    The bound is a consequence of \eqref{eq:cmlsi-estimate-via-gap} and a suitable bound on the Pimsner Popa index in the setting of local Davies generators. Let us denote with $E_A = \lim\limits_{t \to \infty} e^{t\cL_A}$. In \cite{Gao.2022a} the authors showed that 
    \begin{equation}
        C_{\cb}(E_A) \le \norm{\tau^{-1}}_\infty \sum\limits_{i=1}^n d_{\cK_i}^2 \, ,
    \end{equation}
    where 
    \begin{equation}
        E_A(\rho) = \bigoplus_{i = 1}^n \tr_{\cK_i}[P_i \rho P_i] \otimes \tau_i \quad \text{and} \quad  \tau = \bigoplus_{i = 1}^n \Id_{\cH_i} \otimes \tau_i \, . 
    \end{equation}
    is the decomposition of the Davies channel. As $E_A$ only acts non-trivially on $A\partial$, we can conclude that $\sum_{i=1}^n d_{\cK_i} \le d_{A\partial} = d^{|A\partial|}$. By superadditivity of $x \mapsto x^2$, we hence get $\sum_{i = 1}^{n} d_{\cK_i}^2 \le \big(\sum_{i = 1}^n d_{\cK_i}\big)^2 \le d^{2|A\partial|}$. To estimate $\norm{\tau^{-1}}$ note that 
    \begin{equation*}
        E_{A}(d_{A\partial}^{-1} \Id_\Lambda) = \bigoplus_{i = 1}^n \frac{d_{\cK_i}}{d_{A\partial}} \Id_{\cH_i} \otimes \tau_i \le \bigoplus_{i = 1}^n \Id_{\cH_i} \otimes \tau_i = \tau \, ,  
    \end{equation*}
    hence $\norm{\tau^{-1}}_\infty \le d^{|A\partial|}\norm{(E_A(\Id_\Lambda))^{-1}}_\infty \le d^{|A\partial|} e^{2\beta gJ |A\partial|}$. The last inequality follows by the same argument as in the proof of \cref{lem:extended:relation-davies-at-finite-and-infinite-temperature}.
\end{proof}

\begin{thm}[weak MLSI under polynomial gap]\label{thm:extended:wmlsi-under-gap}
    In the setting of \cref{subsec:local-commuting-hamiltonians-and-davies-generators-on-lattices} assume that the local Davies generators satisfy the polynomial local gap assumption for some $C>0, \mu\in\mathbb{N}_0$ and further that the Gibbs state satisfies MCMI decay with constants $K,\xi$ both for a fixed temperature $\beta$, then the semigroup $\{e^{t\cL^D_{\Lambda_L}}\}_{t\geq 0}$ at that temperature satisfies the following weak MLSI for $\epsilon\geq KD2^{D}(2L+1)^D\exp(\frac{r}{\xi}-\frac{L}{D\xi})= \cO(L^De^{-\cO(L)})$.
    \begin{align}
        D(e^{t\cL^D_{\Lambda_L}}(\rho)\|\sigma)\leq e^{-\alpha(\epsilon)t}D(\rho\|\sigma)+\epsilon
    \end{align} 
    with $\frac{1}{\alpha(\epsilon)}\leq \frac{D+1}{C}(5+4\beta gJ+8\log d)\left(2D\left(r+\xi\log\frac{KD^2DN}{\epsilon}\right)\right)^{D(1+\mu)}= \cO\left(\left(\log\frac{N}{\epsilon}\right)^{D(1+\mu)}\right)$, where $\rho_t = e^{t\cL_{\Lambda_L}^D}(\rho)$ for arbitrary $\rho \in \cS(\cH_\Lambda)$.
\end{thm}

\begin{rmk}\label{rmk:extended:very-high-temperature-constant-almost-cmlsi}
    Note that instead of a uniform polynomial local gap, one can also require very high temperature, i.e. $\beta \sim \frac{1}{N}$ leading to a constant correction in \cref{lem:extended:relation-davies-at-finite-and-infinite-temperature}, giving a result analogous to the one above. This means at such very high temperatures, one gets the strengthened \cref{thm:extended:rapid-mixing} only from the uniform decay of MCMI.
\end{rmk}

\begin{proof}[Proof of \cref{thm:extended:wmlsi-under-gap}]
We fix a valid coarse-graining with constants $(k,c,l)$ to be chosen later. Then by \cref{thm:a-weak-approximate-tensorization-for-davies-channels} and the MCMI decay assumed in the statement of the theorem we have
\begin{align*}
    D(\rho\|\sigma)&\leq \sum_{a,i}D(\rho\|E_{C_{a,i}}(\rho))+KD2^DNe^{-\frac{c}{\xi}} \\ &\leq \sum_{a,i}\frac{1}{\alpha(\cL_{C_{a,i}})}\EP_{\cL_{C_{a,i}}}(\rho)+KD2^DNe^{-\frac{c}{\xi}} \\ &\leq
    \frac{D+1}{C}[2\log10+(4\beta gJ+6\log d)\ell^D]\ell^{D\mu} \EP_{\cL_\Lambda}(\rho)+KD2^DNe^{-\frac{c}{\xi}}.
\end{align*}
Now choosing $k=r, c=\xi\log\frac{KD2^DN}{\epsilon}=\cO\left(\log\frac{N}{\epsilon}\right)$, and $\ell=2D(r+c)+1=\cO\left(\log\frac{N}{\epsilon}\right)$ gives a valid coarse-graining iff $c,\ell=2D(k+c)+1\leq 2L+1$. This is equivalent to $c\leq \frac{L}{D}-r$ and thus $\epsilon\geq KD2^DN\exp(-\frac{c}{\xi})\geq KD2^D(2L+1)^D\exp(\frac{r}{\xi}-\frac{L}{D\xi})$.
We get
\begin{align*}
    D(\rho\|\sigma) \leq \alpha(\epsilon)\EP_{\cL_\Lambda}(\rho) + \epsilon,
\end{align*} where
\begin{align*}
    \frac{1}{\alpha(\epsilon)}&=\frac{D+1}{C}[2\log10+(4\beta gJ+6\log d)\ell^D]\ell^{D\mu} \\ 
    &\leq \frac{D+1}{C}\ell^{D(1+\mu)}(5+4\beta gJ+6\log d)\\
    &=  \frac{D+1}{C}(5+4\beta gJ+6\log d)\left(2D(r+\xi\log\frac{KD2^DN}{\epsilon})+1\right)^{D(1+\mu)} \\ &=\cO\left(\left(\log\frac{N}{\epsilon}\right)^{D(1+\mu)}\right),
\end{align*} where we used that $\ell\geq 1$ and $\max\{2\log10,6\log d\} < 8\log d$ with $d\geq 2$. Integration now gives the statement of the theorem. 
\end{proof}

\section{Main results - extended}\label{sec:main-results-extended}
\subsection{\texorpdfstring{$W_1$}{W1}-mixing from MCMI-decay}\label{subsec:the-w1-mixing-from-mcmi-decay}
Here we state and prove the main result of this work concerning the quasi-rapid $W_1$-decay.

\begin{thm}[Quasi rapid Wasserstein mixing]\label{thm:detailed:wassersteinmixing}
Let $\cL^D_\Lambda$ be a Davies Lindbladian at inverse temperature $\beta > 0$ corresponding to a $(\kappa,r)-$local, $J$-bounded, commuting Hamiltonian $H_\Lambda$. Denote the growth constant of $H_\Lambda$ on $\Lambda=\llbracket-L,L\rrbracket^D$ with $g$. Then if the Gibbs state of $H_\Lambda$ with the same temperature (the invariant state of $\cL^D_\Lambda$) satisfies uniform exponential decay of the MCMI with constants $K,\xi>0$ the semi group generated by $\cL^D_\Lambda$ is quasi-rapidly mixing in normalized $W_1$-distance with mixing time
\begin{align*}
t_{\text{mix}}^{W_1}(\varepsilon) &= \frac{(D+1)e^{2gJ}}{\chi^0_{\min}}\\ &\times \exp(4\beta gJ\left[2D\left(2r+\xi\log\left\{\frac{64KD(D+1)2^D}{\varepsilon^2}\left(2D\left(r+\xi\log\frac{8D2^DKN}{\varepsilon^2}\right)+1\right)^{2D}\right\}\right)+1\right]^D) \\ &\times \log\left\{\frac{64(2\beta gJ+\log d)(D+1)}{\varepsilon^2}\left(2D\left(r+\xi\log\left\{\frac{8NKD2^D}{\varepsilon^2}\right\}\right)+1\right)^{2D}\right\} \\ &=\cO(\exp(\poly\left(\log\left(\frac{1}{\varepsilon^2}\log\frac{N}{\varepsilon^2}\right)\right)))
\\ &= \textup{quasi-poly}\left(\frac{1}{\varepsilon^2}\poly\log\frac{N}{\varepsilon^2}\right)= \textup{quasi-poly}(\varepsilon^{-1})_{\varepsilon\to 0}\textup{quasi-log}(N)_{N\to \infty}
\end{align*} 
whenever $\varepsilon\geq 8N\sqrt{(D+1)KD2^D}\exp(\frac{r}{\xi}-\frac{N^{\frac{1}{D}}-1}{4D\xi}) = \cO\left(N\exp(-\cO(^D\sqrt{N}))\right)$ is fixed.
\end{thm}
\begin{rmk}\label{rmk:minimalepsilon}
    Note that the requirement of $\epsilon\geq \cO\left(N\exp(-\cO(^D\sqrt{N}))\right)$ is not of relevance in implying quasi-rapid mixing, since the property of quasi-rapid mixing is only determined by the asymptotic scaling of the mixing time $t^{W_1}_{\text{mix}}(\varepsilon)$ in the system size, for any fixed $\varepsilon$. The asymptotics are, however, not affected by this requirement since they hold eventually for any fixed $\varepsilon>0$ as $\lim_{N\to\infty}\cO(N\exp(-\cO(^D\sqrt{N})))=0$.
    It is also a crude bound on the actual tight one scaling as $\varepsilon\geq \cO\left((\log N)^D\exp(-\cO(^D\sqrt{N}))\right)$, as can be seen in the proof.
\end{rmk}

The proof will follow from the combined application of \cref{thm:extended:weak-transport-cost-inequality} and \cref{thm:extended:weak-mlsi}, taking care of the explicit constants along the way. 
\begin{proof}[Proof of \cref{thm:detailed:wassersteinmixing}]
We set $N := |\Lambda|$ in the following. First, we choose a valid $(k,c,\ell)$ coarse-graining w.r.t which the weak AT \cref{thm:a-weak-approximate-tensorization-for-davies-channels} implies the following weak TC due to \cref{thm:extended:weak-transport-cost-inequality}.    
\begin{align*}
    \|\rho_t-\sigma\|_{W_1}&\leq \max_{a,i}2\sqrt{2}|C_{a,i}|\sqrt{N(D+1)D(\rho_t\|\sigma)}+N\sqrt{2D2^DKN}e^{-\frac{c}{2\xi}} \\
    &\leq 2\sqrt{2}\ell^D\sqrt{D+1}\sqrt{ND(\rho_t\|\sigma)}+N\sqrt{N}\sqrt{2D2^DK}e^{-\frac{c}{2\xi}}.
\end{align*}
Now choosing $k=r, c=\xi\log\frac{2D2^DKN}{\delta^2}$ and $\ell =2D(r+c)+1=2D\left(r+\xi\log\frac{2D2^DKN}{\delta^2}\right)+1=\cO\left(\log\frac{N}{\delta^2}\right)$,
\noindent yields
\begin{align*}
    \|\rho_t-\sigma\|_{W_1} \leq 2\sqrt{2(D+1)}\sqrt{Nl^{2D}D(\rho_t\|\sigma)}+N\delta.
\end{align*}
Inserting the following weak MLSI from \cref{thm:extended:weak-mlsi}, which is implied by the uniform MCMI decay:
\begin{align*}
   D(\rho_t\|\sigma)\leq e^{-\alpha(N\ell^{-2D}\epsilon)t}D(\rho\|\sigma)+\frac{N}{\ell^{2D}}\epsilon \leq  Ne^{-\alpha(N\ell^{-2D}\epsilon)t}(2\beta gJ+\log d)+\frac{N}{\ell^{2D}}\epsilon,
\end{align*} where
\begin{align*}
\frac{1}{\alpha(N\ell^{-2D}\epsilon)}&=\frac{D+1}{\chi^0_{\min}}e^{2gJ}e^{4\beta gJ\left[2D\left(2r+\xi\log\frac{KD2^D\ell^{2D}}{\epsilon}\right)+1\right]^D} \\ &=\frac{D+1}{\chi^0_{\min}}e^{2gJ}e^{4\beta gJ\left[2D\left(2r+\xi\log\frac{KD2^D\left(2D\left(r+\xi\log\frac{2D2^DKN}{\delta^2}\right)+1 \right)^{2D}}{\epsilon}\right)+1\right]^D}.
\end{align*}
Combining this with the weak MLSI above yields
\begin{align*}
    \|\rho_t-\sigma\|_{W_1}\leq 2\sqrt{2}\sqrt{D+1}N\sqrt{e^{-\alpha(N\ell^{-2D}\epsilon)t}\ell^{2D}(2\beta gJ+\log d)+\epsilon}+N\delta \, .
\end{align*} So for times
\begin{align*}
    t&\geq \frac{1}{\alpha(N\ell^{-2D}\epsilon)}\log\frac{(2\beta gJ+\log d)\ell^{2D}}{\epsilon} \\&= \frac{D+1}{\chi^0_{\min}}e^{2gJ}e^{4\beta gJ\left[2D\left(2r+\xi\log\left\{\frac{KD2^D}{\epsilon}\left(2D\left(r+\xi\log\frac{2D2^DKN}{\delta^2}\right)+1 \right)^{2D}\right\}\right)+1\right]^D} \\ &\quad \times \log\left\{\frac{(2\beta gJ+\log d)}{\epsilon}\left(2D\left(r+\xi\log\frac{2D2^DKN}{\delta^2}\right)+1\right)^{2D}\right\} \\
    &=\cO\left(\exp(\poly_D\left(\log\frac{1}{\epsilon}\log\frac{N}{\delta^2}\right))\right),
\end{align*} it holds that
\begin{align*}
    \|\rho_t-\sigma\|_{W_1}\leq 2\sqrt{2(D+1)}N\sqrt{2\epsilon}+N\delta \leq N\varepsilon,
\end{align*} where we set $\epsilon:=\frac{\delta^2}{16(D+1)}$ and $\delta=\frac{\varepsilon}{2}$ completes the proof with the mixing time claimed in the statement of the theorem. The bound on the minimal value of comes from the one in \cref{thm:extended:weak-mlsi}, rewritten in terms of $N=(2L+1)^D$, yielding
\begin{align*}
    \delta &\geq \sqrt{2D2^DKN}\exp(\frac{r}{2\xi}-\frac{^D\sqrt{N}-1}{4D\xi}) \\
    \epsilon &\geq KD2^D\ell^{2D}\exp(\frac{2r}{\xi}-\frac{^D\sqrt{N}-1}{2D\xi}).
\end{align*}
Crudely bounding $\ell^D\leq N$ and $\varepsilon=8\sqrt{D+1\epsilon}= 2\delta$ yields the result, i.e. it is easy to check that the in the theorem claimed bound on $\varepsilon$ satisfies both the inequalities above.
\end{proof}

\subsection{Rapid mixing from MCMI-decay and polynomial local gap}\label{subsec:rapid-mixing-from-mcmi-decay-and-polynomial-local-gap}

\begin{thm}[Rapid mixing and hyper rapid $W_1$ mixing under polynomial local gap]\label{thm:extended:rapid-mixing}
    Let $\cL^D_\Lambda$ be a Davies Lindbladian corresponding to a $(\kappa,r)-$local, $J$-bounded, commuting Hamiltonia $H_\Lambda$, to inverse temperature $\beta>0$. Denote the growth constant of $H_\Lambda$ on $\Lambda=\llbracket-L,L\rrbracket^D$ with $g$. Then if the Gibbs state of $H_\Lambda$ to inverse temperature $\beta$ (the invariant state of $\cL^D_\Lambda$) satisfies uniform exponential decay of the MCMI with constants $K,\xi>0$ and the gap of the local Davies generators are at most polynomially decaying in local region size with degree $\mu\in\mathbb{N}_0$, the semigroup generated by $\cL_\Lambda$ is rapidly mixing in trace-distance with mixing time
    \begin{align*}
        t^1_{\text{mix}}(\varepsilon) &= \frac{D+1}{C}(5+4\beta gJ+8\log d)\left(2D\left(r+\xi\log\frac{2KD^2DN}{\varepsilon^2}\right)\right)^{D(1+\mu)}\log\frac{4(2\beta gJ+\log d)N}{\varepsilon^2} \\ &= \cO\left(\left(\log\frac{N}{\varepsilon^2}\right)^{1+D(1+\mu)}\right)
    \end{align*}
    whenever $\varepsilon \geq \sqrt{2KD2^DN}\exp(\frac{r}{2\xi}-\frac{^D\sqrt{N}-1}{4D\xi})= \cO(\sqrt{N}\exp(-\cO(^D\sqrt{N})))$.
    And it is hyper rapidly mixing in normalized $W_1$ distance with mixing time
    \begin{align*}
        t^{W_1}_{\text{mix}}(\varepsilon) &= 
        \frac{D+1}{C}(5+4\beta gJ+8\log d) \\ &\times\left(2D\left(r+\xi\log\left(\frac{64KD2^D}{\varepsilon^2}\left(2D(r+\xi\log\frac{8D2^DKN}{\varepsilon^2})+1\right)^{2D}\right)\right)+1\right)^{D(1+\mu)} \\ &\times \log\left(\frac{64(2\beta gJ+\log d)}{\epsilon^2}\left(2D(r+\xi\log\frac{8D2^DKN}{\varepsilon^2})+1\right)^{2D}\right) \\ &=\cO\left(\left(\log\left(\frac{1}{\varepsilon^2}\log\frac{N}{\varepsilon^2}\right)\right)^{1+D(1+\mu)}\right)
    \end{align*}
    whenever $\varepsilon\geq 8N\sqrt{(D+1)KD2^D}\exp(\frac{r}{2\xi}-\frac{N^{\frac{1}{D}}-1}{4D\xi}) = \cO\left(N\exp(-\cO(^D\sqrt{N}))\right)$ is fixed.
\end{thm}
Note here equally the remark on the minimal value of $\varepsilon(N)$ as \ref{rmk:minimalepsilon}.
\begin{proof}
The rapid mixing part follows directly from \cref{thm:extended:wmlsi-under-gap}, which is implied by the MCMI decay and uniform local gap and an application of Pinsker's inequality. In detail we get
\begin{align*}
    \|\rho_t-\sigma\|_1&\leq \sqrt{2D(\rho_t\|\sigma)} \leq \sqrt{2e^{-\alpha(\epsilon)t}D(\rho\|\sigma)+\epsilon} \\ &\leq \sqrt{2e^{-\alpha(\epsilon)t}N(2\beta gJ+\log d))+\epsilon}.
\end{align*}
Hence for times 
\begin{align*}
    t&\geq \frac{1}{\alpha(\epsilon)}\log\frac{2(2\beta gJ+\log d)N}{\epsilon} \\ &=  \frac{D+1}{C}(5+4\beta gJ+8\log d)\left(2D\left(r+\xi\log\frac{KD^2DN}{\epsilon}\right)\right)^{D(1+\mu)}\log\frac{2(2\beta gJ+\log d)N}{\epsilon}\\&= \cO\left(\left(\frac{N}{\epsilon}\right)^{1+D(1+\mu)}\right),
\end{align*} it holds that
\begin{align*}
    \|\rho_t-\sigma\|_1 \leq \sqrt{2\epsilon}.
\end{align*}
Replacing $\epsilon$ by $\frac{\varepsilon^2}{2}$ yields the first claim. \\
For the $W_1$ mixing proceed just as in the proof of \cref{thm:detailed:wassersteinmixing}, but with \cref{thm:extended:wmlsi-under-gap} instead of \cref{thm:extended:weak-mlsi}. So fix a valid coarse-graining with constants $(k,c,\ell)$ and consider the weak TC from \cref{thm:extended:weak-transport-cost-inequality} w.r.t this coarse-graining.
\begin{align*}
    \|\rho_t-\sigma\|_{W_1}&\leq \max_{a,i}2\sqrt{2}|C_{a,i}|\sqrt{ND(D+1)D(\rho_t\|\sigma)}+N\sqrt{2KD2^DN}e^{-\frac{c}{2\xi}} \\
    &\leq 2\sqrt{2(D+1)}\ell^D\sqrt{N\left(e^{-\alpha(N\ell^{-2D}\epsilon)t}N(2\beta gJ+\log d)+\frac{N}{\ell^{2D}}\epsilon\right)}+ N\sqrt{N}\sqrt{2D2^DK}e^{-\frac{c}{\xi}}.
\end{align*}
Now setting $k=r, c=\xi\log\frac{2D2^DKN}{\delta^2}=\cO\left(\frac{N}{\delta^2}\right)$ and $l=2D(r+c)+1=\cO\left(\frac{N}{\delta^2}\right)$ yields a valid coarse-graining and it is easily checked that for times
\begin{align*}
    t\geq \frac{1}{\alpha(N\ell^{-2D}\epsilon)}\log\left(\frac{2\beta gJ+\log d}{\epsilon}\ell^{2D}\right) = \cO\left(\left(\log\left(\frac{1}{\epsilon}\right)\right)^{1+D(1+\mu)}\right)
\end{align*}
\begin{align*}
    \|\rho_t-\sigma\|_{W_1}\leq 2\sqrt{2(D+1)}N\sqrt{2\epsilon}+N\delta.
\end{align*}
Setting $\epsilon=\frac{\delta^2}{16(D+1)}$ renaming $\delta=\frac{\varepsilon}{2}$ finishes the proof with the in the theorem claimed mixing time. In the first case, the requirement on the minimal size of $\varepsilon$ comes from the one in \cref{thm:extended:wmlsi-under-gap} rewritten in terms of $N=(2L+1)^D$ and applied to $\varepsilon=\sqrt{2\epsilon}$. In the second case, this is analogous to the proof of \cref{thm:detailed:wassersteinmixing}.
\end{proof}
We want to emphasise again that alternatively to the assumption of the existence of a uniform polynomial gap one can also ask for a very high temperature to again reduce the requirements only to the decay of the MCMI (see \cref{rmk:extended:very-high-temperature-constant-almost-cmlsi})

\subsection{Quasi-Optimal Gibbs state preparation from MCMI-decay}\label{subsec:quasi-optimal-gibbs-state-preparation-from-MCMI-decay}

To convert the prior established mixing time bounds of the Davies semi-group into efficiency results of preparing their fixed points, we require (Davies) Lindblad simulation theorems. These give explicit constructions of quantum circuits, with circuit complexity bounds, which approximate $e^{t\cL}$ in diamond norm. Such suitable circuits are for example constructed in \cite{Cleve.2016, Rall.2023, Li.2022, Chen.2023}, with the latter two being the more efficient ones.

Specifically in \cite[Theorem III.2]{Chen.2023} the authors construct a Lindblad simulation algorithm for which it was shown that the complexity of implementing $e^{t\cL^D_{\Lambda}}$ in terms of two-qubits gates, ancilla qubits, and block-encodings of the dissipative part and Hamiltonian part (see \cite{Chen.2023}) scales linear up to poly logarithmic corrections in $\frac{Nt_{\text{mix}}}{\epsilon}$. Combining this Lindblad simulation algorithm we get the following main result.
\begin{thm}[Quasi-optimal sampling from Gibbs states that satisfy MCMI decay]\label{thm:quasi-optimal-sampling-from-Gibbs-states-that-satisfy-MCMI-decay} Let $\sigma$ be a Gibbs states of $(\kappa,r)-$local, commuting, $J$-bounded Hamiltonian on $\Lambda\subset\mathbb{Z}^D$ which satisfies uniform MCMI decay. Then there exists a quantum algorithm (circuit) that outputs an $\epsilon$-close in normalized $W_1$-distance, state using
\begin{enumerate}
    \item $\cO\left(\textup{poly}\log\left(N\textup{quasi-poly}\left(\frac{1}{\epsilon^2}\poly\log\frac{N}{\epsilon^2}\right)\right)\right)= \cO\left(\poly\log N, \poly\log\frac{1}{\epsilon^2}\right)$ ancilla qubits,
    \item $\cO\left(N\textup{quasi-poly}\left(\frac{1}{\epsilon^2}\poly\log\frac{N}{\epsilon^2}\right)\right) = \cO\left(N\textup{quasi-log}(N),\textup{quasi-poly}(\epsilon^{-2})\right)$ two-qubit gates, block encodings of the Hamiltonian $H_\Lambda$ and the dissipative part of $\cL^D_\Lambda$.
\end{enumerate}
\end{thm}
\begin{rmk}
Such algorithms are usually referred to as optimal if the complexity and/or runtime scales as $\cO(N)$ up to polylogarithmic corrections. Since our scaling is $\cO(N)$ up to quasi-logarithmic corrections we call it \emph{quasi-optimal}, since it still scales better than $\cO(N^2)$. 
In \cite{Chen.2023} they also explicitly construct these block encodings for the dissipative part of $\cL^D_\Lambda$ in terms of simpler gates and the jump operators $S^\omega_{\alpha,k}$.
Note though that any efficient Lindbaldian simulation algorithm, such as the one from \cite{Li.2022}, may be used to construct an efficient algorithm to sample from such Gibbs states.
Our rapid mixing results imply optimal efficient preparation, analogous to the above, yielding circuit complexities and runtimes of $\cO(N)$ up to poly-logarithmic corrections.
\end{rmk}

\begin{proof}
    The above bounds of the number of ancilla quits and necessary gates follow directly from implementing a quantum circuit $\mathcal{C}_\epsilon$ that $\epsilon$ approximates $e^{t\cL^D_\Lambda}$ in diamond norm and \cref{thm:detailed:wassersteinmixing}. The former is done in, e.g. \cite[Theorem III.2]{Chen.2023} and gives the in the theorem mentioned bound when substituting their $t$ for $\cO(N)t^{W_1}_{\mix}(\varepsilon)$, since they consider normalized Lindbaldians, see (1.10) in \cite{Chen.2023}. This suffices since any quantum circuit that is $\epsilon$ close in diamond norm to $e^{t\cL}$ %, such as the one from \cite[Theorem III.2]{Chen.2023} 
    also is $\epsilon$-close in stabilized $1\to \tilde{W_1}$ norm, where $\tilde{W_1}$ denotes the normalized $W_1$ distance. I.e. let $\rho$ be the output of this circuit $\mathcal{C}_\epsilon$, then
    \begin{align*}
        \frac{1}{N}\|\rho-\sigma\|_{W_1} &\leq \frac{1}{N}\|\rho-e^{t^{W_1}_{\textup{mix}}(\varepsilon)\cL^D_\Lambda}(\rho)\|_{W_1}+\frac{1}{N}\|e^{t^{W_1}_{\textup{mix}}(\varepsilon)\cL^D_\Lambda}(\rho)-\sigma\|_{W_1} \\&\leq \|\rho-e^{t\cL^D_\Lambda}(\rho)\|_{1} + \varepsilon \leq \|\mathcal{C}_\epsilon-e^{t^{W_1}_{\textup{mix}}(\varepsilon)\cL^D_\Lambda}\|_{\Diamond}\|\rho\|_1+\varepsilon \leq \epsilon+\varepsilon.
    \end{align*}
    We conclude by setting $\epsilon=\varepsilon$ and rescaling.
\end{proof}

\section{Examples}\label{sec:examples}
In this section, we study systems that exhibit MCMI decay, which consequently leads to quasi-rapid Wasserstein mixing, and, under a uniform local polynomial gap condition further to hyper-rapid Wasserstein mixing and rapid mixing. We begin by deriving MCMI decay from the existence of a strong effective Hamiltonian in \cref{subsec:mcmi-decay-from-strong-effective-hamiltonian}, that exhibits a uniform bound in interaction norm. Next, in \cref{subsec:mcmi-decay-from-commutings-marginals-at-high-temperature}, we assume the system Hamiltonian is marginal commuting, which, based on a result from \cite{Bluhm.2024}, directly implies the existence of a strong effective Hamiltonian at high temperatures with a specified decay rate. By applying the result from \cref{subsec:mcmi-decay-from-strong-effective-hamiltonian}, we thus establish MCMI decay at high temperatures. We know that this commuting marginal assumption is fulfilled for every Hamiltonian composed of commuting Pauli strings (e.g. the Toric code in any dimension)\footnote{This result is shown by Sebastian Stengele in some currently unpublished notes.}. Lastly, note that in all the following results, we will not restate that MCMI decay implies quasi-rapid Wasserstein mixing, which, under a uniform local polynomial gap, can be strengthened to hyper-rapid Wasserstein mixing and supplemented by rapid mixing in trace distance. If wanted the constants of the MCMI decay of all results below could be inserted into the result in \cref{subsec:the-w1-mixing-from-mcmi-decay} and \cref{subsec:rapid-mixing-from-mcmi-decay-and-polynomial-local-gap} to obtain explicit decay rates.

\subsection{MCMI-decay from strong effective Hamiltonian}\label{subsec:mcmi-decay-from-strong-effective-hamiltonian}
The following lemma adapts the proof of \cite[Theorem 4.1]{Bluhm.2024} to show the uniform decay of the MCMI (c.f. \eqref{eq:extended:decay-mcmi}) given the existence of an effective Hamiltonian with a uniform bound on its interaction norm.

\begin{lem}[MCMI-decay from locality of effective Hamiltonian]\label{lem:mcmi-decay-from-effective-hamiltonian}
    In the setting of \cref{subsec:local-commuting-hamiltonians-and-davies-generators-on-lattices} assume the existence of a strong effective Hamiltonian for $H_\Lambda$, with a uniform bound on the interaction norm, i.e.
    \begin{equation}\label{eq:bound-interaction-norm}
        \inorm{\tilde H^A}_{\mu} := \sup\limits_{x \in \Lambda} \sum\limits_{X\subseteq \Lambda : x \in X} \Vert\tilde h_X^A\Vert_\infty e^{\mu \diam(X)} \le \Delta
    \end{equation}
    for $\mu, \Delta \ge 0$ independent of $A \subseteq \Lambda$. Then for any partition $\Lambda = A \sqcup B \sqcup C \sqcup D$
    \begin{equation}
        \norm{\textbf{H}(A:C|D)_\sigma}_\infty \le 4 \min\{|A|, |C|\} \Delta e^{-\mu \dist(A, C)} \, . 
    \end{equation}
\end{lem}
\begin{proof}
    Without loss of generality, we can assume that $|A| \le |C|$. Inserting the definition of the MCMI and using 3. from the definition of a strong effective Hamiltonian in \cref{subsec:the-effective-hamiltonian-as-tool-to-proof-correlation-decay} we get 
    \begin{align*}
        \norm{\textbf{H}(A:C|D)_\sigma}_\infty &= \Big\Vert\sum\limits_{X \subseteq \Lambda}\tilde h_X^{ACD} + \tilde h_X^{D} - \tilde h_X^{AD} - \tilde h^{CD}_X\Big\Vert_\infty \\
        &= \Big\Vert\sum\limits_{\genfrac{}{}{0pt}{}{X \subseteq \Lambda:}{X\cap A \neq \emptyset, X \cap C \neq \emptyset}} \tilde h_X^{ACD} + \tilde h_X^{D} - \tilde h_X^{AD} - \tilde h^{CD}_X\Big\Vert_\infty \\
        &\le \sum\limits_{\genfrac{}{}{0pt}{}{X \subseteq \Lambda:}{X\cap A \neq \emptyset, X \cap C \neq \emptyset}} \norm{\tilde h_X^{ACD}}_\infty + \norm{\tilde h_X^{D}}_\infty + \norm{\tilde h_X^{AD}}_\infty + \norm{\tilde h^{CD}_X}_\infty
    \end{align*}
    The first equality stems from the fact that 
    \begin{equation*}
        \tilde h_X^{ACD} + \tilde h_X^{D} - \tilde h_X^{AD} - \tilde h^{CD}_X = 0
    \end{equation*}
    if $X \cap C =\emptyset$ or if $X\cap A = \emptyset$. Indeed, if $X \cap C = \emptyset$, then $X \cap ACD = X \cap AD$ and $X\cap CD = X \cap D$ which respectively give $\tilde h^{ACD}_X = \tilde h^{AD}_X$ and $\tilde h_X^{CD} = \tilde h_X^{D}$ by 2. of \cref{subsec:the-effective-hamiltonian-as-tool-to-proof-correlation-decay}. The argument is analogous for $X \cap A = \emptyset$ by symmetry in $A$ and $C$. Using \eqref{eq:bound-interaction-norm} we obtain
    \begin{align*}
         \norm{\textbf{H}(A:C|D)_\sigma}_\infty &\le \sum\limits_{\genfrac{}{}{0pt}{}{X \subseteq \Lambda:}{X\cap A \neq \emptyset, X \cap C \neq \emptyset}} e^{-\mu \diam(X)} e^{\mu \diam(X)}\left(\norm{\tilde h_X^{ACD}}_\infty + \norm{\tilde h_X^{D}}_\infty + \norm{\tilde h_X^{AD}}_\infty + \norm{\tilde h^{CD}_X}_\infty\right)\\
         &\le \sum\limits_{x \in A} e^{-\mu \dist(x, C)} \sum\limits_{\genfrac{}{}{0pt}{}{X \subseteq \Lambda:}{x \in X, X \cap C \neq \emptyset}} e^{\mu \diam(X)} \left(\norm{\tilde h_X^{ACD}}_\infty + \norm{\tilde h_X^{D}}_\infty + \norm{\tilde h_X^{AD}}_\infty + \norm{\tilde h^{CD}_X}_\infty\right)\\
         &\le 4 \Delta |A| e^{-\mu \dist(A, C)} \, ,
    \end{align*}
    proving the claim.
\end{proof}

\subsection{MCMI-decay from commuting marginals at high temperature}\label{subsec:mcmi-decay-from-commutings-marginals-at-high-temperature}
Assuming that the system Hamiltonian $H_\Lambda$ is marginal commuting now immediately gives an explicit uniform decay of the MCMI going through the result of \cite[Theorem 3.8]{Bluhm.2024}. This is detailed in the theorem below.

\begin{thm}[MCMI decay for marginal commuting systems at high temperature]\label{thm:MCMI-ecay-for-marginal-commuting-systems-at-high-temperature}
    In the setting of \cref{subsec:local-commuting-hamiltonians-and-davies-generators-on-lattices}, assume that $H_\Lambda$ is marginal commuting (c.f \cref{subsec:the-effective-hamiltonian-as-tool-to-proof-correlation-decay} for a definition). Then, for $\beta < \frac{1}{\kappa g(1 + \kappa g) e^2 g J}$ and for every partition $\Lambda = A \sqcup B \sqcup C \sqcup D$, we have
    \begin{equation*}
        \norm{\textbf{H}_\sigma(A:C|D)}_\infty \le 4 \min\{|A|, |C|\} e^{-\mu \dist(A, C)}
    \end{equation*}
    with $\mu = \frac{1}{r} \log\left(\frac{1}{g \kappa (1 + g\kappa)e^2 g J \beta}\right)$.
\end{thm}

\begin{proof}
    Assume that $\beta < \frac{1}{\kappa g(1 + \kappa g) e^2 g J}$ and set $\mu_\epsilon = \frac{1+\epsilon}{r}\log\left(\frac{1}{g \kappa (1 + g\kappa)e^2 g J \beta}\right) > 0$ for $\epsilon > 0$. Then 
    \begin{equation*}
        \inorm{H_\Lambda}_{\mu_\epsilon} \le g J e^{\mu_\epsilon r} < \frac{1}{g \kappa (1 + g \kappa) e^2 \beta} \, . 
    \end{equation*}
    Reordering the above gives 
    \begin{equation}
        \beta < \frac{1}{g \kappa (1 + g \kappa) e^2 \inorm{H_\Lambda}_{\mu_\epsilon}} \le \frac{1}{\mathfrak{d}(1 + \mathfrak{d}) e^2 \inorm{H_\Lambda}_{\mu_{\epsilon}}} \, .
    \end{equation}
    In the last inequality, we used $\mathfrak{d} \le g \kappa$, where $\mathfrak{d}$ is the degree of the interaction graph (see \cite{Bluhm.2024}). Combining this temperature constraint with the commutativity of the generated algebra and its preservation under partial traces, all conditions of \cite[Theorem 3.8]{Bluhm.2024} are satisfied. We can therefore conclude the existence of a strong effective Hamiltonian with decay 
    \begin{equation*}
        \inorm{\tilde H^A}_{\mu_\epsilon} \le 1
    \end{equation*}
    for every $A \subseteq \Lambda$. Now, by \cref{lem:mcmi-decay-from-effective-hamiltonian}, we immediately conclude the claim with decay rate $\mu_\epsilon$. Taking $\epsilon \to 0$ then concludes the claim with the stated decay rate.
\end{proof}

\end{document}